\documentclass[letter,numberedappendix,appendixfloats,iop,tighten]{emulateapj}
\usepackage{apjfonts}
\usepackage{graphicx}
\usepackage{chngcntr}

\newcommand{\oi}{[O\,{\sc i}]}
\newcommand{\oii}{[O\,{\sc ii}]}
\newcommand{\oiii}{[O\,{\sc iii}]}

\newcommand{\nii}{[N\,{\sc ii}]}

\newcommand{\sii}{[S\,{\sc ii}]}
\newcommand{\siii}{[S\,{\sc iii}]}
\newcommand{\siv}{[S\,{\sc iv}]}

\newcommand{\hei}{He\,{\sc i}}

\newcommand{\neii}{[Ne\,{\sc ii}]}
\newcommand{\neiii}{[Ne\,{\sc iii}]}

\newcommand{\ariii}{[Ar\,{\sc iii}]}

\newcommand{\cliii}{[Cl\,{\sc iii}]}

\newcommand{\feiii}{[Fe\,{\sc iii}]}

\newcommand{\ha}{H$\alpha$}
\newcommand{\hb}{H$\beta$}
\newcommand{\hi}{H\,{\sc i}}

\newcommand{\fii}{[F\,{\sc ii}]}

\usepackage{color}
\newcommand{\kms}{km s$^{-1}$}

\newcommand{\te}{$T_{\epsilon}$}
\newcommand{\Ne}{$n_{\epsilon}$}

\received{}
\revised{}
\accepted{}

\slugcomment{}
\shorttitle{The origin and evolution of the halo PN K648}
\shortauthors{Otsuka et al.}

\begin{document}

\title{Chemical Abundances and Dust in the Halo Planetary Nebula K648 in
M15:\\ Its Origin and Evolution based on an Analysis of Multiwavelength Data}

\author{Masaaki Otsuka\altaffilmark{1,$\dagger$}, Siek Hyung\altaffilmark{2,3}, Akito Tajitsu\altaffilmark{4}}

\affil{$^{1}$Institute of Astronomy and Astrophysics, Academia Sinica
P.O. Box 23-141, Taipei 10617, Taiwan, Republic of China; otsuka@asiaa.sinica.edu.tw}
\affil{$^{2}$School of Science Education (Astronomy), Chungbuk National
University, CheongJu, Chungbuk 361-763, Republic of Korea}
\affil{$^{3}$Department of Astronomy, University of Illinois at Urbana-Champaign, Urbana, IL, 61801, U.S.A.}
\affil{$^{4}$Subaru Telescope, NAOJ, 650 North A'ohoku Place, Hilo,
HI 96720, U.S.A.}
\altaffiltext{${\dagger}$}{Current address: Subaru Telescope, NAOJ, 650 North A'ohoku Place, Hilo,
HI 96720, U.S.A.; otsuka@naoj.org}

\begin{abstract}
We report an investigation of the extremely metal-poor and C-rich
 planetary nebula (PN) K648 in the globular cluster M15 using the UV to
 far-IR data obtained using the Subaru, \emph{HST}, \emph{FUSE},
 \emph{Spitzer}, and \emph{Herschel}. We determined the nebular
 abundances of ten elements. The enhancement of F ([F/H]=+0.96) is
 comparable to that of the halo PN BoBn1. The central stellar abundances
 of seven elements are determined. The stellar C/O ratio is similar to
 the nebular C/O ratios from recombination line and from
 collisionally excited line (CEL) within error,
 and the stellar Ne/O ratio is also close to the nebular CEL Ne/O
 ratio. We found evidence of carbonaceous dust grains and molecules
 including Class B 6-9\,$\mu$m and 11.3\,$\mu$m polycyclic aromatic hydrocarbons and the
 broad 11\,$\mu$m feature. The profiles of these bands are similar to
 those of the C-rich halo PNe H4-1 and BoBn1. Based on the theoretical
 model, we determined the physical conditions of the gas and dust and
 their masses, i.e., 0.048 $M_{\odot}$ and 4.95$\times$10$^{-7}$
 $M_{\odot}$, respectively. The observed chemical abundances and gas
 mass are in good agreement with an asymptotic giant branch
 nucleosynthesis model prediction for stars with an initial 1.25
 $M_{\odot}$ plus a 2.0$\times$10$^{-3}$ $M_{\odot}$ partial mixing zone
 (PMZ) and stars with an initial mass of 1.5 $M_{\odot}$ without a
 PMZ. The core-mass of the central star is approximately 0.61-0.63
 $M_{\odot}$. K648 is therefore likely to have evolved from a progenitor
 that experienced coalescence or tidal disruption during the early
 stages of evolution, and became a $\sim$1.25-1.5 $M_{\odot}$ blue straggler. 
\end{abstract}

\section{Introduction}

Planetary nebulae (PNe) represent a stage in the evolution of
initial $\sim$1-8 $M_{\odot}$ stars. At the end of their evolution,
such stars evolve into asymptotic giant branch (AGB) stars,
then PNe, and finally white dwarves (WD). During this process of evolution, these
stars eject a large fraction of their mass into the interstellar medium.
The history of the progenitors is imprinted in
the central star of the PN (CSPN) and the ejected gas.
An investigation of the CSPN and the ejected material provides useful information to
increase our understanding of stellar evolution, as well as the
chemical evolution of galaxies, i.e., how much of the mass of the star
becomes a PN, which and how much of the elements are synthesized in the inner core
of the progenitor, and how galaxies become chemically rich.
The ejected gas in the PNe consists of both processed and unprocessed matter:
primordial sources of proto-star cluster clouds or intracluster medium,
pollution sources from highly evolved stars AGB and
supernovae (SNe), and the result of stellar evolution processes
(nucleosynthesized elements, molecules, and dust).
Our understanding of the evolution of low-mass stars formed in the early Galaxy, as well as the chemical evolution of the Galaxy, can be enhanced by studying metal-poor PNe
located in the Galactic halo.

Fourteen Galactic halo PNe have been identified since the
discovery of K648 in M15
\citep[e.g.,][]{1997MNRAS.284..465H,Jacoby:1997aa,Pequignot:2005aa,Pereira:2007aa}.
Recently, the number of detections has steadily increased due to
the Sloan Digital Sky Survey (SDSS) \citep{Yuan:2013aa}. Five PNe
are located in the globular clusters (GCs) M15 (K648), M22 (GJJC1
and M2-29), Pal6 (JaFu1), and NGC6441 (JaFu2), and others are
located in the Galactic halo field. The classification of PNe
based on chemical abundances was originally proposed by
\citet{Peimbert:1978aa}, and has recently been revised and
updated, e.g., \citet{Quireza:2007aa}. Halo PNe are classified as
Type\,IV; specifically, \citet{Costa:1996aa} indicated that halo
PNe exhibit a large vertical distance from the Galactic plane
($\langle z \rangle$ = 7.2 kpc) and large peculiar velocity
relative to the rotation of the Galaxy ($\langle{\Delta}V \rangle$
= 173 {\kms}, see their Table~6). Among halo PNe, H4-1
\citep{Tajitsu:2014aa,Otsuka:2013aa}, BoBn1
\citep{2010ApJ...723..658O}, and K648 \citep{Kwitter:2003aa} are
extremely metal-poor and C-rich ($\langle$[Ar/H]$\rangle$ =
--2.03, $\langle$C/O$\rangle$ = 14.49; this work); furthermore,
there is an unresolved issue in terms of the chemical abundances:
how did these progenitors evolve into C-rich PNe? The scientific
backgrounds of these PNe were explained by
\citet{2010ApJ...723..658O} and by \citet{Otsuka:2013aa}. The
progenitors of these three halo PNe were probably $\sim$0.8
$M_{\odot}$ stars, corresponding to the typical mass of turn-off
stars in M15, because the [Ar/H] abundances as a metallicity
indicator are similar to the typical [Fe/H] abundance in M15;
according to \citet{Kobayashi:2011aa}, [Ar/H]$\sim$--2.03
corresponds to [Fe/H]$\sim$--2.3. At least some of the stars
of the Milky Way's  stellar halo were accreted along with
their parent dwarf galaxies. BoBn1, a member of the oldest population in
the Sagittarius dwarf spheroidal galaxy
\citep{2006MNRAS.369..875Z} and H4-1 in the halo field, 
might be younger than the classical Milky Way stellar halo
population.

For low-mass stars to evolve into
C-rich PNe, a third dredge-up (TDU) is essential during
the thermal pulse (TP) AGB phase. TDU conveys the He-shell reaction
products, including C, O, Ne, and neutron ($n$) capture elements, to the
stellar surface. It is widely believed
that $\gtrsim$1-1.5 $M_{\odot}$ stars experience TDU
\citep[e.g.][]{Lattanzio:1987aa,2010MNRAS.403.1413K}.
Recently \citet{Lugaro:2012aa} reported the occurrence of TDU
in initial 0.9 $M_{\odot}$ stars with a metallicity of
$Z$ = 10$^{-4}$, although the minimum mass required for TDU
depends on the model used.
Even if TDU took place in the $\sim$0.9 $M_{\odot}$ progenitors,
the post-AGB evolution of such low-mass stars toward the hot WDs
is very slow. In addition, the ejected mass itself is very small, so
it is difficult to observe them as visible PNe.
Hence, the most likely explanation is that these progenitors gained
mass via binary interactions to
create new conditions for evolving into C-rich PNe.

\begin{figure}
\includegraphics[width = \columnwidth,clip]{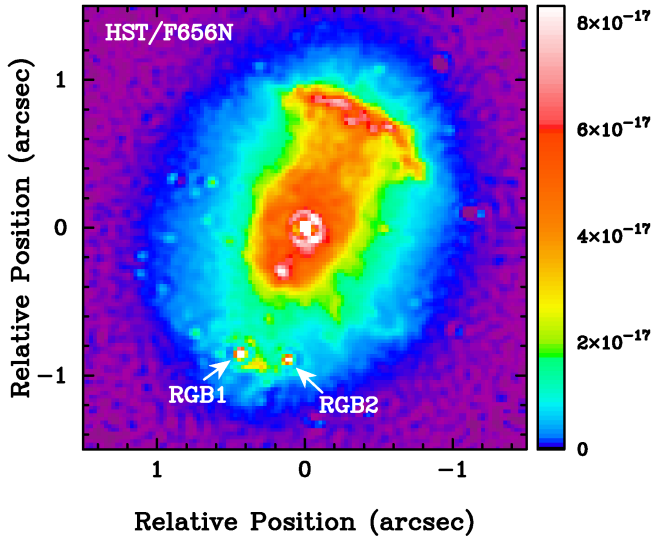}
\caption{The H$\alpha$ image of K648 taken by the \emph{HST}/WFPC2 with
the F656N filter. North is up and east is to the left.
The intensity is in erg s$^{-1}$ cm$^{-2}$ {\AA}$^{-1}$. Bright stars
close to K648 are subtracted out. See Section \ref{hstphot} for details
of the methods employed. The locations of two RGB stars are
indicated by the arrows. The reddening corrected
magnitudes are 17.35 ($B$) and 16.78 ($V$) in RGB1 and 17.54 ($B$) and
17.11 ($V$) in RGB2, respectively.}
\label{hst_ha}
\end{figure}

In view of the internal kinematics and nebular morphology, the progenitor of
K648 appears to be a high-mass star. K648 has bipolar and equatorial outflows
\citep{2006IAUS..234..523T} and asymmetric nebulae \citep{2000AJ....120.2044A}.
In Fig.~\ref{hst_ha}, we show an H$\alpha$ image of K648 obtained using
the \emph{HST}/WFPC2. This image was processed using Lucy-Richardson deconvolution.
K648 is composed of three parts: a very bright inner elliptical
shell, an outer elliptical shell, and a bright arc on the northwestern
limit of the major axis of the nebula, located just inside the edge of the
outer bright elliptical shell. The arc is especially prominent in this
object. A corresponding feature at the other end of the major axis does
not appear to be present, although two fairly bright red giant branch
(RGB) stars are unfortunately superposed at this location, making it difficult to resolve this feature. The locations of these RGB stars are indicated by the
white arrows in the figure. The faint halo surrounding the outer
elliptical shell extends to a radius of $\sim$2.1{\arcsec} \citep[not
shown here, see Fig.~2 of][]{2000AJ....120.2044A}.
The major axis of the inner and outer shells is along
the position angle of --27$^{\circ}$.
\citet{1999ApJ...517..767G} theoretically predicted that bipolar nebulae
can be created in initial $\geq$1.3 $M_{\odot}$ single stars.
We will explore the possibility of a binary system related mass-transfer activity suggested by
\citet{2000AJ....120.2044A}, to solve  the C abundance problem and the
apparent contradiction in the evolutionary timescale.

It would be interesting to study whether an increased mass star would evolve
into a C-rich PN through such an evolutionary route.
In AGB nucleosynthesis models, the predicted abundances, in particular
$n$-capture elements, depend on the TDU efficiency, the number of thermal pulses,
and the $^{13}$C pocket mass. Any $n$-capture elements have
not yet been detected in K648. The Ne abundance is also
sensitive to the amount of $^{13}$C pocket mass \citep{Shingles:2013aa}.
The Ne abundances can be easily determined using atomic gas phase emissions from the
PNe rather than stellar absorption. To obtain a detailed view of the
origin and evolution of K648 through comparison with AGB nucleosynthesis
models, we must accurately determine the abundances
of C, O, Ne, and $n$-capture elements, and estimate the ejected mass.
K648 is an ideal laboratory in which to investigate the evolution of
low-mass metal-poor stars, as well as their nucleosynthesis. The reasons for this are first that the upper mass limit of stars in M15 is known ($\lesssim$1.6 $M_{\odot}$), and second that, because the distances are known with relatively little uncertainty, it is possible to determine the core-mass of the PN as well as of the ejected mass. Study of K648 benefits not only understanding of the
evolution of low-mass metal-poor stars, but also dust production in these stars.

In this paper, we describe detailed spectroscopic analyses of K648
to investigate the origin and evolution of this PN based on
an extensive set of spectroscopic/photometric data from the far-UV to
far-infrared (FIR) regions of the electromagnetic spectrum. The
remainder of the paper is organised as follows. In Section 2, we
describe these observations using the Subaru/HDS, \emph{HST}/WFPC2/FOS/COS, \emph{Spitzer}/IRS/IRAC/MIPS, and \emph{Herschel}/PACS.
In Section 3, we provide the elemental abundances of the nebula and the
CSPN, as well as the physical properties of the CSPN. We determined the abundances of the 10 elements of the nebula of K648, including the first measurements of the $n$-capture element
fluorine (F) in this PN.
Using the spectrum synthesis code {\sc TLUSTY} \citep{Lanz:2003aa}, we determined the abundances of 7
elements of the CSPN and the core-mass of the CSPN.
We also report the C-rich dust features found in the \emph{Spitzer}/IRS
spectrum.
We constructed a self-consistent model, whereby the predicted spectral
energy distribution (SED) fits the observations and accordingly estimated
the mass of ejected gas and dust using the radiative transfer code {\sc
CLOUDY} \citep{Ferland:1998aa}. In Section 4, we compare the elemental
abundances of K648 with those of H4-1 and BoBn1. We discuss the origin and
evolution of K648 by comparing the predicted elemental
abundances, the final core-mass, and the ejected mass reported by \citet{Lugaro:2012aa}
with our determined values. A summary is presented in Section 5.

\section{Observations \& data reduction}

\subsection{HDS observations}

\begin{figure}
\centering
\includegraphics[width = \columnwidth,bb = 36 144 580 684,clip]{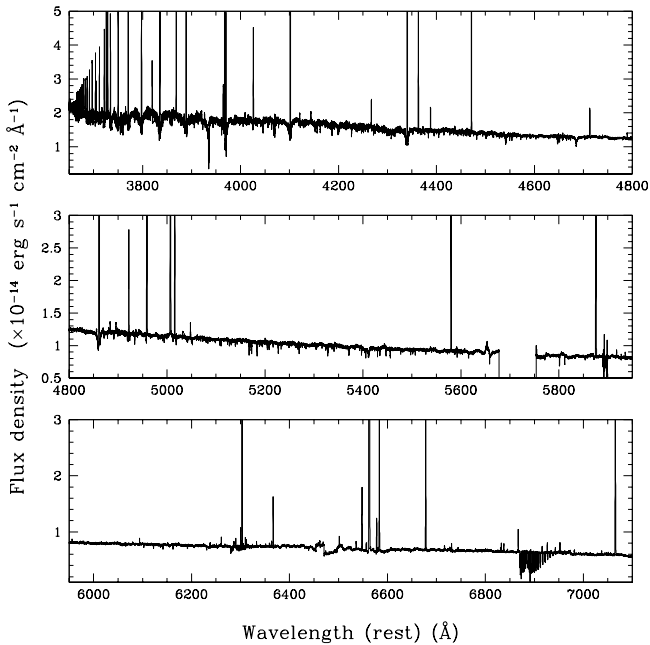}
\caption{The de-reddened HDS spectrum of K648. The wavelength is
corrected to the rest wavelength in air.\label{hds}}
\end{figure}

Optical high-dispersion spectra were obtained using a High-Dispersion Spectrograph
\citep[HDS;][]{2002PASJ...54..855N} attached to the Nasmyth focus of the
8.2-m Subaru telescope on 2012 June 28 (Program ID: S12A-078, PI:
M.~Otsuka).

The weather conditions were stable and clear throughout the night, and
the seeing was $\sim$0.5{\arcsec} measured using the guider CCD. An
atmospheric dispersion corrector (ADC) was used to minimize the differential
atmospheric dispersion throughout the broad wavelength region.
The slit width was set to 1.2{\arcsec} and the slit length was set to 6{\arcsec};
these settings allowed us to reduce contamination from nearby stars.
We selected 2$\times$2 on-chip binning.
The resolving power ($R$) was 33\,500, determined from
the average full-width at half-maximum (FWHM) of over 600 Th-Ar
comparison lines. Blue-cross and the red-cross
dispersers were employed to obtain the 3620-5400\,{\AA} spectrum (blue spectrum) and
the 4320-7140\,{\AA} spectrum (red spectrum), respectively. We set the
position angle to --27$^{\circ}$ using an image de-rotator. The
total exposure times were 7200 s for the blue spectrum and 9000 s for
the red spectrum, respectively. Flux
calibration, blaze function correction and airmass correction were carried out by
observing the standard star BD+28$^{\circ}$ 4211 twice at different
airmasses for each blue and red spectrum.

Data reduction was carried out using the
Echelle Spectra Reduction Package {\sc ECHELLE} in {\sc IRAF}\footnote[5]{IRAF
is distributed by the National Optical Astronomy Observatories,
operated by the Association of Universities for Research in
Astronomy (AURA), Inc., under a cooperative agreement with the National
Science Foundation.}. We generated a single 3620-7140\,{\AA} spectrum by combining the blue and the red spectra after scaling
the flux density of the blue spectrum by a factor of 1.06 to match that of the red
spectrum.
The resulting signal-to-noise (S/N) ratio was $>$50-130 for the continuum of
this single 3620-7140\,{\AA} spectrum.

Figure~\ref{hds} shows the resulting spectrum, which is corrected
for interstellar extinction (see the following section). The observed
wavelength was corrected to the
averaged line-of-sight heliocentric radial velocity of --116.89$\pm$0.41
{\kms} (the root mean square (RMS) of the residuals was 4.15 {\kms}) among all lines detected in
the HDS spectrum (122 lines).

\subsection{Interstellar reddening correction of the HDS spectrum}

The line-fluxes were de-reddened using the follow expression:
\begin{equation}
I(\lambda) = F(\lambda)\times10^{c({\rm H}\beta)(1+f(\lambda))},
\end{equation}
\noindent
where $I$($\lambda$) is the de-reddened line
flux, $F$($\lambda$) is the observed line flux,
$f$($\lambda$) is the interstellar extinction function at $\lambda$
computed by the reddening law reported by \citet{1989ApJ...345..245C} with
$R_{V}$ = 3.1, and $c$(H$\beta$) is the reddening coefficient at H$\beta$.
Our measured $F$({\hb}) was 1.70$\times$10$^{-12}$$\pm$3.14$\times$10$^{-14}$ erg s$^{-1}$
cm$^{-2}$ in the HDS spectrum. Hereafter, X(--Y) corresponds to X$\times$10$^{\rm
-Y}$. We computed $c$({\hb}) by comparing
the observed Balmer line ratios of H$\gamma$ and H$\alpha$ to H$\beta$
with the theoretical ratios reported by \citet{1995MNRAS.272...41S} with an
electron temperature of {\te} = 10$^{4}$ K and an electron density
of {\Ne} = 10$^{4}$ cm$^{-3}$ with the assumptions of Case B.
The values of $c$({\hb}) were 0.121$\pm$0.027 from the
$F$(H$\gamma$)/$F$(H$\beta$) and 0.148$\pm$0.008 from the
$F$(H$\alpha$)/$F$(H$\beta$) ratios. We used an average value of
$c$(H$\beta$) = 0.135$\pm$0.017 for the interstellar reddening correction.

\subsection{Emission-line flux measurements with the HDS spectrum}
The detected emission-lines are given in the Appendix
(see Table~\ref{hdstab}).
For the flux
measurements, we applied multiple Gaussian component fitting.
We list the observed wavelength and de-reddened relative fluxes
for each Gaussian component (indicated by Comp.ID number in the fourth
and the eleventh columns of
Table~\ref{hdstab} in the Appendix), with respect to the
de-reddened H$\beta$ flux of 100. $f$($\lambda$) for each wavelength is
also listed. Most of the line-profiles of the detected
lines can be fitted using a single Gaussian component. For the
lines composed of multiple components, e.g., [O\,{\sc ii}]
$\lambda$\,3726.03\,{\AA}, we list the de-reddened relative fluxes of each
component, as well as the sum of these components (indicated by Tot.).

We supplemented our HDS data with the data given by
\citet{2006IAUS..234..523T} to calculate
the Ar$^{2+}$ abundance using [Ar\,{\sc iii}]\,$\lambda$\,7135\,{\AA},
{\te}({\oii}) and {\Ne}({\oii}) by combining [O\,{\sc
ii}]\,$\lambda\lambda$\,7320/30\,{\AA} with {\oii}\,$\lambda\lambda$\,3726/29\,{\AA}, and
{\te}({\hei}) using He\,{\sc i}\,$\lambda$\,7281\,{\AA}.

\subsection{HST/WFPC2 photometry and the H$\alpha$/H$\beta$ fluxes\label{hstphot}}

\begin{deluxetable*}{@{}lcrccccl@{}}
\centering
\tablecolumns{8}
\tablecaption{\emph{HST}/WFPC2 photometry data for K648. \label{wfpc}}
\tablewidth{\textwidth}
\tabletypesize{\scriptsize}
\tablehead{
\colhead{}&
\colhead{}&
\colhead{}&
\multicolumn{2}{c}{CSPN+Nebula}&
\multicolumn{2}{c}{CSPN}&
\colhead{}\\
\cline{4-5}\cline{6-7}
\colhead{Filter}&
\colhead{$\lambda_{\rm cen.}$}&
\colhead{$\Delta$$\lambda$}&
\colhead{$F_{\lambda}$}&
\colhead{$I_{\lambda}$}&
\colhead{$F_{\lambda}$}&
\colhead{$I_{\lambda}$}&
\colhead{Prop.ID}
\\
\colhead{}&
\colhead{(\AA)}&
\colhead{(\AA)}&
\colhead{(erg s$^{-1}$ cm$^{-1}$ {\AA}$^{-1}$)}&
\colhead{(erg s$^{-1}$ cm$^{-1}$ {\AA}$^{-1}$)}&
\colhead{(erg s$^{-1}$ cm$^{-1}$ {\AA}$^{-1}$)}&
\colhead{(erg s$^{-1}$ cm$^{-1}$ {\AA}$^{-1}$)}&
\colhead{}
}
\startdata
F160BW&1515.16&188.43&1.05(--13)$\pm$2.54(--15)&2.09(--13)$\pm$5.09(--15)&\nodata&\nodata&11975\\
F170W&1820.78&285.52&9.72(--14)$\pm$9.01(--16)&1.89(--13)$\pm$1.75(--15)&9.54(--14)$\pm$7.67(--15)&1.85(--13)$\pm$1.49(--14)&11975\\
F255W&2598.57&171.21&3.01(--14)$\pm$3.95(--16)&5.29(--14)$\pm$6.95(--16)&\nodata&\nodata&10524\\
F300W&2989.04&324.60&2.19(--14)$\pm$2.16(--15)&3.54(--14)$\pm$3.48(--15)&1.43(--14)$\pm$1.28(--15)&2.30(--14)$\pm$2.06(--15)&11975\\
F336W&3359.48&204.49&2.32(--14)$\pm$3.30(--15)&3.56(--14)$\pm$5.06(--15)&1.76(--14)$\pm$4.94(--16)&2.71(--14)$\pm$7.58(--16)&6751\\
F439W&4312.09&202.32&1.11(--14)$\pm$4.19(--15)&1.58(--14)$\pm$5.98(--15)&9.97(--15)$\pm$3.46(--16)&1.42(--14)$\pm$4.93(--16)&6751\\
F547M&5483.88&205.52&4.62(--15)$\pm$9.57(--16)&6.01(--15)$\pm$1.24(--15)&3.84(--15)$\pm$1.04(--16)&4.99(--15)$\pm$1.35(--16)&6751\\
F814W&7995.94&646.13&1.57(--15)$\pm$2.97(--16)&1.84(--15)$\pm$3.47(--16)&1.24(--15)$\pm$5.01(--17)&1.45(--15)$\pm$5.86(--17)&6751\\
F656N&6563.76&53.78&1.04(--13)$\pm$4.33(--16)&1.28(--13)$\pm$5.37(--16)&\nodata&\nodata&6751
\enddata
\tablecomments{$F_{\lambda}$ and $I_{\lambda}$ are the reddened and
de-reddened flux densities, respectively. We used the reddening law reported by \citet{1989ApJ...345..245C}
for interstellar reddening correction with $R_{\rm V}$ = 3.1 and $E(B-V)$ = 0.092.}
\end{deluxetable*}

In the \emph{FOS} UV-spectrum (see the following section),
no H\,{\sc i} or He\,{\sc ii} nebular lines are required to
normalize the C\,{\sc iii}] and [C\,{\sc ii}] fluxes to the {\hb} flux.
Therefore, we measured the {\hb} flux of the entire
nebula and scaled the \emph{FOS} flux density to tune the UV flux
densities at bands including the C\,{\sc iii}]\,$\lambda\lambda$\,1906/09\,{\AA} and
the [C\,{\sc ii}]\,$\lambda$\,2323\,{\AA} lines. The {\hb} flux of the entire
nebula is also necessary to normalize the fluxes of the lines detected in the
\emph{Spitzer}/IRS spectrum (see the following section). Broadband fluxes
are required to estimate the core-mass of the CSPN and to
constrain the incident SED of the CSPN and the emergent spectra predicted by the
nebular model.
For this purpose, we used \emph{HST}/Wide Field and Planetary Camera 2 (WFPC2) photometry using
eight broadband and F656N filters, which are available in the Mikulski Archive for
Space Telescopes (MAST).

We reduced the WFPC2 data (IDs:10524 and 11975, PI:F.~R.~Francesco; ID:6751, PI: H.~E.~Bond) using the standard \emph{HST} pipeline and {\sc
MultiDrizzle} on {\sc PYRAF} to remove cosmic-rays and improve the
angular resolution. First, we removed nearby stars using empirical point-spread functions
(PSFs) generated from {\sc IRAF/DAOPHOT}. We then measured the count rates (cts)
within an aperture radius of 2.1{\arcsec}. We defined the
background sky as being represented by an annulus centered on the CSPN with
inner radius of 3.2{\arcsec} and outer radius of 4.2{\arcsec}.
Finally, we converted the cts into the flux densities using the
PHOTFLAM values in erg s$^{-1}$ cm$^{-2}$ {\AA}$^{-1}$ cts$^{-1}$.
The resulting flux densities $F_{\lambda}$ and the corresponding de-reddened data
$I_{\lambda}$ are listed in the fourth and fifth columns
of Table~\ref{wfpc}.

To measure the H$\alpha$ flux of the entire nebula using the F656N flux density,
it is necessary to remove the contributions of both the local continuum and the [N\,{\sc ii}]\,$\lambda$\,6548
{\AA} line. However, this procedure entails a number of problems (see e.g.,
\citet{Luridiana:2003aa} for a thorough discussion of the pitfalls and
uncertainties in determining line-fluxes from \emph{HST}/WFPC2 images for
the PN NGC6543). An alternative is to use
an intermediate step of computing an equivalent H$\alpha$ flux, that is, using the \emph{Spitzer} H\,{\sc i} Pf$\alpha$ and Hu$\alpha$ recombination lines.
This method also has problems of contamination
due to the 7.7\,$\mu$m PAH feature, as well as the broad 11\,$\mu$m feature, and the
[Ne\,{\sc ii}]\,$\lambda$\,12.80\,$\mu$m line in the \emph{Spitzer}
spectra. Rather than employing the above \emph{HST} H$\alpha$ flux
extraction method, or the intermediate step of using \emph{Spitzer} H\,{\sc
i} lines, we used the \emph{HST}/WFPC2 F656N band flux
density itself, i.e., $F_{\lambda}$(\emph{HST},F656N), which includes the H$\alpha$
flux, the local continuum, and the {\nii}\,$\lambda$6548\,{\AA} line
within the F656N filter band, together with our Subaru/HDS spectrum. The advantage of this
approach is that it is possible to extract the {\ha} flux without
contamination from nebular and stellar continuum and
{\nii}\,$\lambda$6548\,{\AA}. This method was applied in our previous work
on the PN M1-11 \citep{Otsuka:2013ab}.
Taking into account the F656N filter transmission characteristics,
we compared the $F_{\lambda}$(\emph{HST},F656N) with the counterpart
Subaru/HDS scan spectrum, i.e., $F_{\lambda}$(HDS,F656N).
The scaling factor $F_{\lambda}$(\emph{HST},F656N)/$F_{\lambda}$(HDS,F656N) = 1.428 was determined, and was applied to the Subaru/HDS spectral line fluxes
to analyze both the spectra on an equal footing.
After applying the scaling factor, the HDS fluxes should be
$F$(H$\alpha$) = 2.42(--12)$\pm$4.44(--14) erg s$^{-1}$ cm$^{-2}$ and
$F$(H$\beta$) = 7.83(--13)$\pm$1.05(--15) erg s$^{-1}$ cm$^{-2}$. A simple comparison of these
scaled HDS data with the measured HST data shows very small
deviations, i.e., 0.18\%, \& 0.13\%,
corresponding to the {\ha} and {\hb} fluxes measured from {\it
non-scaled} HDS spectrum. The uncertainties of our measurements are much
smaller than the estimated uncertainty of $\sim$10\% reported by
\citet{Luridiana:2003aa}. These reduced errors may be coincidental and the actual errors
could be larger than our estimation; however, the
errors in our analysis appear to be smaller than the estimates reported by
\citet{Luridiana:2003aa}. The scaling factors also give
ratios of Pf$\alpha$ and Hu$\alpha$ with the above H$\beta$ fluxes that are consistent with
the theoretical values (see Section 2.7). Note that
the \emph{Spitzer}/IRS spectra were obtained using a wider slit width, which was sufficient
to cover the entire K648 nebula.

\subsection{HST/FOS UV-spectrum}

\begin{deluxetable}{@{}clcccl@{}}
\centering
\tablecolumns{6}
\tablecaption{The detected lines in the \emph{HST}/FOS spectra.\label{stis}}
\tablewidth{\columnwidth}
\tablehead{
\colhead{$\lambda_{\rm obs}$}&
\colhead{Ion}&
\colhead{$\lambda_{\rm lab}$}&
\colhead{Comp.}&
\colhead{$f$($\lambda$)}&
\colhead{$I$($\lambda$)$^{\rm a}$}\\
\colhead{({\AA})}&
\colhead{}&
\colhead{({\AA})}&
\colhead{}&
\colhead{}&
\colhead{[$I$({\hb}) = 100]}
}
\startdata
1906.83 & C\,{\sc iii}] & 1906/09 & 1 & 1.256 & 334.984$\pm$17.038 \\
2326.45 & [C\,{\sc ii}] & 2323 & 1 & 1.392 & ~~17.091$\pm$1.634
\enddata
\tablenotetext{a}{We used $F$({\hb}) = 7.83(--13) $\pm$ 1.04(--15)
erg s$^{-1}$ cm$^{-2}$, which was measured based on the \emph{HST}/WFPC2
F656N image and the observed
$F$(H$\alpha$)/$F$(H$\beta$) ratio of 3.097 measured using HDS spectra.}
\end{deluxetable}

To calculate the C$^{2+}$ and C$^{+}$ abundances using the C\,{\sc iii}]\,$\lambda\lambda$\,1906/09\,{\AA}
and the [C\,{\sc ii}]\,$\lambda$\,2323\,{\AA} lines, we analyzed the archival
\emph{HST}/FOS spectrum (The Faint Object Spectrograph), which was
obtained on 1993 Nov 18 (Prop.ID: 3196, PI: H.~Ford),
and was downloaded from MAST. We used the data sets Y1C40103P,
Y1C40104T, Y1C40105T, and Y1C40106T.

We scaled the flux density to fit the F160BW, F170W, and F255W
bands listed in the fourth column of Table~\ref{wfpc} using the relevant
transmission curves (scaling factor = 0.837).
Using the $F$({\hb}) = 7.83(--13)$\pm$1.05(--15) erg s$^{-1}$ cm$^{-2}$,
we normalized the C\,{\sc iii}]\,$\lambda\lambda$\,1906/09
{\AA} and the [C\,{\sc ii}]\,$\lambda$\,2323\,{\AA} fluxes.

\subsection{ FUSE and HST/COS UV-spectra}

We analyzed archival UV spectra of K648 from MAST to
calculate the elemental abundances in the photosphere of the CSPN
and determine the parameters required to calculate the stellar radius, surface
gravity $\log$\,$g$, effective temperature $T_{\rm eff}$,
and the current core-mass of the CSPN. The 920-1180\,{\AA} and the
1170-1780\,{\AA} spectra were obtained using the Far
Ultraviolet Spectroscopic Explorer (\emph{FUSE}) on 2004 Nov 1 (data set:
D1570101000, PI: Dixon) for and the \emph{HST}/Cosmic Origins
Spectrograph (COS) on 2013 Nov 13 (data set: LB2402010/20;
Prop-ID:11527, PI: J.~Green). We generated the \emph{FUSE}, \emph{HST}/COS, and HDS
spectra normalized to the flux density at a continuum of 1.0 using {\sc
IRAF/SPLOT}.

\subsection{Spitzer/IRS mid-infrared spectra \label{S:spit}}

\begin{figure}
\includegraphics[width = \columnwidth,clip]{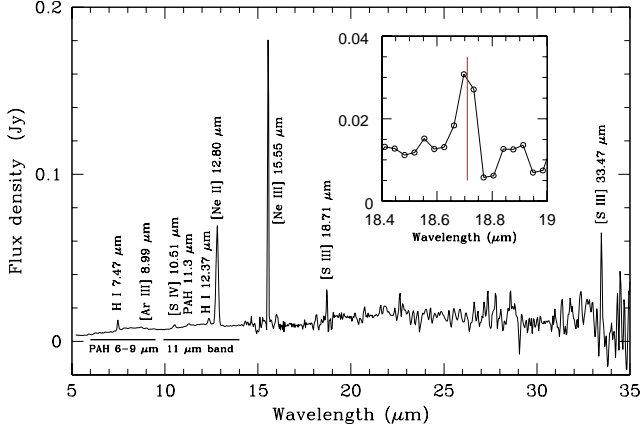}
\caption{The \emph{Spitzer}/IRS spectrum of K648. The detected gas
emission lines listed in Table~\ref{spitzer} are also indicated here.
The 6-9 ${\mu}$m PAH band, the 11.3 ${\mu}$m PAH emission,
and the broad 11 ${\mu}$m band are also indicated. The line-profile of
the {\siii}\,$\lambda$\,18.71\,$\mu$m is zoomed in the inner box.
The position of this line in the laboratory
is indicated by the vertical red line.
\label{spitzer_spec}}
\end{figure}

\begin{deluxetable}{@{}rlccr@{}}
\tablecolumns{5}
\centering
\tablecaption{The detected atomic lines in the \emph{Spitzer} spectra.\label{spitzer}}
\tablewidth{\columnwidth}
\tablehead{
\colhead{$\lambda_{\rm lab}$} &
\colhead{Ion} &
\colhead{$f$($\lambda$)}&
\colhead{$F$($\lambda$)}&
\colhead{$I$($\lambda$)}\\
\colhead{($\mu$m)}&
\colhead{}&
\colhead{}&
\colhead{(erg s$^{-1}$ cm$^{-2}$)}&
\colhead{
[$I$({\hb}) = 100]
}
}
\startdata
7.47&H\,{\sc i}&--0.990&3.23(--14)$\pm$1.56(--16)&3.15$\pm$0.15\\
8.99&[Ar\,{\sc iii}]&--0.959&3.29(--15)$\pm$4.84(--16)&0.32$\pm$0.05\\
10.51&[S\,{\sc iv}]&--0.959&1.08(--14)$\pm$4.46(--16)&1.06$\pm$0.07\\
12.37&H\,{\sc i}&--0.980&1.04(--14)$\pm$5.17(--16)&1.02$\pm$0.07\\
12.80&[Ne\,{\sc ii}]&--0.983&1.53(--13)$\pm$1.08(--14)&14.98$\pm$1.28\\
15.55&[Ne\,{\sc iii}]&--0.985&1.18(--13)$\pm$1.42(--14)&11.54$\pm$1.49\\
18.71&[S\,{\sc iii}]&--0.981&1.36(--14)$\pm$1.98(--15)&1.33$\pm$0.20\\
33.47&[S\,{\sc iii}]&--0.993&6.27(--15)$\pm$1.12(--15)&0.61$\pm$0.11
\enddata
\end{deluxetable}

We reduced the archive data obtained using the Infrared
Spectrograph \citep[IRS;][]{Houck:2004aa} with the SL (5.2-14.5\,$\mu$m and a slit dimension of 3.6{\arcsec}$\times$57{\arcsec}),
SH (9.9-19.6\,$\mu$m, 4.7{\arcsec}$\times$11.3{\arcsec}), and LH (18.7-37.2\,$\mu$m,
11.1{\arcsec}$\times$22.3{\arcsec}) modules (AOR Keys: 15733760 for the SL
and 18627840 for the SH and LH spectra; PIs:
R.~Gehrz and J.~Bernard-Salas, respectively).
We used the data reduction packages {\sc SMART} v.8.2.5 \citep{Higdon:2004aa}
and {\sc IRSCLEAN} provided by the \emph{Spitzer} Science Center.
For the SH and the LH spectra, we subtracted the background sky using the offset spectra.
We scaled the flux density of the SL data to that of the SH \& LH data
in the overlapping wavelength region. The remaining spikes in the spectra
were removed manually.

The resulting spectrum is shown in Fig.~\ref{spitzer_spec}.
\citet{Boyer:2006aa} reported the SL spectrum for K648 only. Therefore,
the spectrum at longer wavelengths (i.e., beyond $\sim$14.5\,$\mu$m)
is shown here for the first time. The line-profile of the
{\siii}\,$\lambda$\,18.71\,${\mu}$m, which is faint in K648 and also
a important diagnostic line, is
also shown in the inner box.

The line fluxes of the detected atomic lines are listed in
Table~\ref{spitzer}. We corrected for the interstellar reddening using Equation (1) and
the interstellar extinction function given by \citet{Fluks:1994aa}.
We computed $c$({\hb}) = 0.12$\pm$0.02 by comparing the theoretical
$I$(H\,{\sc i} 7.47\,$\mu$m)/$I$({\hb}) ratio of 3.15(--2) given by
\citet{1995MNRAS.272...41S} for $T_{\epsilon}$ = 10$^{4}$
K and $n_{\epsilon}$ = 10$^{4}$ cm$^{-3}$ under the assumptions of Case B.
Here, we used $F$({\hb}) = 7.83(--13) erg s$^{-1}$ cm$^{-2}$ (see Section \ref{hstphot}).
This result appears appropriate as the measured $I$(H\,{\sc i}
12.37\,$\mu$m)/$I$({\hb}) = 1.02(--2) is in good agreement with the theoretical data \citep[1.05(--2),][]
{1995MNRAS.272...41S}.

K648 exhibits the 6-9\,$\mu$m polycyclic amorphous carbon (PAH) band and
the
broad 11\,$\mu$m feature. These two features are
frequently seen in C-rich PNe. We will discuss the details on these features in Section \ref{S:dust}.

\subsection{Spitzer/IRAC/MIPS photometry}

\begin{deluxetable}{@{}l@{\hspace{3pt}}rrcl@{}}
\centering
\tablecolumns{5}
\tablecaption{\emph{Spitzer}/IRAC/MIPS and \emph{Herschel}/PACS
photometry of K648. \label{mid-ir}}
\tablewidth{\columnwidth}
\tablehead{
\colhead{Band}&
\colhead{$\lambda_{\rm cen.}$}&
\colhead{$\Delta$$\lambda$}&
\colhead{$F_{\lambda}$}&
\colhead{AORKEY(\emph{Spitzer})/}
\\
\colhead{}&
\colhead{($\mu$m)}&
\colhead{($\mu$m)}&
\colhead{(erg s$^{-1}$ cm$^{-1}$ {$\mu$m}$^{-1}$)}&
\colhead{OBSID(\emph{Herschel})}
}
\startdata
IRAC-ch1 & 3.51 & 0.68 & 1.24(--12)$\pm$1.53(--13)&12030208 \\
IRAC-ch2 & 4.50 & 0.86 & 6.16(--13)$\pm$4.80(--14)&12030208 \\
IRAC-ch3 & 5.63 & 1.26 & 4.78(--13)$\pm$4.51(--14)&12030208 \\
IRAC-ch4 & 7.59 & 2.53 & 4.43(--13)$\pm$2.82(--14)&12030208 \\
MIPS-ch1 & 23.21 & 5.30 & 5.95(--14)$\pm$1.44(--15)&12030464 \\
PACS-B & 68.92 & 21.41 & 1.86(--15)$\pm$3.82(--17)&1342246710/11/12 \\
PACS-R & 153.94 & 69.76 & 3.40(--16)$\pm$4.00(--17)&1342246710/11/12
\enddata
\end{deluxetable}

To provide a constraint in the SED fitting at mid-infrared (MIR)
wavelengths, we reduced archival \emph{Spitzer} MIR images
obtained using the Infrared Array Camera
\citep[IRAC;][]{2004ApJS..154...10F} and the Multiband Imaging
Spectrometer \citep[MIPS;][]{2004ApJS..154...25R}.
We downloaded the basic calibrated data and reduced it using
{\sc MOPEX}, which is provided by the \emph{Spitzer} Science Center, to obtain single
mosaic images for each band.

We carried out PSF fitting photometry of the IRAC images using
{\sc IRAF/DAOPHOT}. We adopted the position of K648 measured in
the \emph{HST}/F656N image and corrected the flux densities measured using PSF
photometry by aperture photometry of the PSF stars. For the MIPS 24\,$\mu$m image, we measured the total count within a 7{\arcsec}
radius region, and subtracted the background represented by the annulus
centered on the PN with 20{\arcsec} inner and 38{\arcsec} outer
radii, respectively. We used an aperture correction factor of 1.61,
as listed in the MIPS instrument hand book. The measured fluxes
are listed in Table~\ref{mid-ir}.

\subsection{Herschel/PACS photometry}

\begin{figure*}
\includegraphics[width = \textwidth,clip,bb = 8 187 814 401]{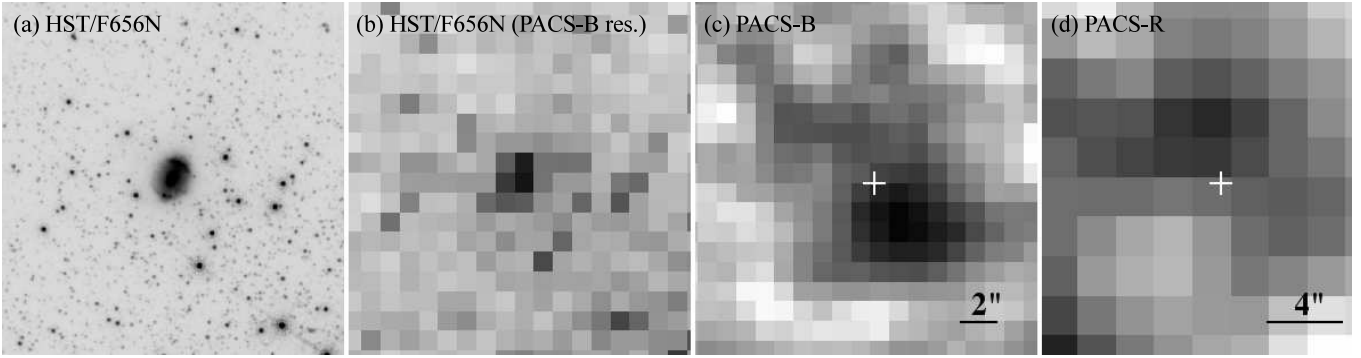}
\caption{The \emph{HST}/F656 and the \emph{Herschel}/PACS images.
The size of each panel is 17${\arcsec}$$\times$17${\arcsec}$. North is
up and east is to the left.
The plate scale of the image in (a)
is 0.025${\arcsec}$ pixel$^{-1}$. The plate scale of the HST/F656N image
in (b) corresponds to that of PACS-B (1${\arcsec}$ pixel$^{-1}$).
K648 is located in the center of each image.
In the PACS-B and PACS-R images, the location of K648 is indicated by
the crosses.
\label{hst_herschel}}
\end{figure*}

By combining the MIR data from the \emph{Spitzer} and the
FIR data from the \emph{Herschel}, we attempted to trace
the ejected mass of K648 during the last TP as accurately as possible.
For this purpose, we analyzed archived 70\,$\mu$m (PACS-B) and 160\,$\mu$m images (PACS-R) obtained using
the \emph{Herschel}/Photodetecting Array Camera and Spectrometer
\citep[PACS;][]{Poglitsch:2010aa}.

We downloaded the reduced PACS data of K648 (OBSID: 1342246710/11/12,
PI: M.~Boyer) from the Herschel Science Archive (HSA).
The PACS images are shown in Fig.~\ref{hst_herschel}. For comparison, we also
show the \emph{HST}/F656N
images. The plate scale of the image shown in Fig.~\ref{hst_herschel}(a) is 0.025${\arcsec}$
pixel$^{-1}$ and that of the \emph{HST}/F656N image shown in Fig.~\ref{hst_herschel}(b)
corresponds to that of PACS-B. Fig.~\ref{hst_herschel}(c) shows PACS-B data at 1${\arcsec}$ pixel$^{-1}$, and Fig.~\ref{hst_herschel}(d) shows the plate scale of PACS-R at 2${\arcsec}$ pixel$^{-1}$.
The most likely position of K648 was determined using the \emph{HST}/F656N image, and
is indicated by the white crosses. The light from K648 is partially
contaminated by nearby stars.

We used {\sc IRAF/DAOPHOT}
to measure the flux densities within a radius of 2 pixels in both the
PACS-B and PACS-R bands. We regarded the median count within the annulus
centered on the PN with an inner radius of four pixels
and an outer radius of five pixels as the background. We corrected the measured
flux densities of K648 using aperture photometry with correction factors
of 4.29 for PACS-B and 3.83 for PACS-R\footnote[6]{These correction
factors were computed using the
table of encircled energy fractions as a function of the radius of the aperture
for the PACS filter bands in the NASA \emph{Herschel} Science
Center.}. The measured flux densities are summarized in Table~\ref{mid-ir}.

\section{Results}

\subsection{Emission-line analysis}
\subsubsection{CEL diagnostics}

\begin{figure}
\centering
\includegraphics[width = \columnwidth, bb = 30 149 563 521,clip]{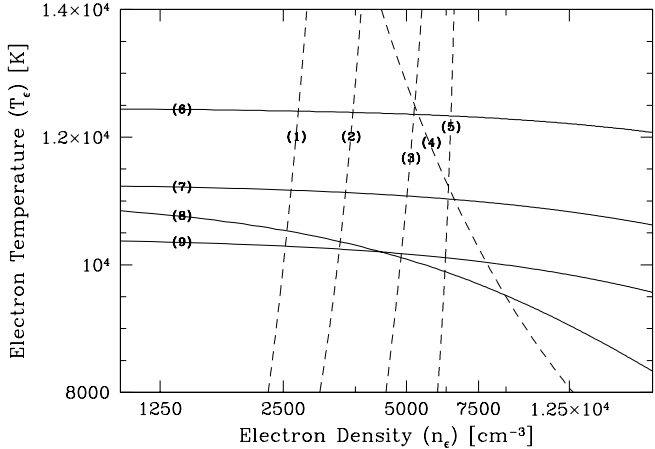}
\caption{An $n_{\epsilon}$-$T_{\epsilon}$ diagram. Each curve is labeled with an ID
number given in Table~\ref{diagno_table}. The solid lines indicate
diagnostic lines of $T_{\epsilon}$, whereas the broken lines indicate diagnostic
lines of $n_{\epsilon}$. \label{diagno_figure}}
\end{figure}

\begin{deluxetable}{@{}clrr@{}}
\tablecolumns{4}
\centering
\tablecaption{Plasma diagnostics.\label{diagno_table}}
\tablewidth{\columnwidth}
\tablehead{
\colhead{ID} &
\colhead{Diagnostic}&
\colhead{Value}&
\colhead{$n_{\epsilon}$ (cm$^{-3}$)}
}
\startdata
(1)&$[$S~{\sc ii}$]$($\lambda$6716)/($\lambda$6731) & 0.658$\pm$0.023 & 2530$\pm$330 \\
(2)&$[$O~{\sc ii}$]$($\lambda$3726)/($\lambda$3729) & 1.852$\pm$0.076 & 3430$\pm$470 \\
(3)&$[$S~{\sc iii}$]$($\lambda$18.7\,$\mu$m)/($\lambda$33.5\,$\mu$m) &2.178$\pm$0.523 &5110$\pm$2100 \\
(4)&$[$O~{\sc
ii}$]$($\lambda$3726/29)/($\lambda$7320/30) & 8.576$\pm$0.108$^{\rm a}$ & 7890$\pm$130 \\
(5)&$[$Cl~{\sc iii}$]$($\lambda$5517)/($\lambda$5537) & 0.749$\pm$0.120 & 7130$\pm$3170 \\
\cline{2-4}
 &Balmer decrement & &7500-10\,000\\
 \hline
 \vspace{-5pt}\\
 \colhead{ID} &
\colhead{Diagnostic}&
\colhead{Value}&
\colhead{$T_{\epsilon}$ (K)}\\
\vspace{-5pt}\\
 \hline
(6)&$[$O~{\sc iii}$]$($\lambda$4959+$\lambda$5007)/($\lambda$4363) & 108.649$\pm$4.425 & 12\,350$\pm$190 \\
(7)&$[$Ne~{\sc iii}$]$($\lambda$15.5\,$\mu$m)/($\lambda$3869+$\lambda$3967) &0.882$\pm$0.115 &11\,090$\pm$450\\
(8)&$[$N~{\sc ii}$]$($\lambda$6548+$\lambda$6583)/($\lambda$5755) & 79.200$\pm$9.333 & 10\,380$\pm$530 \\
(9)&$[$Ar~{\sc iii}$]$($\lambda$8.99\,$\mu$m)/($\lambda$7135) & 0.844$\pm$0.125 & 10\,270$\pm$900 \\
\cline{2-4}
 &He~{\sc i}($\lambda$5876)/($\lambda$4471) & 3.019$\pm$0.033 & 4270$\pm$300 \\
 &He~{\sc i}($\lambda$6678)/($\lambda$4471) & 0.837$\pm$0.012 & 7100$\pm$760 \\
 &He~{\sc i}($\lambda$7281)/($\lambda$5876) & 0.040$\pm$0.001 & 6360$\pm$150 \\
 &He~{\sc i}($\lambda$7281)/($\lambda$6678) & 0.145$\pm$0.003 & 6680$\pm$130 \\
\cline{2-4}
 &(Balmer Jump)/(H11) &0.102$\pm$0.006&11\,650$\pm$950
\enddata
\tablenotetext{a}{The recombination contribution is corrected for the $[$O\,{\sc
ii}$]$\,$\lambda\lambda$\,7320/30\,{\AA} lines.}
\end{deluxetable}

In the following analysis using CELs and RLs,
we used the transition probabilities, collisional impacts, and
recombination coefficients listed in
Tables 7 and 11 of \citet{2010ApJ...723..658O}.

The electron temperatures and densities were determined
using a variety of line diagnostic ratios by calculating the state
populations using a multilevel atomic model.
The observed diagnostic line ratios are listed in
Table~\ref{diagno_table}, where the numbers in the first column indicate
the ID of each curve in the {\Ne}-{\te} diagram shown in Fig.~\ref{diagno_figure}.
The second, third, and final columns in
Table \ref{diagno_table} show
the diagnostic lines, line ratios, and the resulting {\Ne} and {\te},
respectively. We obtained nine diagnostic line ratios with
different ionization potentials (IPs) in the range 10.4 eV ({\sii}) to
41 eV ({\neiii}), and determined a suitable {\te} and {\Ne} combination for
each ion.

For the {\oii}\,$\lambda\lambda$\,7320/30\,{\AA} lines, we eliminated the
recombination contamination due to O$^{2+}$
using the following expression, which is given by \citet{2000MNRAS.312..585L}:
\begin{equation}
\label{roii}
\frac{I_{R}(\rm [O\,{\sc II}]\lambda\lambda7320/30)}{I(\rm H\beta)} =
9.36\left(\frac{T_{\epsilon}}{10^4}\right)^{0.44}\times\frac{\rm O^{2+}}{\rm H^{+}}.
\end{equation}
\noindent
Using the O$^{2+}$ ionic abundances derived from the recombination
O~{\sc ii}\,$\lambda$\,4641.8\,{\AA} line and with {\te} = 11\,650 K, based on the Balmer
jump discontinuity (see the following section), we found that
$I_{R}$({\rm {\oii}}\,$\lambda\lambda$\,7320/30) = 0.12$\pm$0.02.
As we could not detect the N\,{\sc ii} and the pure O~{\sc iii}
recombination lines, we were unable to estimate the contribution
of N$^{2+}$ to the {\nii}\,$\lambda$\,5755\,{\AA} line nor that of O$^{3+}$ to
the {\oiii}\,$\lambda$\,4363\,{\AA} line.

First, we computed {\Ne} with {\te} = 10\,000 K for all
density diagnostic lines. {\te}({\neiii}), \te({\oiii}),
and {\te}({\ariii}) were calculated using {\Ne} = 6100 cm$^{-3}$, which is
the averaged {\Ne} between {\Ne}({\cliii}) and {\Ne}({\siii}).
We calculated {\te}({\nii}) using the {\Ne}({\oii}) determined from the
{\oii} $I$($\lambda$3726)/$I$($\lambda$3729) ratio. We used {\oii}
$I$($\lambda\lambda$3726/29)/$I$($\lambda\lambda$7320/30) as a density indicator for the
$\sim$4500 cm$^{-3}$ region, which is larger than the critical density
of {\oii}\,$\lambda$\,3726\,{\AA}.

Our values of {\te} and {\Ne} are comparable to those
reported by \citet{Kwitter:2003aa}, i.e., {\te}({\oiii}) = 11\,800
K, {\te}({\nii}) = 9200 K, and {\Ne}({\sii}) = 1000 cm$^{-3}$.

\subsubsection{RL diagnostics}

\begin{figure}
\centering
\includegraphics[width = \columnwidth,bb = 40 149 564 551,clip]{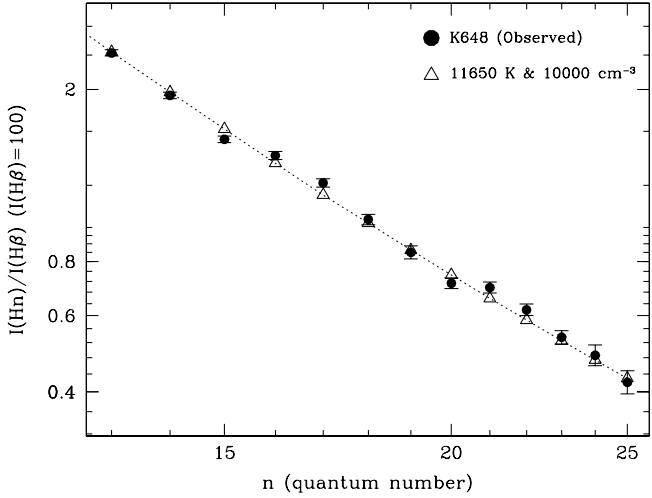}
\caption{
The intensity ratio of the higher-order Balmer lines
to {\hb} (Case B assumption). The theoretical intensity
ratios (dotted curve and triangles) are given for
{\te} = 11\,650 K (determined from the Balmer Jump) and {\Ne} = 10$^{4}$
cm$^{-3}$.
\label{bal}}
\end{figure}

We calculated {\te} using the ratio of the Balmer discontinuity to
$I$(H11). We employed the method reported by \citet{2001MNRAS.327..141L}
to calculate the electron temperature {\te}(BJ).

We calculated the {\hei} electron temperatures using the four different
{\te}({\hei}) line ratios and the emissivities of these {\hei} lines from
\citet{1999ApJ...514..307B}, in the case of {\Ne} = 10$^{4}$ cm$^{-3}$.

The intensity ratio of a high-order Balmer line H$n$ (where $n$ is
the principal quantum number of the upper level) to a lower-order Balmer
line is also sensitive to the electron density.
The ratios of higher-order Balmer lines to
{\hb} are plotted in Fig.~\ref{bal} along with theoretical values from
\citet{1995MNRAS.272...41S} for {\te}(BJ) and {\Ne} = 10\,000 cm$^{-3}$.
We ran small-grid calculations to determine {\Ne} in the range
5000-12\,500 cm$^{-3}$, and found that the models in the range of {\Ne} = 7500-10\,000
cm$^{-3}$ provided the best fit to the observed data.
{\te} and {\Ne} determined using the RL diagnostics are summarized
in Table~\ref{diagno_table}.

\subsubsection{CEL ionic abundances}

\begin{figure}
\includegraphics[width = \columnwidth,clip]{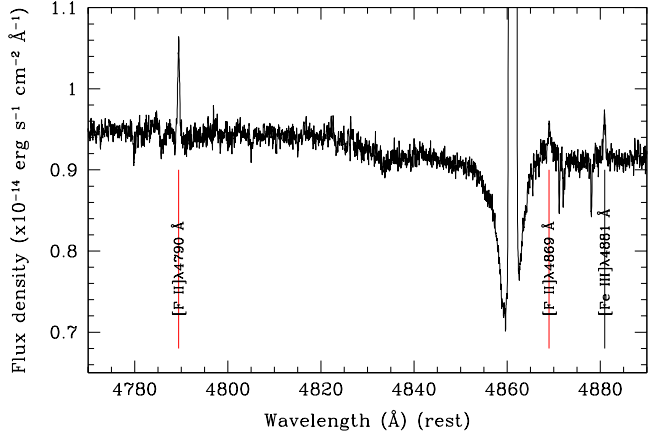}
\caption{The 4770-4890\,{\AA} Subaru/HDS spectrum of K648. The
wavelength is corrected to the rest wavelength in air.
The locations of the {\fii}\,$\lambda\lambda$\,4790/4869\,{\AA} lines in
are indicated by the vertical red lines.
The position of the {\feiii}\,$\lambda$4881\,{\AA} is
indicated by the vertical black line. The narrow absorptions
are from cool stars, possibly two nearby RGB stars (see
Fig.~\ref{hst_ha}).
\label{f2_spec}}
\end{figure}

\begin{deluxetable}{@{}ccl@{}}
\tablecolumns{3}
\centering
\tablecaption{The adopted electron temperatures and densities.\label{tene}}
\tablewidth{\columnwidth}
\tablehead{
\colhead{{\te} (K)}&
\colhead{{\Ne} (cm$^{-3}$)}&
\colhead{Ions}
}
\startdata
10\,380 &2530 &S$^{+}$\\
10\,380 &3430 &C$^{+}$,N$^{+}$,O$^{+}$(3726,29\,{\AA}),F$^{+}$,Fe$^{2+}$\\
10\,380 &7840 &O$^{+}$(7320,30\,{\AA})\\
10\,270 &6100 &C$^{2+}$,Ne$^{+}$,S$^{2+}$,Cl$^{2+}$,Ar$^{2+}$\\
11\,090 &6100 &Ne$^{2+}$\\
12\,350 &6100 &O$^{2+}$,S$^{3+}$
\enddata
\end{deluxetable}

\begin{deluxetable}{lccc}
\tablecolumns{4}
\centering
\tablecaption{Ionic Abundances from CELs.\label{celabund}}
\tablewidth{\columnwidth}
\tablehead{
\colhead{X$^{\rm m+}$} &
\colhead{$\lambda_{\rm lab}$}&
\colhead{$I$($\lambda_{\rm lab}$)} &
\colhead{X$^{\rm m+}$/H$^{+}$}
}
\startdata
C$^{+}$ &2323\,{\AA}& 1.64(+1)$\pm$1.63(0) &{\bf 2.55(--5)$\pm$7.28(--6)} \\
C$^{2+}$ &1906/09\,{\AA}& 3.35(+2)$\pm$1.70(+1)&{\bf 6.91(--4)$\pm$3.23(--4)} \\
N$^{+}$ &5754.64\,{\AA}& 4.31(--2)$\pm$2.44(--3) &4.67(--7)$\pm$1.02(--7) \\
 &6548.04\,{\AA}& 9.01(--1)$\pm$1.23(--2) &4.93(--7)$\pm$5.89(--8) \\
 &6583.46\,{\AA}& 3.18(0)$\pm$4.10(--2) &5.89(--7)$\pm$7.02(--8) \\
& & & {\bf 5.68(--7)$\pm$6.77(--8)} \\
O$^{+}$ &3726.03\,{\AA}& 1.74(+1)$\pm$2.99(--1) &1.34(--5)$\pm$2.50(--6) \\
 &3728.81\,{\AA}& 9.39(0)$\pm$3.50(--1) &1.34(--5)$\pm$2.60(--6) \\
 &7320/30\,{\AA}& 3.12(0)$\pm$3.91(--2)$^{\rm a}$& 1.79(--5)$\pm$4.40(--6) \\
 & & &{\bf 1.34(--5)$\pm$2.54(--6)} \\
O$^{2+}$ &4363.21\,{\AA}& 2.78(0)$\pm$2.56(--2)& 4.03(--5)$\pm$3.29(--6) \\
 &4931.23\,{\AA}& 3.84(--2)$\pm$4.83(--3) &5.12(--5)$\pm$6.80(--6) \\
 &4958.91\,{\AA}& 7.50(+1)$\pm$7.37(0) &3.90(--5)$\pm$4.18(--6) \\
 &5006.84\,{\AA}& 2.27(+2)$\pm$9.46(0) &4.09(--5)$\pm$2.45(--6) \\
 & & &{\bf 4.05(--5)$\pm$2.89(--6)} \\
F$^{+}$ &4789.45\,{\AA}& 1.10(--1)$\pm$3.67(--3)& 6.67(--8)$\pm$1.02(--8) \\
&4868.99\,{\AA}& 2.96(--2)$\pm$3.36(--3) &5.75(--8)$\pm$1.08(--8) \\
 & & &{\bf 6.47(--8)$\pm$1.03(--8)} \\
Ne$^{+}$ &12.80\,$\mu$m& 1.50(+1)$\pm$1.28(0) &{\bf 2.01(--5)$\pm$1.95(--6)} \\
Ne$^{2+}$ &3869.06\,{\AA}& 9.94(0)$\pm$1.24(--1)& 7.28(--6)$\pm$1.00(--6) \\
 &3967.79\,{\AA}& 3.15(0)$\pm$4.35(--2)& 7.65(--6)$\pm$1.06(--6) \\
 &15.55\,$\mu$m& 1.13(+1)$\pm$1.36(0) & 7.48(--6)$\pm$9.79(--7) \\
 & & & {\bf 7.42(--6)$\pm$9.98(--7)} \\
S$^{+}$ &6716.44\,{\AA}& 8.73(--2)$\pm$2.23(--3) & 6.72(--9)$\pm$7.79(--10) \\
 &6730.81\,{\AA}& 1.33(--1)$\pm$3.21(--3) & 6.73(--9)$\pm$7.48(--10) \\
 & & &{\bf 6.72(--9)$\pm$7.60(--10)} \\
S$^{2+}$ &6313.1\,{\AA}& 1.19(--1)$\pm$5.12(--3) & 2.52(--7)$\pm$7.30(--8) \\
 &18.71\,$\mu$m& 1.33(0)$\pm$2.04(--1) & 2.10(--7)$\pm$3.52(--8) \\
 &33.47\,$\mu$m& 6.12(--1)$\pm$1.13(--1) & 2.10(--7)$\pm$4.16(--8) \\
 & & &{\bf 2.12(--7)$\pm$3.93(--8)} \\
S$^{3+}$ &10.51\,$\mu$m&1.06(0)$\pm$6.66(--2) &{\bf 3.35(--8)$\pm$2.11(--9)} \\
Cl$^{2+}$ &5517.72\,{\AA}& 2.12(--2)$\pm$2.71(--3)& 3.15(--9)$\pm$7.86(--10) \\
 &5537.89\,{\AA}& 2.83(--2)$\pm$2.73(--3) & 3.17(--9)$\pm$7.38(--10) \\
 & & &{\bf 3.16(--9)$\pm$7.59(--10)} \\
Ar$^{2+}$ &7135.79\,{\AA}& 3.84(--1)$\pm$6.20(--3) &3.32(--8)$\pm$5.73(--9) \\
 &8.99\,$\mu$m & 3.24(--1)$\pm$4.77(--2) &3.41(--8)$\pm$5.32(--9) \\
& & &{\bf 3.36(--8)$\pm$5.54(--9)} \\
Fe$^{2+}$ &4701.53\,{\AA}& 2.22(--2)$\pm$3.11(--3) & 2.37(--8)$\pm$4.80(--9) \\
 &4881.11\,{\AA}& 5.09(--2)$\pm$3.55(--3) & 2.75(--8)$\pm$4.59(--9) \\
&&&{\bf 2.63(--8)$\pm$4.65(--9)}
\enddata
\tablenotetext{a}{Corrected recombination contribution for the $[$O\,{\sc
ii}$]$\,$\lambda\lambda$\,7320/30\,{\AA} lines.}
\end{deluxetable}

We obtained the following 14 ionic abundances: C$^{+,2+}$, N$^{+}$,
O$^{+,2+}$, F$^{+}$, Ne$^{+,2+}$, S$^{+,2+,3+}$, Cl$^{2+}$, Ar$^{2+}$
and Fe$^{2+}$. The abundances of
F$^{+}$, Cl$^{2+}$, and Fe$^{2+}$ abundances for K648 are reported here
for the first time. The ionic abundances were calculated by
solving the statistical equilibrium equations for more than five levels
with the relevant {\te} and {\Ne}, except for Ne$^{+}$, where we calculated
the abundance using a two-energy level model.
The Fe$^{2+}$ abundances were solved using a 33-level model (from
$^{5}D_{3}$ to $^{3}P_{2}$).
For each ion, we used the electron temperatures and densities
determined using CEL plasma diagnostics. The adopted
{\te} and {\Ne} for each ion are listed in Table~\ref{tene}.

The ionic abundances are listed in Table~\ref{celabund}.
The final column shows the resulting ionic abundances,
X$^{\rm m+}$/H$^{+}$, together with the relevant errors, including errors from
line-intensities, electron temperature, and electron density.
The ionic abundance and the error are listed in the final row for each ion.
These data were calculated based on the weighted mean of the relevant
line-intensity.

In calculation of the C$^{+}$ abundance, we subtracted contamination from
[O~{\sc iii}]\,$\lambda$\,2321\,{\AA} to [C~{\sc ii}]\,$\lambda$\,2323
{\AA} based on the theoretical intensity ratio {\oiii}
$I$($\lambda$\,2326)/$I$($\lambda$\,4363) = 0.236. As described above,
we did not remove the respective contributions from N$^{2+}$ and O$^{3+}$
to the {\nii}\,$\lambda$\,5755\,{\AA} and the {\oiii}\,$\lambda$\,4363
{\AA} line intensities. To determine the
final O$^{+}$ abundance, we excluded data determined using the
{\oii}\,$\lambda\lambda$\,7320/30\,{\AA} lines.

We determined the Ne$^{+}$ abundance of 2.01(--5) using the {\neii}\,$\lambda$\,12.80\,$\mu$m line, which is slightly larger than \citet[][1.53(--5)]{Boyer:2006aa}. This
small disagreement is expected to be mainly due
to the adopted {\hb} flux. \citet{Boyer:2006aa}
calculated the {\hb} flux using the measured H\,{\sc i} lines
at 7.47\,$\mu$m and 12.37\,$\mu$m, using the theoretical ratios
of H\,{\sc i} $I$($\lambda$\,7.47\,$\mu$m,12.37\,$\mu$m)/$I$({\hb}) with
Case B. Their resulting $I$({\hb}) was 1.52$\times$10$^{-12}$ erg
s$^{-1}$ cm$^{-2}$. While, we used
the \emph{HST}/F656N band-pass flux intensity and corresponding HDS
spectral scan to scale the intensities, and find
$I$({\hb}) = 1.07$\times$10$^{-12}$ erg s$^{-1}$ cm$^{-2}$.
\citet{Boyer:2006aa} used {\te} = 10\,000 K and {\Ne} = 1700
cm$^{-3}$ for the Ne$^{+}$ and S$^{2+,3+}$ calculations.
Our plasma diagnostics showed that 10\,000 K is low for
S$^{3+}$, where we used 12\,350 K.

The S$^{2+}$ abundance of 2.17(--7) determined using the two MIR [S\,{\sc iii}]
lines is approximately the same as that calculated from [S\,{\sc iii}]\,$\lambda$\,6312
{\AA}, and is in good agreement with \citet{Kwitter:2003aa}, who
calculated 1.99(--7) using
$I$({\siii}\,$\lambda$\,9532\,{\AA}) = 3.8. However, there was
poor agreement in the S$^{2+}$ abundance between the most recent
measurements by \citet[][2.55(--8)]{Boyer:2006aa} and our data.
\citet{Boyer:2006aa} calculated the S$^{2+}$ abundance using
$I$({\siii}\,$\lambda$\,9532\,{\AA}) = 0.76 measured by
\citet{Barker:1983aa}, because they used the \emph{Spitzer}
SL module spectra only, where no MIR {\siii} lines appear.
We can exclude the possibility that the discrepancy in the
S$^{2+}$ abundance is due to the flux measurements of our
MIR {\siii} and the choice of {\te}. If our flux measurements of the MIR {\siii},
{\siii}\,$\lambda$\,6312\,{\AA} and the {\hb} lines and the {\te}
selection were incorrect, the S$^{2+}$ abundances from two MIR {\siii}
lines would not match that from {\siii}\,$\lambda$\,6312\,{\AA}. The
fine-structure lines are much less sensitive to the electron temperature
compared with the other transition lines. The
auroral lines, such as {\siii}\,$\lambda$\,6312\,{\AA}, were dependent
on the electron temperature (i.e., the S$^{2+}$ abundance determined from
{\siii}\,$\lambda$\,6312\,{\AA} is largely dependent on {\te}). Our
calculated S$^{2+}$ abundances from these three were consistent with
each other, indicating that our flux measurements of the MIR {\siii} and {\hb} lines
and the choice of {\te} for the S$^{2+}$ (and possibly also Ne$^{+}$ and S$^{3+}$)
were appropriate. Therefore, the large discrepancy in S$^{2+}$ between
\citet{Boyer:2006aa} and our data may have been due to the
{\siii}\,$\lambda$\,9532\,{\AA} flux that was used.

It is interesting to note the detection of single isotope
$^{19}$F line candidates [F\,{\sc ii}]\,$\lambda\lambda$\,4789.45/4868.99\,{\AA}, as
shown in Fig.~\ref{f2_spec}.
Together with $^{12}$C and $^{22}$Ne,
$^{19}$F is synthesized in the He-rich intershell during the TP-AGB
phase, and is an $n$-capture element. The observed {\fii}
$I$($\lambda$\,4789.45)/$I$($\lambda$\,4868.99)
of 3.72$\pm$0.44 is in agreement with the
theoretical value of 3.20 calculated using {\te} = 10\,380 K and
{\Ne} = 3430 cm$^{-3}$. We excluded the other candidate
C\,{\sc iv}\,$\lambda$\,4789.65\,{\AA} because no C\,{\sc iv} lines
were detected (e.g., C\,{\sc iv}\,$\lambda$\,5801.35\,{\AA}).
Therefore, we conclude that the lines at 4790 and 4869\,{\AA} are
the [F\,{\sc ii}]\,$\lambda\lambda$\,4789.45/4868.99\,{\AA},
respectively. The detection of F lines is very rare in Galactic PNe
\citep[e.g.,][]{2008ApJ...682L.105O,2005ApJ...631L..61Z,Liu:1998aa}. Among halo
PNe, K648 is the third case of such F line detection reported to date;
NGC4361 \citep{Liu:1998aa}, BoBn1 \citep{2008ApJ...682L.105O}, and
K648 (this work).  We discuss whether these lines are $[$F\,{\sc
ii}$]$\,$\lambda\lambda$\,4789.45/4868.99\,{\AA} using a theoretical
model later in the paper. If the two lines do not originate from the
F$^+$ ion, the prediction cannot fit the fluxes of the two lines
simultaneously.

\subsubsection{RL ionic abundances}

\begin{deluxetable}{@{}llccc@{}}
\tablecolumns{5}
\tablecaption{Ionic abundances from RLs.\label{rlabund}}
\tablewidth{\columnwidth}
\tablehead{
\colhead{X$^{\rm m+}$} &
\colhead{$\lambda_{\rm lab}$} &
\colhead{Multi.}&
\colhead{$I$($\lambda_{\rm lab}$)} &
\colhead{X$^{\rm m+}$/H$^{+}$}
}
\startdata
He$^{+}$&5875.62\,{\AA}&V11&1.48(+1)$\pm$1.38(--1)&1.02(--1)$\pm$6.69(--3)\\
&4471.47\,{\AA}&V14&4.91(0)$\pm$2.78(--2)&9.86(--2)$\pm$6.05(--3)\\
&6678.15\,{\AA}&V46&4.11(0)$\pm$5.48(--2)&9.90(--2)$\pm$6.48(--3)\\
&4921.93\,{\AA}&V48&1.29(0)$\pm$5.28(--3)&9.54(--2)$\pm$5.92(--3)\\
&4387.93\,{\AA}&V51&5.17(--1)$\pm$1.04(--2)&8.34(--2)$\pm$6.66(--3)\\
&&& &{\bf 1.00(--1)$\pm$6.49(--3)}\\
C$^{2+}$&6578.05\,{\AA}&V2&6.92(--1)$\pm$1.08(--2)&8.42(--4)$\pm$1.33(--4)\\
&4267.18\,{\AA}&V6&7.26(--1)$\pm$1.33(--2)&7.32(--4)$\pm$9.10(--5)\\
&6151.27\,{\AA}&V16.04&4.50(--2)$\pm$2.89(--3)&1.04(--3)$\pm$1.31(--4)\\
&6462.04\,{\AA}&V17.04&9.81(--2)$\pm$8.62(--3)&9.67(--4)$\pm$1.63(--4)\\
&&& &{\bf 8.04(--4)$\pm$1.15(--4)}\\
C$^{3+}$&6727.48\,{\AA}&V3&3.71(--2)$\pm$2.74(--3)&2.05(--4)$\pm$1.48(--5)\\
&6742.15\,{\AA}&V3&4.14(--2)$\pm$3.78(--3)&2.75(--4)$\pm$2.52(--5)\\
&6744.39\,{\AA}&V3&6.05(--2)$\pm$2.80(--3)&2.87(--4)$\pm$1.37(--5)\\
&&& &{\bf 2.62(--4)$\pm$1.74(--5)}\\
O$^{2+}$&4641.81\,{\AA}&V1&3.34(--2)$\pm$3.45(--3)&{\bf 1.21(--4)$\pm$1.64(--5)}
\enddata
\end{deluxetable}

The RL ionic abundances are listed in Table~\ref{rlabund}. As we detected C\,{\sc ii,iii} and
O\,{\sc ii} RLs, we can compare the elemental C and O
abundances determined using RLs with those from CELs in K648.

In the abundance calculations, we used the Case B assumption for
lines with levels that have the same spin as the ground state, and
the Case A assumption for lines of other multiplicities.
In the final line of each ion series, we give the
ionic abundance and the error estimated using the line intensity
weighted mean. As the RL ionic abundances
were not sensitive to the electron density with $\lesssim$10$^{8}$
cm$^{-3}$, we used the atomic data in the case of {\Ne} = 10$^{4}$ cm$^{-3}$ for all lines.
To calculate the He$^{+}$ abundances, we used
{\te}({\hei}) = 6710$\pm$350 K, and the average of all {\te}({\hei}) data
listed in Table~\ref{diagno_table}, except for
{\te}({\hei}), where we used the {\hei} $I$($\lambda$\,5876)/$I$($\lambda$\,4471) ratio, which was
smaller than the other data. We used the {\te}(BJ) to calculate the C$^{2+,3+}$ and O$^{2+}$ abundances.

We used the multiplet V1 O\,{\sc ii}\,$\lambda$\,4641.81\,{\AA}
line only, because the observed HDS spectra were partially contaminated
by the absorption lines of the CSPN. According to
\citet{Peimbert:2005aa}, the upper levels of the transitions in the V1 O\,{\sc
ii} line are not in local thermal equilibrium (LTE) for {\Ne} $<$10\,000
cm$^{-3}$. As the value of {\Ne} calculated using the Balmer
decrement method was 7500-10\,000 cm$^{-3}$,
we applied the non-LTE corrections using Equations (8)-(10) in \citet{Peimbert:2005aa}
with {\Ne} = 7500 cm$^{-3}$.

\subsubsection{Nebular ICF abundances \label{S:abund_ICF}}

\begin{deluxetable}{@{}lccl@{}}
\tablecolumns{4}
\tablecaption{The ionization correction factors (ICFs).\label{icf}}
\tablewidth{\columnwidth}
\tablehead{
\colhead{X}&
\colhead{Line}&
\colhead{ICF(X)}&
\colhead{X/H}
}
\startdata
He &RL &$\rm \frac{S^{+}+S^{2+}}{S^{2+}}$&ICF(He)$\rm He^{+}$\\
\noalign{\smallskip}
C &CEL &$\rm \left(\frac{C}{C^{2+}}\right)_{RL}$&C$^{+}$+ICF(C)C$^{2+}$\\
 &RL &$\rm \left(\frac{C^{+}+C^{2+}}{C^{2+}}\right)_{CEL}$&ICF(C)C$^{2+}$+C$^{3+}$\\
\noalign{\smallskip}
N &CEL &$\rm \left(\frac{O}{O^{+}}\right)_{CEL}$&ICF(N)N$^{+}$\\
\noalign{\smallskip}
O &CEL & 1 &O$^{+}$+O$^{2+}$\\
 &RL &$\rm \left(\frac{O}{O^{2+}}\right)_{CEL}$&ICF(O)O$^{2+}$\\
\noalign{\smallskip}
F &CEL &$\rm \left(\frac{O}{O^{+}}\right)_{CEL}$ &ICF(F)F$^{+}$\\
\noalign{\smallskip}
Ne &CEL &1&Ne$^{+}$+Ne$^{2+}$\\
\noalign{\smallskip}
S &CEL &1 &$\rm S^{+}+S^{2+}+S^{3+}$\\
\noalign{\smallskip}
Cl &CEL &$\rm {\left(\frac{Ar}{Ar^{2+}}\right)}$ &ICF(Cl)Cl$^{2+}$\\
\noalign{\smallskip}
Ar &CEL &$\rm \frac{S}{S^{2+}}$&ICF(Ar)$\rm Ar^{2+}$\\
\noalign{\smallskip}
Fe &CEL &$\rm \left(\frac{O}{O^{+}}\right)_{CEL}$ &ICF(Fe)$\rm Fe ^{2+}$
\enddata
\end{deluxetable}

\begin{deluxetable*}{@{}lccrcrc@{}}
\tablecolumns{7}
\centering
\tablecaption{The elemental abundances from CEL and RLs.
 \label{abund}}
\tablewidth{\textwidth}
\tablehead{
\colhead{X} &
\colhead{Types of} &
\colhead{X/H} &
\colhead{log(X/H)+12} &
\colhead{[X/H]} &
\colhead{log(X$_{\odot}$/H)+12} &
\colhead{ICF(X)} \\
\colhead{} &
\colhead{Emissions}
}
\startdata
He &RL &1.04(--1)$\pm$6.82(--3) &11.02$\pm$0.03 &+0.09$\pm$0.03 &10.93$\pm$0.01 &1.04$\pm$0.01\\
C &CEL &9.41(--4)$\pm$3.75(--4) &8.97$\pm$0.17 &+0.58$\pm$0.18 &8.39$\pm$0.04 &1.33$\pm$0.24\\
C &RL &1.10(--3)$\pm$5.54(--4) &9.04$\pm$0.22 &+0.65$\pm$0.22 &8.39$\pm$0.04 &1.04$\pm$0.67\\
N &CEL &2.28(--6)$\pm$5.35(--7) &6.36$\pm$0.10 &--1.47$\pm$0.11 &7.83$\pm$0.05 &4.02$\pm$0.81\\
O &CEL &5.39(--5)$\pm$3.84(--6) &7.73$\pm$0.03 &--0.96$\pm$0.06 &8.69$\pm$0.05 &1.00\\
O &RL &1.61(--4)$\pm$2.72(--5) &8.21$\pm$0.07 &--0.48$\pm$0.09 &8.69$\pm$0.05 &1.33$\pm$0.13\\
F &CEL &2.60(--7)$\pm$6.70(--8) &5.42$\pm$0.11 &+0.96$\pm$0.13 &4.46$\pm$0.06 &4.02$\pm$0.81\\
Ne &CEL &2.75(--5)$\pm$2.19(--6) &7.44$\pm$0.03 &--0.43$\pm$0.11 &7.87$\pm$0.10 &1.00\\
S &CEL &2.53(--7)$\pm$3.93(--8) &5.40$\pm$0.07 &--1.79$\pm$0.08 &7.19$\pm$0.04 &1.00\\
Cl &CEL &3.76(--9)$\pm$1.28(--9) &3.58$\pm$0.15 &--1.92$\pm$0.33 &5.50$\pm$0.30 &1.19$\pm$0.29\\
Ar &CEL &4.00(--8)$\pm$1.17(--8) &4.60$\pm$0.13 &--1.95$\pm$0.15 &6.55$\pm$0.08 &1.19$\pm$0.29\\
Fe &CEL &1.06(--7)$\pm$2.84(--8) &5.02$\pm$0.12 &--2.45$\pm$0.12 &7.47$\pm$0.03 &4.02$\pm$0.81
\enddata
\tablecomments{The types of emission line used to calculate the abundances are
shown in the second column, the number densities of each element
relative to hydrogen are listed in the third column, the fourth
column lists the number densities, where $\log_{10}$\,$n$(H) = 12,
the fifth column lists the logarithmic number densities relative to the solar value,
and the final two columns list the solar abundances and the ICF
values that were used.}
\end{deluxetable*}

\begin{deluxetable*}{@{}lcccccccccc@{}}
\tablecolumns{11}
\centering
\tabletypesize{\footnotesize}
\tablecaption{Comparison of nebular elemental abundances. \label{past}}
\tablewidth{\textwidth}
\tablehead{
\colhead{References}&
\colhead{He}&
\colhead{C}&
\colhead{N}&
\colhead{O}&
\colhead{F}&
\colhead{Ne}&
\colhead{S}&
\colhead{Cl}&
\colhead{Ar}&
\colhead{Fe}
}
\startdata
This work (RL) &11.02&9.04&\nodata&8.21&\nodata&\nodata&\nodata&\nodata&\nodata&\nodata\\
This work (CEL) &\nodata&8.97&6.36&7.73&5.42&7.44&5.40&3.58&4.60&5.02\\
\hline
\citet{Boyer:2006aa} &\nodata&\nodata&\nodata&\nodata&\nodata&7.38&4.63&\nodata&\nodata&\nodata\\
\citet{Kwitter:2003aa} &11.00&\nodata&6.48&7.85&\nodata&7.00&5.30&\nodata&4.60&\nodata\\
\citet{1997MNRAS.284..465H}$^{\rm a}$ &10.98&8.50&6.72&7.61&\nodata&6.57&6.11&\nodata&3.72&\nodata\\
\citet{Henry:1996aa}$^{\rm b}$ &10.92&8.29&6.66&7.62&\nodata&6.47&\nodata&\nodata&\nodata&\nodata\\
\citet{1984MNRAS.207..471A}$^{\rm b}$ &11.02&8.73&6.50&7.67&\nodata&6.70&\nodata&\nodata&\nodata&\nodata\\
\citet{1980ApaSS..71..393A}$^{\rm a}$ &10.90&8.45&6.37&7.53&\nodata&6.40&5.60&\nodata&\nodata&\nodata\\
\citet{Torres-Peimbert:1979aa}&10.99&\nodata&$<$6.39&7.82&\nodata&6.79&$<$6.22&\nodata&$<$5.52&\nodata\\
\citet{Hawley:1978aa} &11.00&\nodata&7.11&7.65&\nodata&6.40&\nodata&\nodata&\nodata&\nodata
\enddata
\tablenotetext{a}{from the photo-ionization models.}
\tablenotetext{b}{The C abundance is from CEL C lines.}
\end{deluxetable*}

To estimate the elemental abundances in the nebula,
it is necessary to correct the ionic abundances that are unseen because of their faintness or
because they lie outside the data coverage.
We used an ionization correction factor, ICF(X), which was based on the
IP. The ICF(X) for each element is listed in Table~\ref{icf}.
The ICF(X)s based on IP are known to be inaccurate, particularly in some cases such as N.

The elemental abundances of the nebula are listed in Table~\ref{abund}.
We referred to \citet{2009ARA&A..47..481A} for N and Cl, and
\citet{2003ApJ...591.1220L} for the other elements.

The RL C abundance was almost identical to that of the CEL C, that is,
the C abundance discrepancy factor (ADF) $n$(C)$_{\rm RL}$/$n$(C)$_{\rm
CEL}$ = 1.17$\pm$0.75, whereas the O ADF was large,
$n$(O)$_{\rm RL}$/$n$(O)$_{\rm CEL}$ = 2.99$\pm$0.55.
The RL C abundance is greater than the RL O abundance.
The RL C/O ratio of 17.46$\pm$7.07 agrees with the CEL C/O
ratio of 6.83$\pm$3.63 within error. The
(C/O)$_{\rm RL}$/(C/O)$_{\rm CEL}$ ratio is 2.56$\pm$1.71.
It follows that these C/O ratios indicate that K648 is a C-rich PN.

The aforementioned O ADF value in K648 is approximately the same as the
O$^{2+}$ ADF = 2.99$\pm$0.46.
The O$^{2+}$ ADF has been reported for other
C-rich halo PNe, i.e., H4-1 \citep[1.75$\pm$0.36,][]{Otsuka:2013aa} and BoBn1
\citep[3.05$\pm$0.54,][]{2010ApJ...723..658O}. The smaller ADF of the
O$^{2+}$ found in H4-1 may be due to temperature fluctuations proposed
by \citet{1967ApJ...150..825P}; however, the
relatively large ADF of O$^{2+}$ found in K648 is too large to be explained by
temperature fluctuations.
Therefore, as with BoBn1, we should seek other
plausible solutions to explain the O$^{2+}$ abundance discrepancy in
K648. For BoBn1, \citet{2010ApJ...723..658O} suggested that
the bi-abundance pattern may solve the O$^{2+}$ and Ne$^{2+}$ abundance
discrepancy. The RL abundances in the nebula may correspond to
abundances in the stellar wind, as seen in the C-rich PN IC418
\citep{Morisset:2009aa}. It should be noted that the stellar C and O abundances
for K648 examined by \citet{2002AaA...381.1007R} are closer to our RL C
and O abundances (C = 9.00 and O = 9.00).
In the following section, we determine the stellar
abundances of K648 using the \emph{FUSE}, \emph{HST}/COS, and HDS spectra,
and check for correlations with the stellar abundances
of the nebular RL C and O values in K648.

Table~\ref{past} lists the nebular elemental
abundances of K648. These data were determined using the semi-empirical ICF
method, except for \citet{1997MNRAS.284..465H} and
\citet{1980ApaSS..71..393A}, who obtained the abundances using
photo-ionization (P-I) models. We determined the abundances of Ne, S, and Ar, as well as that of CEL C, and added those of RL O, and the CEL F, Cl, and Fe using the HDS and \emph{Spitzer}/IRS spectra
for many ionization stages. Our measurements show good
agreement with those reported previously, with the exception
of those for C and N. Scatter in the CEL
C abundance may be due to the use of {\te} for the C$^{2+}$
abundance and/or the {\hb} flux measurements, because the emissivity
of the C\,{\sc iii}$]$ lines is very sensitive to {\te}. Note that the
observation window of the international ultraviolet explore
(\emph{IUE}) is very large for K648 (window dimension: 10.3$\times$23 arcsec$^{2}$
elliptical shape). The scatter of N abundance may be due to the use of
ICF(N). We will check the CEL C and N abundances in the
P-I model in Section \ref{sedm}. The F abundance is comparable to that in BoBn1
\citep[F/H = 5.98,][]{2010ApJ...723..658O}. The Ne
abundance reported by \citet{Boyer:2006aa} was performed by adding the
Ne$^{+}$ abundance determined from the \emph{Spitzer}/IRS
spectrum, whereas others did not calculate the Ne$^{+}$
abundance. For this reason, our Ne abundance is larger than
has been reported previously, except for \citet{Boyer:2006aa}.

\subsection{Absorption line analysis \label{cspnfit}}

We employed a spectral synthesis fitting method to investigate
the elemental abundances in the photosphere of the CSPN of K648 using O-type star grid models (OStar2002 grid)
based on {\sc TLUSTY} \citep{Lanz:2003aa}, which considers 690 metal line-blanketed,
non-LTE, plane-parallel, and hydrostatic model atmospheres. We considered
the 8 elements He, C, N, O, Ne, S, P, S, Fe, and Ni, together with approximately 100\,000
individual atomic levels from 45 ions \citep[see Table~2 of][]{Bouret:2003aa}.

\subsubsection{Modeling process \label{S:absmodel}}

We found [Ar,Fe/H] abundances of --1.96 and --2.45
from the nebular line analysis, respectively,
and sorted models with a metallicity of $Z$ = 0.01 and 0.001
$Z_{\odot}$ from the OStar2002 grid models. All of the initial abundances in these models
(except He) were set to [X/H] = --2 (0.01 $Z_{\odot}$) models and --3
(0.001 $Z_{\odot}$). The initial ratio of He/H abundances was set to
0.1 in both models.

Following \citet{Bouret:2003aa} and \citet{2002AaA...381.1007R}, we
determined $\log$\,$g$, $T_{\rm eff}$, and the He/H abundance ratio,
which are the basic parameters used for characterizing the photosphere. First, we
generated models with [X/H] = --2.3, corresponding to 0.005 $Z_{\odot}$, by
interpolating between the 0.01 and 0.001 $Z_{\odot}$ grid models using
the {\sc IDL} programs {\sc INTRPMOD} and {\sc INTRPMET}. We set
the microturbulent velocity to 5 {\kms} and the rotational
velocity to 20 {\kms}, because models with these values were found to fit
the absorption line profiles in the \emph{FUSE} and the \emph{HST}/COS spectra, as well as the
HDS spectrum. Before attempting to determine $T_{\rm eff}$ and $\log$\,$g$ using
the stellar absorption lines, we ran the SED models using {\sc
CLOUDY} \citep{Ferland:1998aa} to find the ranges of $T_{\rm eff}$
and $\log$\,$g$ in the {\sc TLUSTY} models. These {\sc CLOUDY}
and SED models maintain the initial photosphere
abundances (i.e., He/H = 0.1 and 0.005 $Z_{\odot}$).
We found that the models can reproduce the observed
\emph{HST}/WFPC2 F547M flux density and the emission line fluxes if we use
the incident SED generated using the {\sc TLUSTY} model
with 0.005 $Z_{\odot}$, $T_{\rm eff}$$\sim$34\,000-40\,000 K, and $\log$\,$g$$\sim$3.5-4.1 cm
s$^{-2}$.

Using the 0.005-$Z_{\odot}$ grid models, we determined
$\log$\,$g$ and He/H by monitoring the chi-squared value of the
HDS He\,{\sc ii}\,$\lambda$\,4541\,{\AA} and the synthesized line profiles of
this line. We ran grid models with $T_{\rm eff}$ = 35\,000-41\,000 K (in 100-K steps),
$\log$\,$g$ = 3.5-4.1 cm s$^{-2}$ (in 0.01-cm s$^{-2}$ steps), and
He/H = 10.98-11.06 (in steps of 0.01). We used {\sc SYNSPEC}
to generate synthesized spectra. We set the spectral
resolution to $R$ = 33\,500 and used a heliocentric radial velocity of
--125.30 {\kms} determined using the He\,{\sc ii}\,$\lambda$\,4541\,{\AA}
absorption line before running {\sc SYNSPEC}. We monitored
the spectrum in the range 4535-4547\,{\AA}. The best fit
was given by $\log$\,$g$ = 3.96$\pm$0.02 cm s$^{-2}$ and
He/H = 11.05$\pm$0.02. In this process, we estimated $T_{\rm
eff}$ = 37\,000 K. Our data are in good agreement with those of
\citet{2002AaA...381.1007R}, who reported $\log$\,$g$ = 3.9$\pm$0.3 cm
s$^{-2}$, $T_{\rm eff}$ = 39\,000$\pm$2000 K and He/H = 10.9$\pm$0.3
obtained using their non-LTE model.

We determined $T_{\rm eff}$ and the abundance of C
assuming that $\log$\,$g$ = 3.96 cm s$^{-2}$ and He/H = 11.05. Here,
we used the C\,{\sc iii} and C\,{\sc iv} lines in the \emph{FUSE}
and the \emph{HST}/COS spectra, including C\,{\sc iii}\,$\lambda$\,1246/47\,{\AA},
C\,{\sc iv}\,$\lambda$\,1107/08\,{\AA} and C\,{\sc iv}\,$\lambda$\,1230/31\,{\AA}.
At approximately $T_{\rm eff}$ = 35\,000-41\,000 K, the strengths of the C\,{\sc iii} lines were
sensitive to $T_{\rm eff}$, whereas those of the C\,{\sc iv} lines were not.
Therefore, we can determine $T_{\rm eff}$
accurately and the abundance of C simultaneously using a plot of
the C abundance as a function of $T_{\rm eff}$. We find
$T_{\rm eff}$ = 36\,360$\pm$700 K and C = 9.38$\pm$0.10.

Using $\log$\,$g$ = 3.96 cm s$^{-2}$ and $T_{\rm
eff}$ = 36\,360 K, we determined the N, O, Ne, P, and Fe abundances to match the observed line profiles. We used {\sc SPTOOL}\footnote[7]{{\sc SPTOOL} is a software package for analyzing
high-dispersion stellar spectra (i.e., line identification,
determination of radial velocity, investigation of the atmospheric parameters, such as
turbulent velocities or elemental abundances), developed
by Youichi Takeda. We also used
the {\sc ATLAS9}/{\sc WIDTH9} packages written by R.~L.~Kurucz.}
for line identification. The N abundance was obtained using
N\,{\sc iii}\,$\lambda$\,1243\,{\AA} and
N\,{\sc iv}\,$\lambda$\,1719\,{\AA}.
The O abundance was found from the many O\,{\sc iii} lines
around 3774\,{\AA} in the HDS spectrum and at $\lambda\lambda$\,1149/51\,{\AA},
O\,{\sc iv}\,$\lambda\lambda$\,1342/44, and O\,{\sc v}\,$\lambda\lambda$\,1371\,{\AA}. The Ne abundance was
found from Ne\,{\sc iii}\,$\lambda$\,1257\,{\AA} only. The P abundance was
determined from the P\,{\sc v}\,$\lambda\lambda$\,1118/28\,{\AA}, and the Fe abundance from
Fe\,{\sc v}\,$\lambda\lambda$\,1448/56\,{\AA}.

\subsubsection{Comparisons between stellar and nebular abundances}

\begin{figure}
\includegraphics[width = \columnwidth,clip,bb = 20 220 530 700,clip]{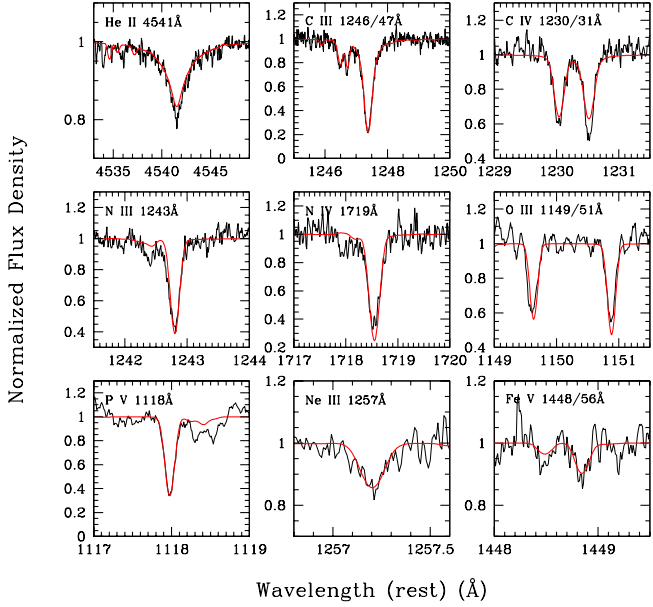}
\caption{The line-profiles of the selected lines observed in the \emph{FUSE}
and \emph{COS} spectra (black lines) and the synthesized spectrum calculated using the
{\sc TLUSTY} model (red lines). The wavelength is shifted to the
 wavelength in vacuum. \label{tmodel_fig}}
\end{figure}

\begin{deluxetable}{@{}lr@{}}
\tablecolumns{2}
\centering
\tablecaption{The central star properties determined using the {\sc TLUSTY} model. \label{tmodel}}
\tablewidth{\columnwidth}
\tablehead{
\colhead{Parameters} &
\colhead{Values}
}
\startdata
\multicolumn{2}{c}{Basic Parameters}\\
\hline
$T_{\rm eff}$ &36\,360$\pm$700 K\\
$\log$\,$g$ &3.96$\pm$0.02 cm s$^{-2}$\\
\hline
\multicolumn{2}{c}{Photosphere Abundances ($\log$(H) = 12)}\\
\hline
He/H ([He/H]) &11.05$\pm$0.02 (+0.12$\pm$0.02)\\
C/H ([C/H]) &9.38$\pm$0.02 (+0.99$\pm$0.04)\\
N/H ([N/H]) &6.53$\pm$0.10 (--1.30$\pm$0.11)\\
O/H ([O/H]) &8.36$\pm$0.10 (--0.33$\pm$0.11)\\
Ne/H ([Ne/H]) &8.21$\pm$0.10 (+0.34$\pm$0.14)\\
P/H ([P/H]) &3.64$\pm$0.10 (--1.82$\pm$0.11)\\
Fe/H ([Fe/H]) &5.23$\pm$0.10 (--2.24$\pm$0.10)
\enddata
\end{deluxetable}

The resulting spectrum synthesized using the {\sc TLUSTY} and the observed
\emph{FUSE} and \emph{HST}/COS spectra are shown in
Fig.~\ref{tmodel_fig}. The parameters, including the elemental
abundances, are listed in Table~\ref{tmodel}.
The derived stellar abundances are also listed in Table~\ref{past} for comparison.
We were unable to detect any F absorption lines, e.g.,
F\,{\sc v}\,$\lambda$\,1082/87/88\,{\AA}
due to the low S/N ratio. The detection of the single isotope $^{31}$P is
interesting because phosphorus (along with fluorine) is an $n$-capture element that is synthesized in the
He-rich intershell during the TP-AGB phase.

We found that the stellar C and O abundances were close to the nebular RL
abundances; however, the stellar He and Ne abundances were larger than the nebular
abundances. The stellar N and Fe abundances were comparable to the
nebular abundances. We may expect slightly higher stellar C, O, and Ne
abundances than the nebular abundances, as the former
are indicative of more recent products of AGB
nucleosynthesis. These three elements are synthesized in the He-rich
intershell during the AGB phase, and are then brought up to
the stellar surface via the TDU.
Note that the stellar
C/O and the Ne/O ratios (10.75$\pm$2.43 and 0.75$\pm$0.23,
respectively) are in good agreement with nebular ratios
C/O (17.46$\pm$7.07 in RL and 6.83$\pm$3.63 in CEL) and Ne/O (0.51$\pm$0.06)
in CEL. Although it is difficult to determine whether the RL or CEL abundance
represents the nebular C and O chemical abundances in K648,
the similarity of the C/O ratios determined from the RLs and CELs
indicates a positive correlation with the stellar abundance.

K648 shows large stellar and nebular CEL [O/Fe]
abundances (1.91$\pm$0.15 dex versus 1.49$\pm$0.13 dex). The
[Ne/Fe] abundances were also large (2.01$\pm$0.16 dex in the CEL and
2.58$\pm$0.18 dex in the stellar region). It has been
reported that metal-poor stars in the Milky Way exhibit large
[$\alpha$/Fe] abundances, where the $\alpha$-elements include O, Ne, Mg, Si,
and Ca. The effect is greatest for the most metal-poor
populations, such as members of the stellar halo and, in particular, in the
[O/Fe] \citep[see, for example,][]{McWilliam:1997aa,Feltzing:2013aa}.
This is interpreted as a consequence of time delay in Fe
production from Type~Ia SNe relative to the $\alpha$-elements from
core-collapse SNe. The $\alpha$-elements are mainly produced by
Type~II SNe. Both types of SNe should produce Fe in the proportions
of $\sim$1/3 for Type~II and $\sim$2/3 for Type~Ia SNe. For
M15, \citet{Sobeck:2011aa} reported that
three red giant branch (RGB) stars ($\langle$[Fe/H]$\rangle$ = --2.55),
exhibited O abundance of 6.75-7.03 and the [O/Fe] of +0.62 to
+0.85 ($\langle$[O/Fe]$\rangle$ = +0.75). If we observe an RGB star with
[Fe/H] = --2.3, [O/Fe] should be +0.50, which corresponds to
an O abundance of 6.89.

The difference between the [O/Fe] of M15 RGB stars reported by
\citet{Sobeck:2011aa} and that
of K648 suggests that, in K648, O synthesis was $\gtrsim$0.9 dex during the
TP-AGB phase. The TP-AGB phase nucleosynthesis process can contribute to
enhancement of O and Ne abundances in the helium convective zone with
$^{13}$C formed from mixed protons as an $n$-source using a nuclear
network from H through S. The abundance of $^{16}$O may increase in
proportion to the square root of the amount of mixed $^{13}$C until
it reaches a significant fraction of $^{12}$C, whereas the abundance
of $^{22}$Ne may increase in proportion to the amount of mixed
$^{13}$C, and attains half of the mixed $^{13}$C
\citep{Nishimura:2009aa}.
Indeed, \citet{Lugaro:2012aa} demonstrated that [Fe/H] = --2.19
AGB stars can synthesize significant quantities of O and Ne (see Section \ref{S:agb}).

\subsubsection{The core-mass of the CSPN \label{S:core mass}}
The core-mass of the CSPN can impose a significant constraint
on the initial mass of the progenitor.
Through construction of a {\sc TLUSTY} model atmosphere, we obtained the
$H_{\lambda}$ spectrum of the stellar photosphere. Using the
$H_{\lambda}$ and the observed \emph{HST}/WFPC2 F547M flux density
$I_{\lambda}$ listed in Table~\ref{wfpc}, we
determined the core-mass of the CSPN $M_{c}$ using Equation (1) of
\citet{Shipman:1979aa}, i.e.,
\begin{eqnarray}
I_{\lambda} & = & 4\,\pi\,H_{\lambda}\,R^{2}\,D^{-2} \label{m1}, \\
g & = &G\,M_{c}\,R^{-2} \label{m2},
\end{eqnarray}
\noindent where $R$ is the radius of the CSPN, $D$ is the distance to
K648 from us, $g$ is the surface gravity of the CSPN, and $G$ is the
gravitational constant.

Using the synthesized spectrum from {\sc TLUSTY} model atmosphere fitting, we found that
$H_{\lambda}$ = 4.53(+7) erg s$^{-1}$ cm$^{-2}$
{\AA}$^{-1}$ at $\lambda$\,5483.88\,{\AA} by taking the transmission
curve of the \emph{HST} WFPC2/F547M band into account.
Recent measurements of the distance to M15 have been reported by \citet[12.3$\pm$0.6 kpc]{Reid:1996aa},
\citet[9.98$\pm$0.47 kpc]{McNamara:2004aa}, and \citet[10.3$\pm$0.4
kpc]{van-den-Bosch:2006aa}. We find $\log$\,$g$ = 3.96$\pm$0.02 cm
s$^{-2}$, as determined in Section \ref{S:absmodel}.

If we use the average distance amongst these distance measurements, i.e., 10.9$\pm$0.5 kpc.
we obtain $M_{c}$ = 0.68$\pm$0.07 $M_{\odot}$
and $R$ = 1.43$\pm$0.08 $R_{\odot}$ using Equations
(\ref{m1}) and (\ref{m2}).
Using the most recent data, i.e., $D$ = 10.3$\pm$0.4 kpc,
the values of $M_{c}$ and $R$ are
$M_{c}$ = 0.61$\pm$0.06 $M_{\odot}$ and 1.35$\pm$0.08 $R_{\odot}$,
which are
in agreement with \citet{2001AJ....122.1538B} and
\citet{2002AaA...381.1007R}.
We found that $R$ = 1.3 $R_{\odot}$
and $M_{c}$ of 0.62$\pm$0.10 $M_{\odot}$ with $D$ = 10.3 kpc and
$\log$\,$g$ = 4.0 cm s$^{-2}$.
\citet{2002AaA...381.1007R} calculated $M_{c}$ = 0.57 $M_{\odot}$ from the
theoretical $T_{\rm eff}$-$\log$\,$g$ diagram. The exact value of
the $M_{c}$ is still dependent on the choice of distance. We will
discuss the initial mass of K648 in section \ref{S:agb}.

\subsection{Dust features in the \emph{Spitzer}/IRS spectrum \label{S:dust}}

As discussed in Section \ref{S:spit}, K648 exhibits the 6-9\,$\mu$m PAH
band, the 11.3\,$\mu$m PAH band, and the broad 11\,$\mu$m feature.
These PAH bands are sometimes seen in C-rich PNe, such as
BD+30$^{\circ}$ 3639
\citep[C/O = 1.59,][]{2003AaA...406..165B,Waters:1998aa}, as well as O-rich
PNe such as NGC6302
\citep[C/O = 0.43,][]{2011MNRAS.418..370W,Molster:2001aa}. Both
BD+30$^{\circ}$ 3639 and NGC6302 exhibit strong crystalline
silicate features at 23.5, 27.5, and
33.8\,$\mu$m, which have never been observed in K648. In addition,
the 9 and 18\,$\mu$m features attributed to the amorphous silicate
were also not seen in K648. Therefore, we concluded that K648 is a
C-rich gas-and-dust PN.

\begin{figure}
\includegraphics[bb = 25 316 572 681,width = \columnwidth,clip]{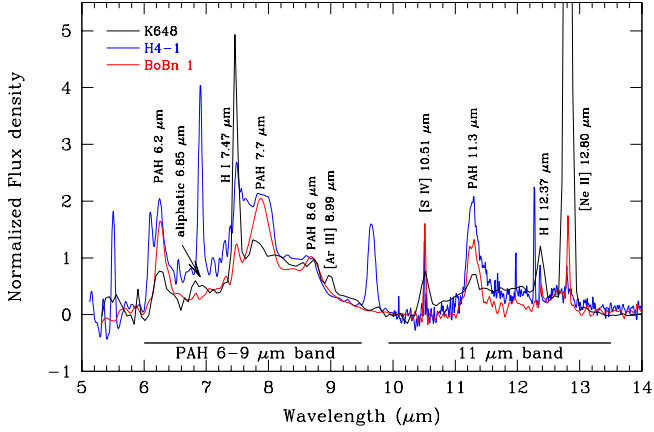}
\caption{The 5-15\,$\mu$m
spectra of C-rich halo PNe K648 (black), H4-1 (blue,
from \citealt{Tajitsu:2014aa}),
and BoBn1 (red, from \citealt{2010ApJ...723..658O}).
The local continuum was subtracted from the observed flux,
which was then normalized to that at 8.6\,$\mu$m.
The data for K648 are obtained from
from the SL module. The 5-10\,$\mu$m spectra of H4-1 and BoBn1 were obtained from
the SL module and the remaining data from the SH module.
\label{spitzer_spec2}}
\end{figure}

Figure~\ref{spitzer_spec2} shows the 5-15\,$\mu$m spectrum, where the local
dust continuum was subtracted by fourth-order
spline fitting, using the same technique as applied for C-rich PNe by
\citet{Otsuka:2014aa}. The flux density was then normalized to the
intensity of the 8.6\,$\mu$m PAH band. For comparison, we also show the
\emph{Spitzer}/IRS spectra of the C-rich halo PNe H4-1
\citep{Tajitsu:2014aa}, as well as that of BoBn1 \citep{2010ApJ...723..658O}.
We discuss the dust features in more detail below.

\subsubsection{The 6-9\,$\mu$m and 11.3\,$\mu$m PAH bands}

\begin{deluxetable}{@{}lccc@{}}
\tablecolumns{4}
\centering
\tablecaption{The results of Gaussian fittings to the PAH bands,
the {\it possible} 6.85\,$\mu$m aliphatic feature, and the broad 11\,$\mu$m band.\label{t:pah}}
\tablewidth{\columnwidth}
\tablehead{
\colhead{$\lambda_{c}$} &
\colhead{FWHM} &
\colhead{$F$($\lambda$)}&
\colhead{$I$($\lambda$)}\\
\colhead{($\mu$m)}&
\colhead{($\mu$m)}&
\colhead{(erg s$^{-1}$ cm$^{-2}$)}&
\colhead{
[$I$({\hb}) = 100]
}
}
\startdata
~~6.25$\pm$0.01 &0.22$\pm$0.02  &2.25(--14)$\pm$1.88(--15)&  ~~2.20$\pm$0.21\\
~~6.47$\pm$0.01$^{\rm a}$ &0.13$\pm$0.03 &4.24(--15)$\pm$9.82(--16)& ~~0.41$\pm$0.10\\
~~6.85$\pm$0.02$^{\rm b}$ &0.21$\pm$0.04 &6.01(--15)$\pm$1.65(--15)& ~~0.59$\pm$0.16\\
~~7.83$\pm$0.01 &0.25$\pm$0.03  &1.34(--14)$\pm$2.12(--15)&
~~1.31$\pm$0.22\\
&&&~~2.17$\pm$0.13$^{\rm d}$\\
~~8.73$\pm$0.01 &0.16$\pm$0.04  &5.37(--15)$\pm$1.55(--15)&  ~~0.53$\pm$0.15\\
11.31$\pm$0.01  &0.25$\pm$0.01  &1.00(--14)$\pm$6.46(--16)&  ~~0.98$\pm$0.08\\
11.81$\pm$0.03$^{\rm c}$ &1.97$\pm$0.14 &1.23(--13)$\pm$1.09(--14)&12.10$\pm$1.21
\enddata
\tablenotetext{a}{He\,{\sc i} 6.47\,$\mu$m.}
\tablenotetext{b}{The complex of the {\it possible} aliphatic 6.85
\,$\mu$m and the [Ar\,{\sc
ii}]\,$\lambda$\,6.99\,$\mu$m lines. See Section \ref{alip} regarding the respective intensities.}
\tablenotetext{c}{The broad 11\,$\mu$m band.}
\tablenotetext{d}{The extrapolated value using the PAH $I$($\lambda$\,7.7
$\mu$m)/$I$($\lambda$\,8.6\,$\mu$m) ratio in BoBn1.}
\end{deluxetable}

The 6-9\,$\mu$m and 11.3\,$\mu$m PAH band profiles are remarkably
similar to those of H4-1 and BoBn1, although the intensity peak of the 7.7\,$\mu$m
PAH in K648 is smaller than those of BoBn1 and H4-1, which is attributed to noise
around 7.7\,$\mu$m.

We measured the central wavelength $\lambda_{c}$, FWHM, flux $F(\lambda)$, and relative
intensity $I(\lambda)$ of each PAH band by single Gaussian
fitting, and the results are shown in Table~\ref{t:pah}.
For the 6.2\,$\mu$m band, we employed a double Gaussian
component fit, where one component corresponds to the 6.25\,$\mu$m PAH band and the other to
the He\,{\sc i}\,$\lambda$\,6.47\,$\mu$m.

\citet{Peeters:2002aa} examined the profiles of the 6.2, 7.7, and 8.6\,$\mu$m
PAH bands using \emph{ISO}/SWS spectra, and classified the spectra into Classes A, B, and
C according to the peak positions of each PAH feature. Class B
PAHs are frequently seen in C-rich PNe, including BoBn1 and H4-1, and
have a peak in the range 6.235-6.28\,$\mu$m,
a stronger component at $\sim$7.8\,$\mu$m than at 7.6\,$\mu$m,
and a peak at $>$8.62\,$\mu$m. The 6.2, 7.7, and
8.6\,$\mu$m features in K648 satisfy the definition of a Class B PAH
spectrum.

According to the classification of the 11.3\,$\mu$m PAH profiles by
\citet{van-Diedenhoven:2004aa}, the 11.3\,$\mu$m PAH in K648 falls under
Class B$_{11.2}$, with a peak at $\sim$11.25\,$\mu$m. Many C-rich
PNe in the Magellanic Clouds also have this class of PAH \citep{2009ApJ...699.1541B}.

\subsubsection{The 6.85\,$\mu$m aliphatic feature? \label{alip}}

K648 exhibits a weak broad feature at 6.85\,$\mu$m, which might be
a combination of the 6.85\,$\mu$m aliphatic feature (CH$_{2,3}$ asymmetric deformation)
and [Ar\,{\sc ii}]\,$\lambda$\,6.99\,$\mu$m. \citet{Otsuka:2014aa} established a relationship
between $T_{\rm eff}$ and the $I$({\ariii}\,$\lambda$\,8.99\,$\mu$m)/$I$([Ar\,{\sc ii}]\,$\lambda$\,6.99\,$\mu$m) ratio in C-rich PNe
based on P-I  models with {\sc Cloudy} code.
Using their Equation (A1) and $T_{\rm eff}$ = 37\,100 K (see
Section \ref{cspnfit}),
we found that the 6.85\,$\mu$m aliphatic feature and the [Ar\,{\sc ii}]\,$\lambda$\,6.99\,$\mu$m
intensities are 0.37$\pm$0.09 and 0.21$\pm$0.13,
respectively, where
the {\hb} intensity is 100.

Following \citet{Li:2012aa}, we estimated the number ratio of C-atoms in aliphatic form relative to those in
aromatic form using the 6-9\,$\mu$m PAH band, i.e., $N_{\rm C, aliph}$/$N_{\rm
C,arom}$. As we underestimated the 7.7\,$\mu$m PAH flux in
K648, we extrapolated a 7.7\,$\mu$m PAH intensity of
2.17$\pm$0.13 using the PAH $I$(7.7\,$\mu$m)/$I$(8.6\,$\mu$m) ratio of 4.11$\pm$0.13 measured
in BoBn1.
Our derivation is $N_{\rm C, aliph}$/$N_{\rm C,arom}$ of $\sim$0.1-0.4,
indicating that $<$29\% of the C-atoms exists in aliphatic form in K648.

For a more accurate estimate of the number of C-atoms in the aliphatic
form, $L$-band spectroscopy is useful to check for the existence of
the 3.4\,$\mu$m aliphatic feature, as well as $I$(3.3\,$\mu$m PAH)/$I$(3.4\,$\mu$m aliphatic).

\subsubsection{The broad 11\,$\mu$m band\label{S:11um}}

\begin{figure}
\includegraphics[bb = 45 170 512 714,width = \columnwidth]{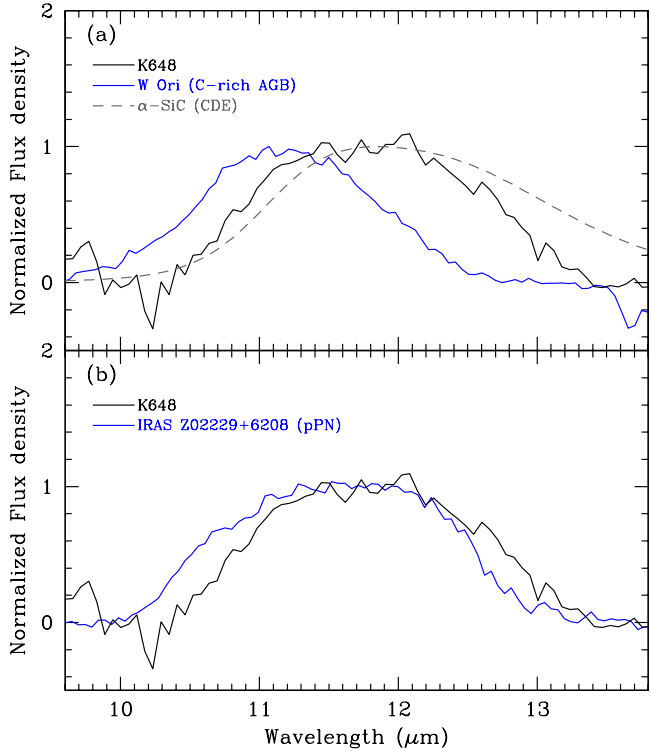}
\caption{The broad 11\,$\mu$m band profiles. The 11.3\,$\mu$m PAH and the
atomic lines were subtracted out via Gaussian fitting in both (a)
and (b). The flux density is shown normalized to the intensity peak.
The spectral resolution of the C-rich AGB star W Ori and the proto PN
IRAS Z02229+6208 was adjusted to match that of K648.
\label{spitzer_spec3}}
\end{figure}

K648 exhibits the broad 11\,$\mu$m feature, which is frequently seen in Galactic and Magellanic C-rich PNe
\citep[e.g.,][]{Otsuka:2014aa,2009ApJ...699.1541B,Stanghellini:2012aa}.
The band profile appears to show an almost flat portion in the
range 11.4-12.2\,$\mu$m.
However, as shown in Fig.~\ref{spitzer_spec3}(a), the resulting band profile did not exhibit
a flat top after removal of the 11.3\,$\mu$m PAH band and the
atomic lines.

The results of Gaussian fitting are also listed
in Table~\ref{t:pah}. The FWHM of the 11\,$\mu$m band is comparable to
those for H4-1 (2.08$\pm$0.05\,$\mu$m) and BoBn1 (1.85$\pm$0.39\,$\mu$m);
however, $\lambda_{c}$ was slightly blue-shifted
(12.28$\pm$0.06\,$\mu$m in H4-1 and 12.30$\pm$0.08\,$\mu$m in BoBn1).
Our results corroborate those of \citet{2009ApJ...699.1541B}, who reported
that the profile and the central wavelength of the 11\,$\mu$m band in
MC PNe differ from source to source.

There is some debate regarding the origin of the broad 11\,$\mu$m feature.
Silicon carbide (SiC) is one possible explanation for the feature at
11\,$\mu$m in C-rich MC PNe \citep{2009ApJ...699.1541B}.
In Fig.~\ref{spitzer_spec3}(a), as a SiC template, we show a comparison
with the 11\,$\mu$m band profile of the
Galactic solar metallicity C-rich AGB star W Ori, extracted from the archive \emph{ISO}/SWS
spectrum. These data were downloaded from \citet{Sloan:2003aa}. \citet{Abia:2002aa}
reported a C/O ratio of 1.005, and a metallicity of [M/H] = +0.05. The $\lambda_{c}$
(11.2\,$\mu$m) and FWHM (1.51\,$\mu$m) in the 11\,$\mu$m band of W Ori
differ significantly from those measured for K648. The 11\,$\mu$m band profile
in W Ori may be fitted to an absorption efficiency $Q_{\lambda}$ of a spherical
$\alpha$-SiC grain (or 6-H SiC, hexagonal unit cell) calculated from \citet{Pegourie:1988aa}, which
peaks sharply at $\sim$11.2\,$\mu$m and has an FWHM of $\sim$1.2\,$\mu$m. However, we must be careful with $\lambda_{c}$ in
W Ori. \citet{Leisenring:2008aa} demonstrated how the C$_{2}$H$_{2}$ absorption band around 13.7\,$\mu$m, as well as the SiC
self-absorption band around 10\,$\mu$m affect the central wavelength of
SiC in AGB stars such as W Ori. They argued that the C$_{2}$H$_{2}$ absorption band suppressed
the long-wavelength part of the feature at 11\,$\mu$m, and caused the central wavelength
to be blue-shifted.
This may be the case for W Ori.
We could fit neither $\lambda_{c}$ nor the FWHM of the 11\,$\mu$m band, even with a continuous distribution of ellipsoids \citep[CDE,
e.g.,][]{Bohren:1983aa,Min:2003aa} of $\alpha$-SiC; we find $\lambda_{c}\sim11.8 \mu$m and a FWHM of $\sim$2\,$\mu$m using the $Q_{\lambda}$ for the CDE $\alpha$-SiC, as shown by the by the gray line in Fig.~\ref{spitzer_spec3}(a).

\citet{Kwok:2001aa} argued that a collection of out-of-plane
bending modes of aliphatic side groups attached to an aromatic ring
could result in to the broad 11\,$\mu$m feature. Indeed, we found
the feature at 6.85\,$\mu$m in K648, corresponding to {\it possibly} aliphatic C.
Figure~\ref{spitzer_spec3}(b) shows a
comparison of the 11\,$\mu$m band
profiles of K648 and the proto-PN IRAS Z02229+6208. The [C/H] and
[M/H] abundances of this proto-PN are +0.29 and --0.50, respectively, \citep{Reddy:1999aa}.
The measured values of $\lambda_{c}$ and the FWHM of IRAS Z02229+6208 are 11.68$\pm$0.02\,$\mu$m and 1.81$\pm$0.05\,$\mu$m, respectively.
The 11\,$\mu$m band profile in IRAS Z02229+6208
exhibits a good fit to that of K648, except for the $\lesssim$11\,$\mu$m part
of the 11\,$\mu$m band. According to \citet{Kwok:2001aa}, K648 may have a
few cyclic alkanes, which contribute to the 9.5-11.5\,$\mu$m part
of this band \citep[see Fig.~4 of][]{Kwok:2001aa}.

The low metallicity of K648 implies a very low abundance of Si. Indeed, we did not detect
any lines corresponding to Si in either the nebula or the central star. Therefore,
we expect that the broad 11\,$\mu$m band profile in K648 is
attributable to a wide variety of alkane and alkene groups attached
to hydrogenated aromatic rings, rather than to SiC.

\subsection{Radiative transfer modeling and SED fitting\label{sedm}}

We constructed an SED model to investigate the physical
conditions of the gas and dust grains and derive their
masses using {\sc Cloudy} c10.00. The quantity of dust
mass formed in extremely metal-poor objects such as K648 is of
interest. The gas mass as well as the core-mass of the CSPN are
required to unveil the origin and evolution of K648 via a
comparison of these parameter values with the results of AGB nucleosynthesis models.

\subsubsection{Modeling approach}

\begin{figure}
\includegraphics[bb = 40 179 566 593,width = \columnwidth,clip]{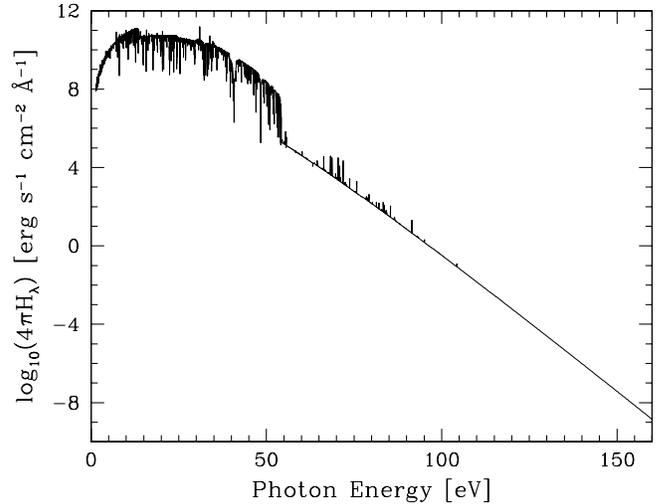}
\caption{The SED of the CSPN of K648 synthesized using {\sc TLUSTY}.
This SED was adopted in the dust+gas SED model using {\sc Cloudy} code as the incident SED.
\label{sed_cspn}}
\end{figure}

\begin{figure}
\centering
\includegraphics[width = \columnwidth,bb = 38 340 565 696,clip]{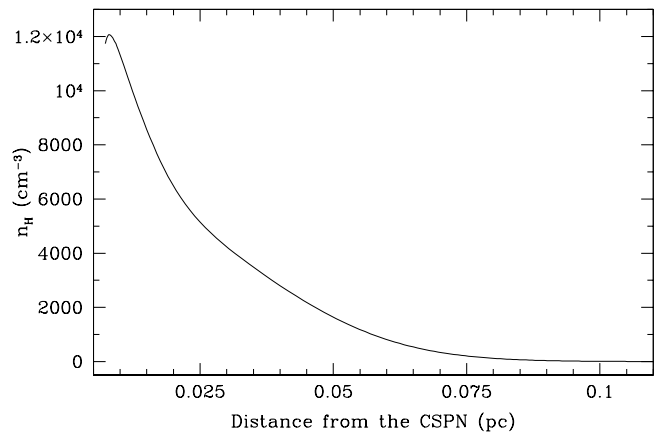}
\caption{The hydrogen density radial profile used in the {\sc
Cloudy} P-I SED modeling.
\label{density}}
\end{figure}

We attempted to fit the observed SED in the range 0.1-160\,$\mu$m,
assuming that the dust in K648 is composed of PAH molecules and
amorphous carbon (AC) grains. No SiC grains were considered
to fit the broad 11\,$\mu$m feature.

The distance to K648 is required to compare our model values with the observed
fluxes; here we assumed a distance of 10.9 kpc.
For the incident SED from the central star, we used the synthesized
spectrum of the central star of K648 using the {\sc TLUSTY} model,
as discussed in Section \ref{cspnfit}. The input SED is shown in
Fig.~\ref{sed_cspn}. We adjusted the input SED to match the de-reddened absolute
$V$-band magnitude of --0.528 measured from \emph{HST}/F547M photometry
of the CSPN. The number of Lyman continuum photons $N_{\rm Lyc}$
with $>$13.5 eV was 6.92(+45) s$^{-1}$, as determined from
the synthesized spectrum of the CSPN.

The P-I model construction with the {\sc Cloudy} and other codes
involves an ad hoc nebular geometry and central stellar property
modification. Until it gives a right prediction to the line intensities
and continuum, one must adjust not only the chemical abundances but also
the model nebular geometry with a new value close to the observation
indication. In order to tune the other diagnostically indicated physical
properties, e.g., electron temperature, one even needs to consider other
chemical elements which were not observed at all. We employed the observed values of the gas-phase elemental abundances
listed in Table~\ref{abund} as initial estimates, and refined
these to match the observed line intensities of each element.
We considered the RL and CEL C line fluxes and the observed CEL O line
fluxes to determine the nebular C and O
abundances, respectively. We revised the transition probabilities and
collisional impacts of C~{\sc iii}], [N~{\sc
ii}], [O~{\sc ii, iii}], [F\,{\sc ii,iv}], [Ne~{\sc ii,iii,iv}], [S~{\sc
ii,iii}], [Cl\,{\sc ii,iii}] and [Ar~{\sc ii,iii,iv}], which were the
same as those used in our semi-empirical ICF abundance calculations
 (i.e., using IP coincidence method). 
The abundances of other elements were fixed to be constant, with [X/H] = --2.3.

We determined the radial hydrogen density profile of the nebula based on the
radial intensity profile of the \emph{HST}/WFPC2 F656N image
using Abel transformation, and assuming spherical symmetry. We
fixed the outer radius to $R_{\rm out} = 2.1''$ (0.11 pc) and the inner radius to
$R_{\rm in} = 0.14''$ (0.0072 pc). We used a constant filling
factor of $\epsilon = 0.5$. The hydrogen density radial profile
is shown in Fig.~\ref{density}. The $R_{\rm out}$ that was used
corresponds to the Str\"{o}mgren radius (0.11 pc), assuming
{\te} = 10$^{4}$ K, $n({\rm H^{+}})$ = {\Ne} = 3000
cm$^{-3}$, and the same values of $\epsilon$ and $N_{\rm Lyc}$

We assume that both PAH molecules and AC grains exist in the nebula, and that
the observed IR-excess from MIR to FIR wavelengths is
due to the thermal emission from these species.
We assumed spherical AC grains and PAH molecules. The optical
constants were taken from \citet{2007ApJ...657..810D} for PAHs and
from \citet{1991ApJ...377..526R} for the AC grains. For the PAHs, we
assumed that the radius was in the range of 0.0004-0.0011\,$\mu$m
(i.e., 30-500 C atoms) with an $a^{-3.5}$ size distribution. For the AC grains,
we used the standard interstellar dust grain size distribution reported by
\citet{Mathis:1977aa}, i.e., an $a^{-3.5}$ size distribution, but with a
smaller radius of 0.0005-0.010\,$\mu$m, which was determined by running several test models.

To evaluate the degree of accuracy of the model fitting, we calculated the
chi-square ($\chi^{2}$) value from the 39 gas emission fluxes,
10 gas-phase abundances, and the five broad band fluxes, as well as the 15
flux densities of the features of interest from UV to FIR wavelengths.

\subsubsection{Modeling results and SED fitting}

\begin{figure*}
\centering
 \includegraphics[bb = 26 170 566 593,width = 0.75\textwidth,clip]{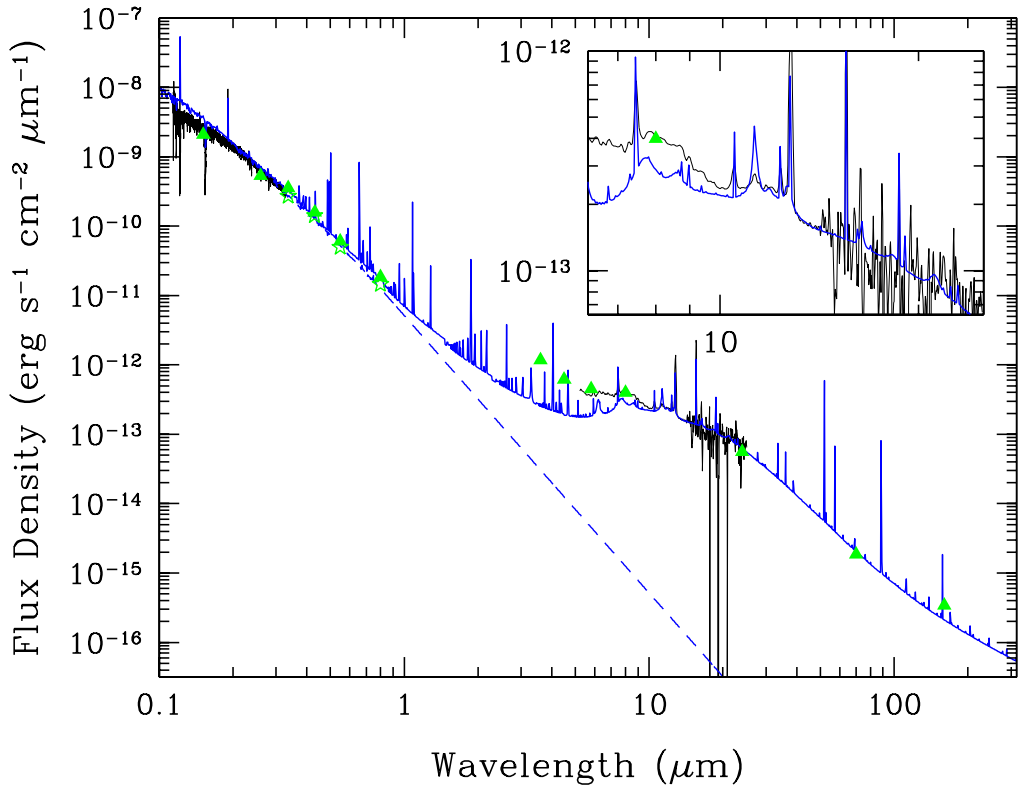}
\caption{The SED of K648. The blue solid and
broken lines are the predicted SED of the sum of the CSPN and the nebula and
the incident SED of the CSPN, respectively. The black lines are
the observed spectra obtained from the \emph{HST}/FOS and
\emph{Spitzer}/IRS. The Subaru/HDS spectrum is not shown because
its appeared to be partially contaminated by foreground stars. Instead,
we plotted the \emph{HST}/WFPC2 photometry, indicated by the green
asterisks (CSPN) and triangles (CSPN+PN). The green triangles in
the MIR and FIR (3.6/4.5/5.8/8.0/24/70/100/160\,$\mu$m)
are \emph{Spitzer}/IRCS/MIPS and \emph{Herschel}/PACS photometry.
\label{sed}}
\end{figure*}

\begin{deluxetable}{@{}ll@{}}
\tablecolumns{2}
\centering
\tablecaption{The properties from the {\sc Cloudy} P-I model. \label{model}}
\tablewidth{\columnwidth}
\tablehead{
\colhead{Parameters} &
\colhead{Values}
}
\startdata
&Central Star\\
\hline
$M_{\rm V}$ &--0.528, measured from \emph{HST}/F547M obs\\
$L_{\ast}$ &3076 $L_{\odot}$\\
$T_{\rm eff}$ &36\,360 K \\
$\log\,g$ &3.96 cm s$^{-2}$\\
Distance &10.9 kpc\\
\hline
 &Nebula\\
\hline
Abundances$^{\rm a}$ &He:11.00, C:8.71, N:6.96, O:7.82,\\
($\log$$n$(X)/$n$(H)+12) &F:5.41, Ne:7.02, S:5.48, Cl:3.44\\
 &Ar:4.44, Fe:5.48, Others:[X/H] = --2.3\\
Geometry &Spherical\\
Shell size &$R_{\rm in}$ = 0.14 {\arcsec} (0.0072 pc),
$R_{\rm out}$ = 2.1 {\arcsec} (0.125 pc)\\
$n_{\rm H}$ &See Fig.~\ref{density}\\
filling factor &0.50\\
$\log I$({\hb}) &--11.972 erg s$^{-1}$ cm$^{-2}$ (de-redden)\\
$m_{\rm g}$ &4.81(--2) $M_{\odot}$\\
\hline
 &Dust in Nebula\\
\hline
Composition &PAHs, amorphous carbon (AC)\\
Grain size &0.0005-0.010\,$\mu$m for AC\\
 &0.0004-0.011\,$\mu$m for PAH\\
$T_{d}$(PAHs) &140-472 K\\
$T_{d}$(AC) &99-290 K \\
$m_{d}$(Tot.)$^{\rm b}$&4.95(--7) $M_{\odot}$ \\
$m_{d}$(Tot.)/$m_{\rm g}$ &1.029(--5)
\enddata
\tablenotetext{a}{The relative error of the gas-phase elemental abundances
is within 0.2 dex. }
\tablenotetext{b}{The total dust mass of the PAH and AC grains.}
\end{deluxetable}

\begin{deluxetable}{@{}lcrrr@{}}
\tablecolumns{5}
\tabletypesize{\footnotesize}
\centering
\tablecaption{Comparison between the results of the P-I model and the
observations. \label{model2}}
\tablewidth{\columnwidth}
\tablehead{
\colhead{Ion} &
\colhead{$\lambda$} &
\colhead{$I$(P-I)} &
\colhead{$I$(Obs)} &
\colhead{$\Delta$$I$/$I$(Obs)} \\
\colhead{} &
\colhead{({\AA}/$\mu$m)} &
\colhead{[$I$(H$\beta$) = 100]} &
\colhead{[$I$(H$\beta$) = 100]} &
\colhead{($\%$)}
}
\startdata
C\,{\sc iii}$]$ & 1906/09 & 468.701 & 334.984 & 39.92 \\
$[$C\,{\sc ii}$]$ & 2323 & 25.134 & 17.091 & 47.06 \\
{\oii} & 3726 & 17.646 & 17.383 & 1.51 \\
{\oii} & 3729 & 9.385 & 9.387 & 0.02 \\
{\neiii} & 3869 & 10.255 & 9.939 & 3.18 \\
{\neiii} & 3968 & 3.091 & 3.147 & 1.78 \\
C\,{\sc ii} & 4267 & 0.492 & 0.726 & 32.23 \\
H$\gamma$ & 4340 & 46.974 & 46.674 & 0.64 \\
{\oiii} & 4363 & 2.273 & 2.782 & 18.31 \\
{\hei} & 4388 & 0.640 & 0.517 & 23.84 \\
{\hei} & 4471 & 5.218 & 4.914 & 6.18 \\
{\fii} & 4791 & 0.105 & 0.110 & 4.94 \\
{\fii} & 4870 & 0.033 & 0.030 & 8.83 \\
{\feiii} & 4881 & 0.054 & 0.051 & 4.92 \\
{\hei} & 4922 & 1.383 & 1.290 & 7.23 \\
{\oiii} & 4931 & 0.032 & 0.038 & 14.87 \\
{\oiii} & 4959 & 79.089 & 74.974 & 5.49 \\
{\oiii} & 5007 & 238.058 & 227.263 & 4.75 \\
{\cliii} & 5518 & 0.022 & 0.021 & 3.14 \\
{\cliii} & 5538 & 0.027 & 0.028 & 3.50 \\
{\nii} & 5755 & 0.064 & 0.052 & 22.17 \\
{\hei} & 5876 & 15.676 & 14.834 & 5.67 \\
{\siii} & 6312 & 0.147 & 0.119 & 23.55 \\
{\nii} & 6548 & 0.979 & 0.901 & 8.70 \\
H$\alpha$ & 6563 & 282.047 & 282.399 & 0.12 \\
{\nii} & 6584 & 2.890 & 3.180 & 9.12 \\
{\hei} & 6678 & 4.182 & 4.114 & 1.65 \\
{\sii} & 6716 & 0.069 & 0.087 & 20.98 \\
{\sii} & 6731 & 0.110 & 0.133 & 17.23 \\
{\ariii} & 7135 & 0.389 & 0.384 & 1.21 \\
{\oii} & 7323 & 1.573 & 1.792 & 12.20 \\
{\oii} & 7332 & 1.256 & 1.450 & 13.39 \\
H\,{\sc i} & 7.47 & 3.192 & 3.150 & 1.35 \\
{\ariii} & 9.00 & 0.296 & 0.324 & 8.64 \\
{\siv} & 10.51 & 1.095 & 1.064 & 2.95 \\
{\neiii} & 15.55 & 8.736 & 11.545 & 24.33 \\
{\neii} & 12.80 & 3.645 & 14.980 & 75.67 \\
{\siii} & 18.71 & 2.343 & 1.332 & 75.93 \\
{\siii} & 33.47 & 0.893 & 0.612 & 45.89 \\
IRS-1 & 8.55 & 13.365 & 15.859 & 15.73 \\
IRS-2 & 9.825 & 4.809 & 4.560 & 5.46 \\
IRS-3 & 12.03 & 4.973 & 4.531 & 9.75 \\
IRS-4 & 14.00 & 2.672 & 2.322 & 15.08 \\
IRS-5 & 25.50 & 9.044 & 9.136 & 1.01 \\
\hline
Band &$\lambda$&$F_{\nu}$(P-I) &$F_{\nu}$({\sc Obs}) & $\Delta$$F_{\nu}$/$F_{\nu}$(Obs)\\
 &({\AA}/$\mu$m) &(mJy)&(mJy) &($\%$)\\
\hline
F160BW & 1515 & 26.647 & 15.993 & 66.61 \\
F170W & 1820 & 25.808 & 20.886 & 23.57 \\
F255W & 2599 & 15.952 & 11.907 & 33.97 \\
F300W & 2989 & 13.823 & 10.543 & 31.11 \\
F336W & 3360 & 12.211 & 13.393 & 8.82 \\
F439W & 4312 & 9.554 & 9.793 & 2.43 \\
F547M & 5484 & 5.810 & 6.025 & 3.56 \\
F814W & 7996 & 3.701 & 3.921 & 5.62 \\
IRAC-1 & 3.51 & 1.273 & 5.096 & 75.01 \\
IRAC-2 & 4.50 & 1.499 & 4.158 & 63.95 \\
IRAC-3 & 5.63 & 2.035 & 5.046 & 59.67 \\
IRAC-4 & 7.59 & 4.734 & 8.510 & 44.37 \\
MIPS-1 & 23.21 & 11.300 & 10.684 & 5.77 \\
PACS-B & 68.93 & 3.207 & 2.950 & 8.71 \\
PACS-R & 153.9 & 1.804 & 2.680 & 32.69 \\
\hline
${\chi}^{2}$& & & & 88.12
\enddata
\tablecomments{The data in the IRS-1, 2, 3, 4, and 5 bands
are the integrated fluxes between the following wavelengths: 8.26-8.84\,$\mu$m,
9.7-9.95\,$\mu$m, 11.9-12.16\,$\mu$m, 13.9-14.1\,$\mu$m and 24.5-26.5
$\mu$m, respectively. Data are shown with two or three decimal places
to avoid rounding errors.}
\end{deluxetable}

\begin{deluxetable*}{@{}lccccc@{}}
\tablecolumns{6}
\centering
\tablecaption{Comparison of the observed elemental abundances and those
 predicted using the P-I model. \label{cel-pi}}
\tablewidth{\textwidth}
\tablehead{
\colhead{X}&
\colhead{Obs$^{\rm a}$}&
\colhead{{\rm P-I}$^{\rm b}$}&
\colhead{$\triangle^{\rm c}$}&
\colhead{ICF(X$_{\rm Obs}$)$^{\rm d}$}&
\colhead{ICF(X$_{\rm P-I}$)$^{\rm e}$}\\
\colhead{}&
\colhead{log(X/H)+12} &
\colhead{log(X/H)+12} &
\colhead{log(X$_{\rm Obs}$/X$_{\rm P-I}$)}
}
\startdata
He &11.02$\pm$0.03 &10.99$\pm$0.20 &+0.03$\pm$0.20 &1.04$\pm$0.01&~~1.00\\
C &~~8.97$\pm$0.17 &~~8.71$\pm$0.20 &+0.26$\pm$0.26 &1.33$\pm$0.24&~~1.00\\
N &~~6.36$\pm$0.10 &~~6.96$\pm$0.20 &--0.60$\pm$0.22 &4.02$\pm$0.81&20.56\\
O &~~7.73$\pm$0.03 &~~7.82$\pm$0.20 &--0.09$\pm$0.20 &1.00&~~1.00\\
F &~~5.42$\pm$0.11 &~~5.41$\pm$0.20 &+0.01$\pm$0.23 &4.02$\pm$0.81&~~5.22\\
Ne &~~7.44$\pm$0.03 &~~7.02$\pm$0.20 &+0.42$\pm$0.20 &1.00&~~1.00\\
S &~~5.40$\pm$0.07 &~~5.48$\pm$0.20 &--0.08$\pm$0.21 &1.00&~~1.00\\
Cl &~~3.58$\pm$0.15 &~~3.44$\pm$0.20 &+0.14$\pm$0.25 &1.19$\pm$0.29&~~1.05\\
Ar &~~4.60$\pm$0.13 &~~4.44$\pm$0.20 &+0.16$\pm$0.24 &1.19$\pm$0.29&~~1.04\\
Fe &~~5.02$\pm$0.12 &~~5.48$\pm$0.20 &--0.46$\pm$0.23 &4.02$\pm$0.81&~~5.15
\enddata
\tablenotetext{a}{From Table~\ref{abund}. We used the RL He and the
CEL C/N/O/F/Ne/S/Ar/Cl/Fe abundances in the P-I model. }
\tablenotetext{b}{Determined from the P-I model.}
\tablenotetext{c}{Elemental abundance difference between the observed
and the model predicted abundances.}
\tablenotetext{d}{from Table~\ref{abund}.}
\tablenotetext{e}{calculated from the P-I model.}
\end{deluxetable*}

Figure~\ref{sed} shows the predicted SED, the observed
spectra, and the band flux densities.
The predictions were taken at the matter-bounded
radius near the Str\"{o}mgren edge (or at the
radius close to the ionization-bounded radius of the P-I model nebula).
This provides an appropriate level of nebular excitation, e.g.,
for O$^{2+}$/(O$^{+}$+O$^{2+}$). Note that the observed and predicted nebular
ratios O$^{2+}$/(O$^{+}$ + O$^{2+}$)$\sim$0.75 \citep[0.92 in BoBn1 and
0.67 in H4-1,][]{2010ApJ...723..658O,Otsuka:2013aa} were large despite
the cool CSPN of K648. Such a high ratio indicates that K648 could be a
matter-bounded nebula, where the edge of the mass distribution falls
inside the Str\"{o}mgren edge, rather than an ionization-bounded
nebula, and also it might be related to the small nebula mass.

The fitted elemental abundances,
gas mass $m_{g}$, dust mass $m_{d}$, and dust temperatures
$T_{d}$ are listed in Table~\ref{model}. The third and fourth columns of Table~\ref{model2} show a
comparison of the predicted fluxes and flux densities with the observed
data. The discrepancies of each flux and each flux density between
the observation and model are listed in the final column. In the SED fitting
for the MIR wavelengths, we place emphasis on the band fluxes (IRS-1,2,3,4,5)
and flux densities (IRAC-4 and MIPS-1) rather than the atomic
line fluxes, because our interest in SED modeling is in calculating the gas and dust
masses. Therefore, there are some discrepancies in the MIR atomic lines between the observed and calculated data.
The $\chi^{2}$ values are listed in the bottom line of Table~\ref{model}. The chi-square analysis implies that, within
1-$\sigma$, there was no difference between the predicted and the
observed flux densities/band fluxes, but rather a slight (negligible) disagreement between the calculated
and observed fluxes, owing to the C\,{\sc iii}$]$\,$\lambda\lambda$\,1906/09
{\AA} flux. Without the C\,{\sc iii}$]$\,$\lambda\lambda$\,1906/09\,{\AA} flux,
$\chi^{2}$ = 34.75 indicates that the modeled
flux densities and band fluxes are in excellent agreement with the observations.

The discrepancy between the observed calculated C\,{\sc iii}$]$\,$\lambda\lambda$\,1906/09\,{\AA} line
fluxes appears to result from fluctuations in the structure of {\te}. The C\,{\sc iii}$]$ lines are the most
sensitive to the {\te} among those considered in the
model; the excitation energy difference between the upper and the
lower levels ($\chi$) is 6.5 eV and the excitation temperature is
75\,380 K (=$\chi$/$k$, where $k$ is the Boltzmann constant).
We used {\te} = 10\,270 K in the calculations of C$^{2+}$, whereas the
volume-averaged {\te}(C$^{2+}$) in the model was 11\,090 K. With
a constant {\Ne}, but a difference of only 820 K, the volume emissivity of this complex
line at 11\,090 K became $\sim$1.65 times larger than that with 10\,270 K.
Accordingly, we obtained C\,{\sc iii}$]$\,$\lambda\lambda$\,1906/09\,{\AA}
fluxes that were greater than those of the observations
by a factor of $\sim$1.65. The {\hb} emissivity at {\te} = 10\,270 K was
1.07 times greater than that at {\te} = 11\,090 K. Therefore, the
modeled C$^{2+}$ abundance was smaller than the observation by $\sim$--0.10
dex. Taking the differences in the structure of {\Ne} and {\te} between
the model and the observed data into account, we estimate that the accuracy of the elemental
abundances calculated using the was within $\sim$0.2 dex.

As the model provides decent predictions for the [S\,{\sc
iii}]\,$\lambda\lambda$\,18.7/33.5\,$\mu$m, [S\,{\sc
iii}]\,$\lambda$9532\,{\AA} ($I$(P-I)) of [S\,{\sc
iii}]\,$\lambda$9532\,{\AA} = 5.760, which are not listed in
Table~\ref{model2}), and [S\,{\sc iii}]\,$\lambda$\,6312\,{\AA}
simultaneously, we may assume that the two MIR [S\,{\sc iii}] lines are
not spurious, but rather genuine features of the spectra.

Our P-I model with the {\sc Cloudy} code was not able to fit [Ne\,{\sc ii}]\,$\lambda$\,12.80\,$\mu$m,
whereas the prediction of the other atomic lines of similar IPs,
i.e., ions such as {\fii} (see below), {\siii}, {\ariii}, and
{\cliii}, is in good agreement with the observations. Many P-I models
using {\sc Cloudy} have been used to fit the [Ne\,{\sc
ii}]\,$\lambda$\,12.80\,$\mu$m in PNe; however, to our knowledge,
there has been little success \citep[e.g.,][]{Pottasch:2011aa,Pottasch:2009aa}.
In our model, we monitored the chi-square values to obtain the
best fitting parameters. With an almost constant radial density
profile, the [Ne\,{\sc ii}]\,$\lambda$\,12.80\,$\mu$m could be modeled;
however, the other line-fluxes and band fluxes/flux densities
exceeded the observed values, i.e., chi-square increased.
The recombination rates for some heavy element ions (e.g., S$^{+}$)
are uncertain, so that photo-ionization models may give
line fluxes that are in poor agreement with measured data. Therefore, the lack of agreement may
be due to the uncertainties in the atomic data for Ne$^{+}$.
Improvements in these data, however, are beyond the scope of this paper.

Our P-I model  predictions   provide good fits to two of the {\fii} line
intensities. If both lines are not [F\,{\sc ii}] lines but other elemental
lines, the P-I model cannot fit these two lines simultaneously.   
Therefore, we conclude that the detected {\fii} lines are likely to be real.
The two observed {\fii} line fluxes and the calculated elemental
abundance of F using the ICF(F) are in good agreement with the predictions of the model.

The second and third columns of Table~\ref{cel-pi} list a comparison
of the nebular elemental abundances determine using the semi-empirical ICF method and the
P-I model. As we mentioned above, the accuracy of the elemental
abundances determined using the P-I model was within $\sim$0.2 dex.
Careful treatment for the Ne abundance is necessary for the reasons
discussed above. Therefore, we excluded the Ne abundance from the
following discussion. The difference between the two data sets, ($\triangle$),
is listed in the fourth column of Table~\ref{cel-pi}. The agreement
between the He, C, O, F, S, Cl, and Ar abundances between the ICF method
and the P-I model is generally good.

However, poor agreement is found for N and
Fe ($|\triangle|$$\geq$0.2 dex). The final two columns of
Table~\ref{cel-pi} list the ICF values used in
Section~\ref{S:abund_ICF} and those predicted by the P-I model.
The ICF values from the P-I model were generally in
agreement with those of the semi-empirical  methods, except for N.
This is because most fractional ionizations occurred
in other ionic stages: the P-I prediction suggests 5$\%$ for N$^+$ and 95~$\%$ for N$^{2+}$.
Poor agreement for N between the ICF and P-I models is often found for
PNe and in the O-rich halo PN DdDm1 \citep[see
e.g.,][]{2009ApJ...705..509O,Delgado-Inglada:2014ab}. In many cases,
including DdDm1, N$^{+}$ abundances determined from optical spectra alone have been
used to determine the elemental N abundance, because it is difficult to detect
the N$^{2+}$ forbidden lines, which appear in the UV or FIR
spectra. Based on grid models using {\sc
Cloudy}, \citet{Delgado-Inglada:2014ab} proposed that the N/O ratio for
PNe showing no-He\,{\sc ii} lines can be estimated by the following equations,
\begin{eqnarray}
 {\rm \frac{N}{O}} &=& {\rm ICF(N)_{GI14}\cdot\frac{N^{+}}{O^{+}}},\\
 {\rm ICF(N)_{GI14}} &=& {\rm 10^{0.64\cdot\frac{O^{2+}}{O^{+}+O^{2+}}}.} 
\end{eqnarray}
In the case of K648, the ICF(N)$_{\rm GI14}$ and the N/H abundance
using the observed O$^{+,2+}$, N$^{+}$ abundances
(Table~\ref{celabund}) and the elemental O
abundance (Table~\ref{abund}) were estimated to be 3.40$\pm$2.42 and 7.77(--6)$\pm$5.82(--6). The
$\log_{10}$(N/H)+12 of 6.89$\pm$0.33 is very close to the model predicted
value (6.96$\pm$0.20), although we should note that the ionic and
elemental abundances would depend on the density structure, incident
ionization source, ionization boundary condition, gas metallicity,
dust grains/molecules, and so on.
We should keep in our mind that we were unable to detect N~{\sc
iii}$]$\,$\lambda$1750\,{\AA} in K648. The predicted line-intensity around
1750~{\AA} \emph{FOS} spectrum with
S/N = 1 was $\sim$5.3 (i.e., a detection limit), which is approximately three times larger than
the prediction of the P-I model (i.e., 1.7). The N\,{\sc iii}$]$\,$\lambda$\,1750\,{\AA} line was
too faint to detect using \emph{FOS}. As the P-I model provided a value
of ICF(N) that was too large, in this paper we prefer to use the
semi-empirically  determined N abundance, i.e., the ICF abundance (not ICF(N)$_{\rm GI14}$).
The P-I model indicates that Fe ions are also concentrated in other ionic stages,
rather than the observed ionic stage, i.e.,
20{\%} Fe$^{2+}$ and 80{\%} Fe$^{3+}$. The availability of the
N$^{2+}$ and Fe$^{3+}$ lines would be expected to
improve the accuracy of the abundance calculation. FIR observations
using \emph{SPICA}/SAFARI would be helpful to detect
{\nii}\,$\lambda$\,121.3\,$\mu$m and $[$N\,{\sc
iii}$]$\,$\lambda$57.3\,$\mu$m lines and verify ICF(N) in PNe.

Note that we used the ICF abundances rather than the P-I results.
Analysis with the P-I results, however,
is not expected to alter the conclusions. The P-I results
should be carefully examined in a more sophisticated future study.

The nebula is fully ionized, so that the ionized gas mass is consistent with
$m_{g}$. \citet{1995A&A...301..537B} determined an ionized gas mass of
0.05-0.09 $M_{\odot}$, and \citet{Kingsburgh:1992aa} determined it to be
0.042 $M_{\odot}$. Both authors assumed a constant density profile.
Although $m_{g}$ (and $m_{d}$) depends on the
distance used, $m_{g}$ in this work is consistent with these estimates.

Here, we estimate $m_{d}$ and the dust-to-gas mass ratio
$m_{d}$/$m_{g}$. If we assume that the AC dust grains have a radius of
0.0005-0.25\,$\mu$m and a size
distribution that follows $a^{-3.5}$, we obtain $m_{d}$ = 9.74(--7) $M_{\odot}$
and $m_{d}$/$m_{g}$ = 2.07(--5). However, this model does not fit the flux density at the
above-mentioned wavelengths. To our knowledge, the value of
$m_{d}$ = 4.95(--7) $M_{\odot}$ for K648 is the smallest mass among known PNe, and is
approximately one order of magnitude smaller than that of BoBn1, where the
$m_{d}$ = 5.78(--6) $M_{\odot}$. For BoBn1, \citet{2010ApJ...723..658O} used grains with a radius of 0.001-0.25\,$\mu$m with an $a^{-3.5}$ size distribution in the SED
model, whereas for K648 we used much smaller grains to match the observed
SED at wavelengths in the range 10-15\,$\mu$m. We found the ratio $m_{d}$/$m_{g}$ = 1.03(--5),
which was much lower than that for BoBn1 (5.84$\times$10$^{-5}$).
For H4-1, \citet{Tajitsu:2014aa} reported that
$m_{g}$ = 0.3 $M_{\odot}$, $m_{d}$ = 7.34(--4) $M_{\odot}$, and $m_{d}$/$m_{g}$ = 2.48(--3); however, H4-1 contains abundant cold dust and hydrogen-rich molecules. Among
these C-rich halo PNe, where the metallicity is similar to that of K648,
we could not find dependence of the metallicity on the dust mass.

\subsection{Expansion velocities and the time since the AGB phase\label{s:vexp}}

\begin{deluxetable}{@{}lcrcc@{}}
\tablecolumns{5}
\centering
\tablecaption{Expansion velocities of K648.  \label{vexp}}
\tablewidth{\columnwidth}
\tablehead{
Ion &Type of &
I.P. &
Num. of &
$\langle{V_{\rm exp}}\rangle$ \\
&
lines&
(eV)&
sample lines&
({\kms})
}
\startdata
{\oi} & CEL & 0.00 & 2 & 10.12$\pm$0.27 \\
{\sii} & CEL & 10.36 & 2 & 14.70$\pm$0.28 \\
H\,{\sc i} & RL & 13.59 & 26 & 15.07$\pm$0.58 \\
{\oii} & CEL & 13.62 & 2 & 15.99$\pm$0.18 \\
{\nii} & CEL & 14.53 & 3 & 13.81$\pm$0.43 \\
{\feiii} & CEL & 16.18 & 2 & 12.70$\pm$1.05 \\
{\fii} & CEL & 17.42 & 2 & 13.85$\pm$0.90 \\
{\siii} & CEL & 23.33 & 1 & 15.96$\pm$0.57 \\
{\cliii} & CEL & 23.81 & 2 & 16.12$\pm$1.55 \\
C\,{\sc ii} & RL & 24.38 & 6 & 19.37$\pm$0.99 \\
He\,{\sc i} & RL & 24.59 & 17 & 15.71$\pm$0.56 \\
{\ariii} & CEL & 27.63 & 1 & 14.25$\pm$0.16 \\
N\,{\sc ii} & RL & 29.60 & 2 & 15.72$\pm$1.49 \\
{\oiii} & CEL & 35.12 & 4 & 15.75$\pm$0.52 \\
O\,{\sc ii} & RL & 35.12 & 1 & 12.65$\pm$0.99 \\
{\neiii} & CEL & 40.96 & 2 & 13.75$\pm$0.09 \\
C\,{\sc iii} & RL & 47.89 & 3 & 15.54$\pm$0.63
\enddata
\tablecomments{The third and fourth columns list the IP and the
number of sample lines used in the calculation of the average $V_{\rm exp}$ of
each ion, respectively.}
\end{deluxetable}

We employed a multiple Gaussian
fitting method for the flux measurements, except
for the strong lines {\oii}\,$\lambda\lambda$\,3726/29\,{\AA}, {\oiii}\,$\lambda\lambda$\,4959/5007\,{\AA}, {\oi}\,$\lambda$\,6300\,{\AA},
{\ha}, and {\nii}\,$\lambda$\,6583\,{\AA}, because these lines have a weak broad tail
component or a small offset velocity component. Here, we focus on
the nebula expansion velocity $V_{\rm exp}$ determined from the main
Gaussian component of each line.

We measured $V_{\rm exp}$ using the following relation:
\begin{equation}
V_{\rm exp} = 1/2~(V_{\rm FWHM}^2 - V_{\rm therm}^2 - V_{\rm instr}^2)^{1/2},
\label{expf}
\end{equation}
\noindent where $V_{\rm FWHM}$ is the FWHM of the velocity,
$V_{\rm therm}$ is the thermal broadening velocity, and $V_{\rm
instr}$ is the instrumental velocity
\citep[e.g.,][]{2010ApJ...723..658O,2009ApJ...705..509O,2001AJ....122.1538B}. Here,
$V_{\rm therm}$ is represented by
21.4$(T_{\epsilon}{\times}10^{-4}/A_{r})^{1/2}$, where $A_{r}$ is the relative
atomic mass of the target ion. For CELs, we used the {\te} listed in Table~\ref{tene}.
For RLs, we used {\te}(BJ) for H\,{\sc i}, C\,{\sc ii,iii}, {\nii}, and
O\,{\sc ii} and the {\te}(He\,{\sc i}) = 6710 K for the He\,{\sc i} lines.
We measured $V_{\rm instr}$ for all the identified lines listed in
Table~\ref{hdstab} in the Appendix
using the nearby Th-Ar lines, i.e., 4.3 {\kms} for [Ar\,{\sc
iii}]\,$\lambda$\,7135\,{\AA}, He\,{\sc i}\,$\lambda$\,7281\,{\AA}, and
{\oii}\,$\lambda\lambda$\,7320/7330\,{\AA} (the resolving power of these
lines was $\sim$69\,000), and 8.8-9.0 {\kms} for the
others . We did not include the turbulent velocity, because these velocities
have been measured in $\sim$100 Galactic PNe by e.g.,
\citet{Acker:2002aa} and \citet{2003A&A...400..957G}, who found no
turbulent velocities in PNe with non-WC type central stars, such as K648.

The resulting $V_{\rm exp}$ are summarized in Table~\ref{vexp}.
We measured the $V_{\rm exp}$ of over 100 lines selected from
the lines listed in
Appendix Table~\ref{hdstab}.
For each ion, we excluded the
measurements far from the average using 1-$\sigma$ clipping. We then
calculated the average expansion velocity, $\langle{V_{\rm exp}}\rangle$ of, the 18
ions listed in the final column of Table~\ref{vexp}.

Our measurements showed good agreement with those of \citet{2001AJ....122.1538B},
who measured the $\langle{V_{\rm exp}}\rangle$({\hi}) of 16.7 {\kms}
using {\ha} and $\langle{V_{\rm exp}}\rangle$({\nii}) of 11.9 {\kms},
using both {\nii}\,$\lambda$\,6548\,{\AA} and {\nii}\,$\lambda$\,6583
{\AA} lines, and with a constant electron temperature of {\te} = 10\,000 K for the thermal broadening
velocities. We used 26 H\,{\sc i} lines for the
$\langle{V_{\rm exp}}\rangle$({\hi})
and 3 lines for the $\langle{V_{\rm exp}}\rangle$({\nii}) calculations
with {\te}(BJ) and {\te}({\nii}), respectively. The slight differences between our data and those reported by \citet{2001AJ....122.1538B} can be attributed to the value of {\te} used and the number of sample lines.

The correlation between the $\langle{V_{\rm exp}}\rangle$ and IPs is given by
\begin{equation}
\langle{V_{\rm exp}}\rangle = (5.23~\pm~4.03)\cdot10^{-3}~{\rm IP} + 13.54~\pm~1.01.
\end{equation}
The correlation factor was 0.32. Assuming that K648 has a standard
ionized structure, i.e., high-intensity IP lines are emitted from regions close to the central
star and low-intensity IP lines are emitted from regions far from the central star, the expansion
velocity of the nebula may be slowing with an almost constant value of $r$. In general, the PN
shell is known to follow a Hubble type expansion, i.e., acceleration of
the expanding gas shell. Perhaps it did not gain its impulsion from the
CSPN yet.

As we found for BoBn1
\citep{2010ApJ...723..658O}, $V_{\rm exp}$ for {\oii} was at least 1.5
{\kms} smaller than that for {\oiii}.

The apparent outer radius of K648 is 2.1$''$, which corresponds to
0.125 pc at 10.9 kpc. The wind velocity of K648 during the AGB mass-loss
phase is unknown. We used an expansion
velocity of $\langle{V_{\rm exp}}\rangle$(H\,{\sc i}) = 15.07 {\kms}, and
estimated the dynamical
age of the K648 nebula to be 8110$\pm$490 years since the AGB
phase. \citet{2002AaA...381.1007R}
estimated the post-AGB age of
6800$^{+3500}_{-2000}$ yrs by plotting their derived luminosity and
surface gravity on theoretical evolutional tracks. However, the
evolutionary age after the AGB phase with more precision is unknown at this moment.
\citet{McCarthy:1990} investigated the disagreement
between evolutionary and dynamical time scales for the evolution of the
CSPNe using the results of high resolution spectra of about 23 CSPNe. According
to them, the AGB-CSPN evolutionary transition times could have been increased
by small additional amounts of
residual envelope material remaining after the superwind mass-loss
phase.

\section{Discussion}

\subsection{Comparison with the AGB nucleosynthesis model \label{S:agb}}

\begin{deluxetable*}{@{}llcccccc@{}}
\tablecolumns{8}
\centering
\tablecaption{
Comparison of the observed nebular abundances with the predictions of the [Fe/H] = --2.19 AGB model. \label{modelabun}}
\tablewidth{\textwidth}
\tablehead{
&
\multicolumn{6}{c}{Models}&Obs\\
\cline{2-7}
&Initial mass ($M_{\odot}$)&
0.9 &
0.9 &
1.25 &
1.25 &
1.5 &
 \\
Elements
&PMZ mass ($M_{\odot}$)&
0 &
2(--3) &
0&
2(--3) &
0
}
\startdata
He && 11.00 & 11.00 &11.00& 11.00 & 11.01 & 11.02$\pm$0.03 \\
C && ~~9.07 & ~~9.04 &~~8.94 & ~~8.90 & ~~9.26 & ~~8.97$\pm$0.17 \\
N && ~~7.55 & ~~7.53 &~~6.67 & ~~6.68 & ~~6.76 & ~~6.36$\pm$0.10 \\
O && ~~7.48 & ~~7.63 &~~7.31 & ~~7.47 & ~~7.56 & ~~7.73$\pm$0.03 \\
F && ~~4.74 & ~~5.03 &~~4.35 & ~~4.78 & ~~5.08 & ~~5.42$\pm$0.11 \\
Ne&& ~~7.33 & ~~7.87 &~~6.95 & ~~7.65 & ~~7.68 & ~~7.44$\pm$0.03 \\
P$^{\rm a}$ && ~~3.48 & ~~3.64 &~~3.36 & ~~3.54 & ~~3.49 & ~~3.64$\pm$0.10\\
\hline
ejected mass during &&2.0(--3)&
2.0(--3)&8.0(--3)&8.0(--3)&5.99(--1)&4.8(--2)$^{\rm b}$\\
last TP ($M_{\odot}$)\\
core-mass ($M_{\odot}$) &&0.77 &0.77 &0.66 &0.66 &0.66 &0.61-0.63\\
envelope mass ($M_{\odot}$)&&0.04 &0.04 &0.02 & 0.03 &0.18 &\nodata
\enddata
\tablecomments{The initial He, C, N, O, F, Ne and P abundances in all models are 10.92, 6.29, 5.69, 6.55, 2.24, 5.79 and 3.23, respectively.}
\tablenotetext{a}{The P abundance of the CSPN measured in the \emph{FUSE} spectrum.}
\tablenotetext{b}{The mass estimated using the {\sc Cloudy} SED model.}
\end{deluxetable*}

In Section~\ref{S:core mass}, we determined the core-mass of the central
star as 0.61-0.68 $M_{\odot}$, depending on the choice of the distance to M15. The initial-final
mass relation has been studied using solar metallicity for young ($\sim$1-2
Gyr) open clusters \citep[e.g.,][]{Kalirai:2008aa}; however, it has
been not studied using metal-poor old clusters. Semi-empirical initial-final mass
relations are only available for the chemical composition of the solar
neighborhood and for Magellanic Clouds \citep[see,][]{Prada-Moroni:2007aa}.
The mass-loss and the dredge-up efficiency (depending on the core-mass,
metallicity, and total mass of the star) during the AGB phase determine
the fate of stars. From these reasons, we utilized the theoretical initial-final
masses for $Z$ = 10$^{-4}$ stars reported by \citet{Prada-Moroni:2007aa}
to estimate the initial mass of K648. From polynomial fitting to the
initial-final masses listed in Table~1 of \citet{Prada-Moroni:2007aa}, we found that
core-masses of 0.61, 0.63, 0.66 and 0.68 $M_{\odot}$ correspond to the
initial masses of 1.15, 1.60, 1.76, and 1.87 $M_{\odot}$, respectively.
As the upper limit of the mass of stars in M15 is $\sim$1.6
$M_{\odot}$, the current core-mass and initial mass of K648 would be $\sim$0.61-0.63
$M_{\odot}$ and $\sim$1.15-1.6 $M_{\odot}$, based on \citet{Prada-Moroni:2007aa}.

For comparison, we discuss the results of \citet{Lugaro:2012aa} for
0.9, 1.25, and 1.5 $M_{\odot}$ stars with an initial [Fe/H] = --2.19.
\citet{Lugaro:2012aa} used
scaled solar abundances as the initial composition
for all elements from Li to Pb [X/Fe]$\simeq$0 and [He/Fe] = +2.18.
The initial conditions and the mass loss formulae
used in \citet{Lugaro:2012aa} were discussed by
\citet{2010MNRAS.403.1413K}. Table~\ref{modelabun} lists the predicted
abundances after the final TP.
Here, we used the nebular abundances, except for P, where we used
the stellar abundance. The C and O
abundances used for K648 were the values from the CELs.
The 0.9, 1.25, and 1.5 $M_{\odot}$ stars would,
theoretically, experience 38, 15 and 18 TPs, respectively. The final three lines of Table~\ref{modelabun} lists the
ejected mass during the final TP, the final core-mass, and the envelope
mass.

Our estimated core-mass agrees with the predictions for 1.25 and 1.5 $M_{\odot}$ reported by \citet{Lugaro:2012aa}. As the 0.9 $M_{\odot}$
models experienced many TPs, the final core-mass was larger than that
predicted by the 1.25 and 1.5 $M_{\odot}$ models. \citet{Lugaro:2012aa}
included a partial mixing
zone (PMZ), which is formed in a mixing zone from the H-rich
envelope down to the layer at the top of the He-rich intershell \citep{Shingles:2013aa}.
The PMZ produces a $^{13}$C (as well as a $^{14}$N) pocket during the interpulse
period. The $^{13}$C releases additional free neutrons ($n$) via
$^{13}$C($\alpha$,{\it n})$^{16}$O, resulting in further $n$-process elements, such
as $^{19}$F and $^{31}$P.
The available mass of $^{13}$C mainly affects
the final compositions of K648, in particular, C, N, O, Ne, and F, which
are synthesized in the He-rich
intershell. \citet{Shingles:2013aa} showed that the Ne abundance
increases as the mass of the PMZ increases. They argued that the
Ne enhancement is due to $^{22}$Ne production via
double $\alpha$-particle capture by $^{14}$N.

For the reason given above, we checked the abundances of N and Ne.
First, the 0.9 $M_{\odot}$ and 1.25
$M_{\odot}$ model with no PMZ were excluded. The former could explain the Ne abundance,
but not the N abundance. In addition,
the final core-mass appeared larger than the predictions of these models. The latter
model could not explain the Ne abundance either. The remaining two, i.e., the 1.25 $M_{\odot}$+2(--3)
$M_{\odot}$ PMZ and 1.5 $M_{\odot}$ models, provided reasonable agreement
with not only the N and Ne abundances, but also the He, C, O, and F
abundances. The P abundance of the CSPN was comparable to the predicted value of 3.49 using this model.

Our estimate of the core-mass of the CSPN is in good agreement with both
the 1.25 $M_{\odot}$ and 1.5 $M_{\odot}$ models; however, the resulting
gas masses could not be explained using either model.
Therefore, we expect that models for
stars with initial masses in the range of 1.25-1.5 $M_{\odot}$ can
explain the ejected mass, as well as elemental abundances and the core-mass of the CSPN.

\subsection{Was the progenitor a blue straggler?}

We found that even the lower core-mass of 0.61 $M_{\odot}$,
which corresponds to an initial mass of 1.15 $M_{\odot}$,
exceeds the mass of turn-off stars in M15; however, the
uncertainty of $\sim$0.03 $M_{\odot}$ should be noted. The 1.25 $M_{\odot}$+2(--3)
$M_{\odot}$ PMZ model can also explain observed nebular abundances.
Hence, it is possible that
the progenitor of K648 is a binary system. Indeed, K648 has long
been suspected to have undergone binary evolution \citep{Jacoby:1997aa}.

During the evolution of the progenitor of K648, if it efficiently
gained mass and nucleosynthesized products via mass-transfer
from an evolved massive primary, it could evolve into a C-rich PN. Although
it initially appears difficult to accept that binarity may be responsible
for many cases of anomalous composition, there is now evidence
of radial velocity variations or bright equatorial disk structures,
signaling a binary orbit, and binary interactions are regarded
as the explanation of a wide range of C-rich stellar classes,
barium stars, CH stars, and CEMP stars \citep[Carbon-Enhanced Metal-Poor stars, see, e.g.,][]{Beers:2005aa,Masseron:2010aa,Bisterzo:2012aa}.

In the case of the C-rich halo PNe H4-1, \citet{Otsuka:2013aa}
proposed that the H4-1 chemistry may be the
evolutionary result of a $\sim$0.8-0.9 $M_{\odot}$ star that had been
affected by mass-transfer from a more massive AGB companion in a
binary system; however, \citet{Otsuka:2013aa} did not evaluate the
core-mass of the central star due to the lack of observation data of the central
star. The traditional evolution theory that the progenitors of PNe are
the remnants of single stars at the end of the AGB phase
does not provide a natural explanation for the non-spherical morphologies observed
for the great majority of PNe. Although the binary interaction model
explains some of the anomalies associated with the observed PN population,
the number of PN central stars with known binary companions is very
small and carrying out programs to detect such objects are extremely difficult (see \citealt{Jacoby:2013} and references therein).

 H4-1 and K648 both appeared to exhibit evidence of
binary evolution structures, such as bright
equatorial disk structures and a bipolar nebula
\citep{Tajitsu:2004aa}.
\citet{2000AJ....120.2044A} did not detect any time variation in the magnitude of the central star using \emph{HST}/WFPC2.
They argued that
 the failure to detect a current binary companion lends support to the
 picture of a complete merger, as opposed to more modest mass transfer,
 because in the latter case there would still be a remnant companion
 (possibly a helium-rich white dwarf). We neither detected any variation in the radial velocity in our HDS
spectra. Although the failure to detect a companion is not conclusive proof,
it is worthwhile to re-examine whether the central star of K648 is
a binary or a merger.

K648 might be not a ``typical'' PN. For example, the ionized gas mass
is unusually small, which may indicate that it formed via a non-standard
mechanism. Based on a comparison with AGB yields of a single star, as
discussed above, we propose that the merging of two stellar bodies occurred, or
a large mass fraction was transferred from a companion.
In a close binary system, the gravity of one component can induce a
significant tidal force in the other. The dissipation of this tidal force may
synchronize the rotation and circularize the orbit, leading to coalescence (or
consumption the outer envelope of its companion)
in extreme cases. Although there have been many theoretical analyses and simulations
of binary coalescence of neutron stars or black holes, there have been no reports
of closely related work in binary systems in PNe such as K648
(e.g., \citealt{Zhang:2013aa} and references therein).

The notion that K648 and other globular clusters may arise
from coalescence of binary systems was proposed by \citet{2000AJ....120.2044A}
and \citet{Jacoby:1997aa}. \citet{2000AJ....120.2044A}
argued that the progenitor of K648 experienced mass
augmentation in a close binary merger, and evolved as a higher
mass star to become a PN. Such a high-mass star would be a blue
straggler (BS). A number of possible BS candidates
(20-69 objects) have been found in M15
\citep{Dieball:2007aa,Diaz-Sanchez:2012aa}.

There are several ways in which stars may evolve into BSs, i.e.,
MS-MS collisions, WD-MS collisions, and close binary transfer or mergers
\citep[e.g.,][]{Umbreit:2008aa,2009Natur.462.1028F}.
The progenitor of K648 may have been formed via a close orbital activity
of a binary with a large mass inflow from its companion during the MS stage.
one possibility is a close binary system that consists of two stars of $\sim0.9
M_{\odot}$ with slightly different masses, or with significantly different masses
as proposed by \citet{2010ApJ...723..658O}, and that one star consumed
a large mass fraction of its companion so that the mass of this star would
approach $\sim$1.6 $M_{\odot}$, relegating its
companion to the position of an accessory.

To date, 20-69 BSs (including candidates) have been identified in M15 \citep{Dieball:2007aa,Diaz-Sanchez:2012aa}.
The typical PNe lifetime is $\sim$25\,000 years \citep[e.g.,][]{Moe:2006aa,Feldmeier:2003aa}.
When we use a BS lifetime of $\sim$1.2 Gyrs \citep{Sills:1997aa},
the expected number of PNe in M15 is 0.0004-0.0014 PNe per a BS (=25\,000 years / 1.2 Gyrs
$\times$ 20-69). If we assume a birth rate of 2.5-5.0(--8) BSs yr$^{-1}$ via this process \citep{Umbreit:2008aa}, and given that the age of M15 is 13.5 Gyrs,
the estimated number of PNe formed in M15 is
0.135-0.675 PNe. Therefore, K648 would be a rare PN evolved from a BS.
The central star of K648 may be a BS of higher mass
determined by us (i.e., 1.5 $M_{\odot}$) or close to the limit in M15
($\sim$1.6 $M_{\odot}$).
Other BSs found in M15 may well evolve into PNe, similar to K648.
If new evidence of much lower stellar mass is reported, the milder binary
interaction scenario must be explored accordingly, i.e., mass-transfer
from a more massive AGB companion in a binary system. However, this
scenario may be not appropriate to explain the relatively large mass of K648.

\subsection{Comparison of K648 with BoBn1 and H4-1}

\begin{figure}
\includegraphics[bb = 8 14 444 310,width = \columnwidth,clip]{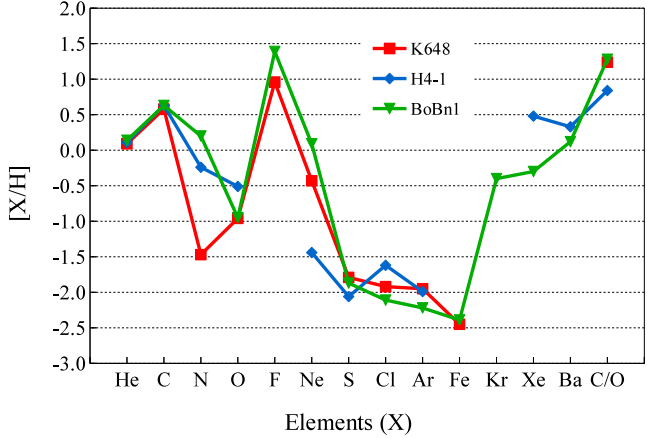}
\caption{The relative elemental abundances of K648, H4-1 and
BoBn1 compared with those of the Sun. The solar Kr, Xe, and Ba
abundances are from \citet{2003ApJ...591.1220L}. See Table~\ref{halopn_abund} for the elemental abundances of each PN.
\label{halopn_hist}}
\end{figure}

\begin{deluxetable}{@{}lcccc@{}}
\tablecolumns{5}
\centering
\tablecaption{
Comparison of the abundances for K648, H4-1, and
BoBn1. \label{halopn_abund}}
\tablewidth{\columnwidth}
\tablehead{
Elements&
K648&
H4-1&
BoBn1&
Average
}
\startdata
He &11.02$\pm$0.03 & 11.03$\pm$0.15 & 11.07$\pm$0.01 & 11.04$\pm$0.06 \\
C & ~~8.97$\pm$0.17 & ~~9.02$\pm$0.18 & ~~9.02$\pm$0.08 & ~~9.00$\pm$0.14 \\
N & ~~6.36$\pm$0.10 & ~~7.59$\pm$0.04 & ~~8.03$\pm$0.10 & ~~7.69$\pm$0.09 \\
O & ~~7.73$\pm$0.03 & ~~8.18$\pm$0.02 & ~~7.74$\pm$0.03 & ~~7.94$\pm$0.02 \\
F & ~~5.42$\pm$0.03 &\nodata & ~~5.85$\pm$0.09 & ~~5.68$\pm$0.09 \\
Ne & ~~7.44$\pm$0.03 & ~~6.43$\pm$0.10 & ~~7.96$\pm$0.02 & ~~7.60$\pm$0.03 \\
S & ~~5.40$\pm$0.07 & ~~5.13$\pm$0.03 & ~~5.32$\pm$0.16 & ~~5.30$\pm$0.09 \\
Cl & ~~3.58$\pm$0.15 & ~~3.88$\pm$0.13 & ~~3.39$\pm$0.07 & ~~3.66$\pm$0.12 \\
Ar & ~~4.60$\pm$0.13 & ~~4.56$\pm$0.12 & ~~4.33$\pm$0.04 & ~~4.51$\pm$0.10 \\
Fe & ~~5.02$\pm$0.12 & \nodata & ~~5.08$\pm$0.13 & ~~5.05$\pm$0.13 \\
Kr & \nodata &\nodata & 2.88 & 2.88 \\
Xe & \nodata & $>$2.70 & $<$2.97 & 2.86 \\
Ba & \nodata & \nodata $<$2.51 & 1.97 & 2.32 \\
\hline
C/O & 17.47$\pm$7.07 & 6.93$\pm$2.96 & 19.06$\pm$3.75 & 14.49$\pm$4.59
\enddata
\tablecomments{The abundances of all elements except He were determined from the CEL lines; that of He was determined from RL lines.
The elemental abundances are in the form of $\log_{10}$(X/H)+12, where H is 12. The C/O number density
ratio is linear value. The values of H4-1 and BoBn1 were taken from
\citet{Otsuka:2013aa} and \citet{2010ApJ...723..658O}. }
\end{deluxetable}

\citet{2010ApJ...723..658O} reported a similar comparative study of their data for BoBn1
with earlier analyses of K648. Based on incomplete observations, e.g., no detection of {\fii} lines in K648,   \citet{2010ApJ...723..658O}
argued that BoBn1 might have undergone binary evolution
with a 0.75 $M_{\odot}$+1.5 $M_{\odot}$ system, whereas  K648
might be an object that went through either a binary evolution with 0.75 $M_{\odot}$+1.5 $M_{\odot}$
or a single 1.8 $M_{\odot}$ stellar evolution, ignoring the upper
limit of mass for stars in M15.  Here, we refined the earlier
guessing based on the abundances of 10 elements. In contrast to BoBn1, no evidence of Ba and Xe
was observed in K648; however, there are similarities between BoBn1 and K648 which are not shared with H4-1. 
Here, we discuss similarities between K648, BoBn1 and H4-1.

Figure~\ref{halopn_hist} shows the elemental abundance patterns of
K648, as well as those of H4-1 and BoBn1. The abundances of all elements except He were
determined based on CEL lines; the abundance of He was determined from the RL lines.
Data for H4-1 and BoBn1 are from \citet{Otsuka:2013aa}
and \citet{2010ApJ...723..658O}, respectively. Discrepancies in the C
and O abundances in H4-1 and BoBn1 are discussed in these reports.
A comparison of the RL C and O
abundances among K648, H4-1 and BoBn1 is beyond the scope of this
paper.
The elemental abundances for each PN and the average abundance
of each element are listed in Table~\ref{halopn_abund}.

The He and C abundances were the same for the three systems to within
error.
The $\alpha$-elements Ar and S, and Cl were not synthesized in significant quantities
in the PN progenitors. For example, with the 1.25 $M_{\odot}$
+2.0(--3) PMZ model of \citet{Lugaro:2012aa}, an increase of only $\sim$0.02-0.03 dex
was found compared with the initial abundances. Therefore, we can regard
these three elements as mostly SN products.
Some fractions of O and Ne are synthesized in the He-rich
intershell during the TP-AGB phase. Indeed, the above
1.25 $M_{\odot}$ model \citet{Lugaro:2012aa} predicted that such stars
can increase +0.94 and +1.88 dex from the initial O and Ne abundances,
respectively.

We determined gas-phase abundance of [Fe/H]=--2.45$\pm$0.12 using the
ICF(Fe) in K648, which is comparable to the typical [Fe/H] abundance in a M15
star (\citealt{Sobeck:2011aa}, [Fe/H]$\sim$--2.3). There are a number
of other possible forms of Fe in the solid phase in laboratory
experiments.
According to the Jena Database of Optical
Constants\footnote[8]{http://www.astro.uni-jena.de/Laboratory/Database/jpdoc/index.html}, for
example, FeO, FeS, magnesium-iron oxides, magnesium-iron silicates,
and olivine. \citet{Delgado-Inglada:2014aa} employed a more
sophisticated ionization correction factor scheme than that used here.
They reported that the highest depletion factors are found in C-rich
objects, exhibiting SiC around 11\,$\mu$m or a broad 30\,$\mu$m feature in the infrared
spectra. Note that the carriers of these features are under debates.
The central positions of the SiC features in the sample of
\citet{Delgado-Inglada:2014aa} were not reported, however, and
K648, H4-1, and BoBn1 were not included in their sample.
  According to \citet{Delgado-Inglada:2014aa}, the Fe abundance
  ratio depletion detected in most PNe might be due to situation that
  less than 10~{\%} of the Fe is in the gas phase with more than 90~{\%}
  in the solid phase. Instead of using the semi-empirical ICF scheme, we have an alternative method of
estimating the [Fe/H] abundance, i.e., using the {\sc Cloudy} P-I model.
The [Fe/H] abundance predicted using our P-I model was [Fe/H]=--1.99$\pm$0.2,
which is close to that determined using the ICF method. The
agreement was significantly better than that reported by
\citet{Delgado-Inglada:2014aa}. The resulting ICF [Fe/H] value for K648 is likely
to represent the abundance of Fe, because this value was consistent
with that for typical M15 metallicity, and we did not find features corresponding to amorphous silicate,
crystalline silicates, or SiC (see section \ref{S:11um}).

There have been no reports of the
detection of these features in H4-1, BoBn1, or K648.
The presence of MgS and FeS is known to result in broad
features around 30\,$\mu$m (although the carrier of this feature remains the subject
of some debate). We did not detect any other refractory element lines,
such as Si and Mg, to estimate their ionized abundances in these
PNe. The 1.25 $M_{\odot}$+PMZ 2(--3) $M_{\odot}$ model of
\citet{Lugaro:2012aa} predicted that the [Si/H] and [Mg/H] abundances
are --2.15 and --1.67, respectively. If some of the Si and Mg-atoms might
exist as dust grains, the gas-phase abundances of these two elements
would become smaller, so it would be difficult to detect ionized
emission-lines of these elements.
The large fraction of Fe and S cannot be due to dust grains such as MgS
and FeS. Therefore, the S and Fe abundances represent S and Fe in these halo
PNe, and these elements are expected to exist mostly in the gas phase.
However, we cannot completely exclude the possibility that some fraction
of Fe resides  in other solid forms.

The abundances of S, Ar, Cl, and Fe for K648 are approximately the same as those for H4-1
and BoBn1. At first, it nay be expected that
all three progenitors were born in the same chemical environment during
the same epoch. However, there appear to be subtle differences in the
birth environments as well as the evolutionary histories.

The enhanced abundances of O, Ne, and the $n$-capture elements provide clues regarding the chemical
environment where the progenitors originated and the nucleosynthesis in the inner core of the progenitors.
Intrinsically larger [O/Fe] appears in metal-poor stars, which
is known to be the result of the time delay effects. However, note that
the observed O abundance in our sample is the sum of SN and AGB
nucleosynthesized values. The O abundances in K648 and BoBn1 are
approximately equal. For H4-1, however, after a detail discussion, \citet{Otsuka:2013aa} argued that 0.2-0.3
dex of the observed $\alpha$-elements are SN products.
The C/O ratio of K648 (the C-richness
indicator) is very similar to that of BoBn1.
The similarities of these two elements in both PNe may be
explained as binary evolution and chemical enrichment
during the AGB phase. Due to the O-richness,
the C/O ratio of H4-1 is lower than that for the other
two PNe.

The Ne abundances vary significantly amongst these halo PNe.
The very small Ne abundance for H4-1 indicates that the progenitor of this
PN has no PMZ. A PMZ may have been formed in the He-rich intershells of K648 and
BoBn1. For BoBn1, the Ne enhancement would be due to $^{14}$N in the large PMZ and
in H-burning ashes, and the $^{22}$Ne enhancement via double
$\alpha$-capturing by $^{14}$N. The enhancements of Ne and
N in BoBn1 are similar ([N/H] = +0.20$\pm$0.11 and
[Ne/H] = +0.09$\pm$0.10), although AGB models do not yet reproduce the N and
F overabundances in BoBn1. Nonetheless,
the models successfully explain both the N and F abundances in
K648. K648 therefore is expected to have been born in an similar chemical environment as BoBn1, but
the progenitors experienced different nucleosynthesis.

\citet{Otsuka:2013aa} argued that the abundance of
Xe in H4-1 appears heavily polluted due to the $r$-process in Type II
SNe, whereas the Xe abundance in
BoBn1 is close to the theoretically predicted amount via $s$-processes in
AGB nucleosynthesis; therefore, \citet{2010ApJ...723..658O}
concluded that the Xe in BoBn1 is a product of $s$-processes.
Therefore, the chemical environments where H4-1 and BoBn1 were formed were very different.

\section{Summary}

We have described observations of the PN K648 in M15 and investigated chemical
abundances in the nebula, the CSPN, and dust-based regions, using
multiwavelength data. We determined 10 elemental
abundances for the nebula, including those for F, Cl, and Fe, which
are reported here for K648 for the first time. The F enhancement in K648
is comparable to that for the C-rich halo PN BoBn1. We determined the C and O abundances from
both CELs and RLs. The RL C abundance was consistent with the
CEL value, whereas the RL O abundance was approximately three times larger
than that of the O CELs. We attempted to obtain Ne abundance more accurately by
adding the Ne$^{+}$ abundance determined using the \emph{Spitzer} data. We determined the abundances of He, C, N, O, Ne, P and Fe, as well as the
physical parameters of the CSPN, by employing a spectral synthesis fitting method.
We found that the C/O and Ne/O ratios of the CSPN are roughly consistent with 
 those of the nebula determined from the CELs and RLs within the error. 
The similar C/O ratios might indicate that the nebular abundances are reflective of the most recent stellar wind ejection from the central stellar surface.
\emph{Spitzer}/IRS shows the Class
B 6-9\,$\mu$m and 11.3\,$\mu$m PAHs, as well as the broad 11\,$\mu$m
feature in K648.

We constructed a {\sc Cloudy}  radiative transfer P-I model to investigate physical conditions
of the gas and dust in a self-consistent manner,
and estimated the respective masses. The observed chemical
abundances and core-mass of K648 are in agreement with AGB
nucleosynthesis models for initial 1.25 $M_{\odot}$+PMZ =
2$\times$10$^{-3}$ $M_{\odot}$
stars, as well as initial 1.5 $M_{\odot}$ stars without PMZ.
Our simulation result confirms a possibility that  
K648 had evolved from a star with a mass in the range of
1.25-1.5 $M_{\odot}$. Perhaps the progenitor of K648
experienced coalescence (or a large mass-transfer from its companion)
during the early stages of evolution, and became
a $\sim$1.25-1.5 $M_{\odot}$ blue straggler (BS).
If K648 is a PN that evolved from such a BS in M15, it would be a very rare or the first
such case identified among BS stars in M15, given that the expected number of PNe that
evolved from BSs to date is only 0.135-0.675.

We performed the  analysis  of all observational data available, across
a wide range of instruments and telescopes from the UV to the infrared
for K648 with the help of P-I model construction. Based on our analysis,
we proposed that K648 could be evolved from a BS. The  BS evolution
scenario into a C-rich PN is still at the speculative stage.
A detailed hydrodynamic simulation may help to visualize
the population-based chemical evolution,
or assist in understanding the evolution of the progenitor.
The most appealing scenario for K648 is that the progenitor was
a close binary system that experienced
coalescence or tidal disruption while both stars were in the MS stages and
one emerged as a new star with a mass of
$\leq$1.6 $M_{\odot}$, which then started a new life as the progenitor of
K648. This progenitor passed through the AGB phase stage, and finally
became the presently observable C-rich PN K648.

\section*{Acknowledgments}
This work is largely based on data collected using the Subaru Telescope, which is operated by
the National Astronomical Observatory of Japan (NAOJ). This work also
uses \emph{HST} and \emph{FUSE} archive data downloaded from MAST,
as well as archival data obtained
using the \emph{Spitzer} Space Telescope, which is operated by the
Jet Propulsion Laboratory, California Institute of
Technology, under a contract with NASA. Support for this
work was provided by an award issued by JPL/Caltech.
We thank the anonymous
referee for the helpful comments that make the manuscript more
consistent and readable. MO thanks fruitful discussions with Dr. Francisca Kemper and ICSM group
members in IAA. A part of this work is based on the use of the IAA
clustering computing system. SH would like to acknowledge support from the Basic
Science Research Program through the National
Research Foundation of Korea (2014R1A1A4A01006509).

\appendix
\section{Appendix }
\subsection{ HDS optical spectra}

\setcounter{table}{0}
\renewcommand*\thetable{\Alph{table}}

\begin{deluxetable}{@{}clccrrr|clccrrr@{}}
\tablecolumns{14}
\centering
\tabletypesize{\scriptsize}
\tablecaption{Detected nebular lines and identification of the HDS spectra.\label{hdstab}}
\tablewidth{\textwidth}
\tablehead{
\colhead{$\lambda_{\rm obs}$}&
\colhead{Ion}&
\colhead{$\lambda_{\rm lab}$}&
\colhead{Comp.}&
\colhead{$f$($\lambda$)}&
\colhead{$I$($\lambda$)}&
\colhead{$\delta$$I$($\lambda$)}&
\colhead{$\lambda_{\rm obs}$}&
\colhead{Ion}&
\colhead{$\lambda_{\rm lab}$}&
\colhead{Comp.}&
\colhead{$f$($\lambda$)}&
\colhead{$I$($\lambda$)}&
\colhead{$\delta$$I$($\lambda$)}\\
\colhead{({\AA})}&
\colhead{}&
\colhead{({\AA})}&
\colhead{}&
\colhead{}&
\colhead{}&
\colhead{}&
\colhead{({\AA})}&
\colhead{}&
\colhead{({\AA})}&
\colhead{}&
\colhead{}&
\colhead{}&
\colhead{}
}
\startdata
3655.29 & H37 & 3656.66 & 1 & 0.336 & 0.043 & 0.010 & 4879.13 & [Fe\,{\sc iii}] & 4881.11 & 1 & --0.005 & 0.051 & 0.004 \\
3655.88 & H36 & 3657.27 & 1 & 0.336 & 0.074 & 0.013 & 4920.01 & He\,{\sc i} & 4921.93 & 1 & --0.016 & 1.290 & 0.005 \\
3656.55 & H35 & 3657.92 & 1 & 0.336 & 0.031 & 0.011 & 4929.28 & [O\,{\sc iii}] & 4931.23 & 1 & --0.019 & 0.038 & 0.005 \\
3657.20 & H34 & 3658.64 & 1 & 0.336 & 0.114 & 0.015 & 4956.85 & [O\,{\sc iii}] & 4958.91 & 1 & --0.026 & 26.803 & 4.845 \\
3657.98 & H33 & 3659.42 & 1 & 0.336 & 0.130 & 0.016 & 4957.06 & [O\,{\sc iii}] & 4958.91 & 2 & --0.026 & 48.170 & 5.549 \\
3658.89 & H32 & 3660.28 & 1 & 0.335 & 0.202 & 0.018 & & & & Tot. & & 74.974 & 7.367 \\
3659.77 & H31 & 3661.22 & 1 & 0.335 & 0.190 & 0.012 & 4970.02 & [O\,{\sc iii}] & 4958.91 & 1 & --0.029 & 0.011 & 0.003 \\
3660.84 & H30 & 3662.26 & 1 & 0.335 & 0.275 & 0.023 & 4975.43 & O\,{\sc v}? & 4977.25 & 1 & --0.030 & 0.055 & 0.003 \\
3661.97 & H29 & 3663.40 & 1 & 0.335 & 0.379 & 0.021 & 5004.78 & [O\,{\sc iii}] & 5006.84 & 1 & --0.038 & 74.252 & 5.601 \\
3663.26 & H28 & 3664.68 & 1 & 0.334 & 0.337 & 0.025 & 5004.96 & [O\,{\sc iii}] & 5006.84 & 2 & --0.038 & 153.010 & 7.626 \\
3664.66 & H27 & 3666.10 & 1 & 0.334 & 0.437 & 0.023 & & & & Tot. & & 227.263 & 9.462 \\
3666.27 & H26 & 3667.68 & 1 & 0.334 & 0.366 & 0.023 & 5029.97 & C\,{\sc ii} & 5032.13 & 1 & --0.044 & 0.071 & 0.005 \\
3668.03 & H25 & 3669.46 & 1 & 0.334 & 0.421 & 0.026 & 5033.87 & C\,{\sc ii} & 5035.94 & 1 & --0.045 & 0.035 & 0.004 \\
3670.04 & H24 & 3671.48 & 1 & 0.333 & 0.486 & 0.027 & 5045.78 & He\,{\sc i} & 5047.74 & 1 & --0.048 & 0.187 & 0.003 \\
3672.31 & H23 & 3673.76 & 1 & 0.333 & 0.535 & 0.019 & 5059.64 & N\,{\sc iv}? & 5061.62 & 1 & --0.051 & 0.039 & 0.003 \\
3674.93 & H22 & 3676.36 & 1 & 0.332 & 0.619 & 0.019 & 5119.81 & C\,{\sc ii} & 5121.83 & 1 & --0.065 & 0.033 & 0.004 \\
3677.91 & H21 & 3679.35 & 1 & 0.332 & 0.697 & 0.020 & 5515.51 & [Cl\,{\sc iii}] & 5517.72 & 1 & --0.145 & 0.021 & 0.003 \\
3681.36 & H20 & 3682.81 & 1 & 0.331 & 0.713 & 0.020 & 5535.63 & [Cl\,{\sc iii}] & 5537.89 & 1 & --0.149 & 0.028 & 0.003 \\
3685.38 & H19 & 3686.83 & 1 & 0.330 & 0.840 & 0.029 & 5752.41 & [N\,{\sc ii}] & 5754.64 & 1 & --0.185 & 0.043 & 0.002 \\
3690.11 & H18 & 3691.55 & 1 & 0.329 & 1.001 & 0.028 & 5873.37 & He\,{\sc i} & 5875.62 & 1 & --0.203 & 14.834 & 0.138 \\
3695.70 & H17 & 3697.15 & 1 & 0.328 & 1.216 & 0.026 & 5907.09 & Si\,{\sc i} & 5909.37 & 1 & --0.208 & 0.013 & 0.002 \\
3702.42 & H16 & 3703.85 & 1 & 0.327 & 1.407 & 0.031 & 5977.50 & S\,{\sc ii} & 5979.76 & 1 & --0.218 & 0.016 & 0.001 \\
3703.59 & He\,{\sc i} & 3705.14 & 1 & 0.327 & 0.722 & 0.026 & 6148.93 & C\,{\sc ii} & 6151.27 & 1 & --0.242 & 0.045 & 0.003 \\
3710.52 & H15 & 3711.97 & 1 & 0.325 & 1.534 & 0.027 & 6234.73 & C\,{\sc i} & 6237.23 & 1 & --0.254 & 0.022 & 0.001 \\
3720.48 & H14 & 3721.94 & 1 & 0.323 & 1.937 & 0.033 & 6236.13 & Fe\,{\sc ii} & 6238.39 & 1 & --0.254 & 0.014 & 0.001 \\
3724.63 & [O\,{\sc ii}] & 3726.03 & 1 & 0.322 & 7.729 & 0.156 & 6236.39 & Ne\,{\sc ii} & 6238.92 & 1 & --0.254 & 0.023 & 0.001 \\
3724.61 & [O\,{\sc ii}] & 3726.03 & 2 & 0.322 & 9.654 & 0.255 & 6257.02 & C\,{\sc ii} & 6259.56 & 1 & --0.257 & 0.009 & 0.003 \\
 & & & Tot. & & 17.383 & 0.299 & 6257.37 & C\,{\sc ii} & 6259.56 & 1 & --0.257 & 0.013 & 0.001 \\
3727.40 & [O\,{\sc ii}] & 3728.81 & 1 & 0.322 & 3.047 & 0.166 & 6297.90 & [O\,{\sc i}] & 6300.30 & 1 & --0.263 & 0.225 & 0.008 \\
3727.36 & [O\,{\sc ii}] & 3728.81 & 2 & 0.322 & 6.340 & 0.309 & 6298.44 & [O\,{\sc i}] & 6300.30 & 2 & --0.263 & 0.041 & 0.007 \\
 & & & Tot. & & 9.387 & 0.350 & & & & Tot. & & 0.266 & 0.011 \\
3732.90 & H13 & 3734.37 & 1 & 0.321 & 2.428 & 0.040 & 6309.60 & [S\,{\sc iii}] & 6313.10 & 1 & --0.264 & 0.119 & 0.005 \\
3748.69 & H12 & 3750.15 & 1 & 0.317 & 2.968 & 0.043 & 6361.31 & [O\,{\sc i}] & 6363.78 & 1 & --0.271 & 0.068 & 0.003 \\
3769.16 & H11 & 3770.63 & 1 & 0.313 & 3.934 & 0.052 & 6459.20 & C\,{\sc ii} & 6462.04 & 1 & --0.284 & 0.076 & 0.007 \\
3796.41 & H10 & 3797.90 & 1 & 0.307 & 5.291 & 0.067 & 6459.61 & C\,{\sc ii} & 6462.04 & 2 & --0.284 & 0.022 & 0.005 \\
3818.13 & He\,{\sc i} & 3819.60 & 1 & 0.302 & 0.911 & 0.015 & & & & Tot. & & 0.098 & 0.009 \\
3833.89 & H9 & 3835.38 & 1 & 0.299 & 7.105 & 0.089 & 6522.88 & Ne\,{\sc ii} & 6525.59 & 1 & --0.293 & 0.014 & 0.001 \\
3867.22 & [Ne\,{\sc iii}] & 3869.06 & 1 & 0.291 & 9.939 & 0.124 & 6545.56 & [N\,{\sc ii}] & 6548.04 & 1 & --0.296 & 0.901 & 0.012 \\
3887.32 & H8 & 3889.05 & 1 & 0.286 & 22.451 & 0.456 & 6559.94 & H3 & 6562.82 & 1 & --0.298 & 119.758 & 3.906 \\
3917.39 & C\,{\sc ii} & 3918.97 & 1 & 0.279 & 0.078 & 0.011 & 6560.45 & H3 & 6562.82 & 2 & --0.298 & 162.641 & 4.167 \\
3919.12 & C\,{\sc ii} & 3920.68 & 1 & 0.279 & 0.167 & 0.012 & & & & Tot. & & 282.399 & 5.712 \\
3925.02 & He\,{\sc i} & 3926.54 & 1 & 0.277 & 0.137 & 0.013 & 6575.48 & C\,{\sc ii} & 6578.05 & 1 & --0.300 & 0.692 & 0.011 \\
3963.20 & He\,{\sc i} & 3964.73 & 1 & 0.267 & 0.780 & 0.015 & 6580.77 & [N\,{\sc ii}] & 6583.46 & 1 & --0.300 & 1.488 & 0.032 \\
3965.89 & [Ne\,{\sc iii}] & 3967.79 & 1 & 0.267 & 3.147 & 0.043 & 6580.93 & [N\,{\sc ii}] & 6583.46 & 2 & --0.300 & 1.692 & 0.025 \\
3968.60 & H7 & 3970.07 & 1 & 0.266 & 10.703 & 0.167 & & & & Tot. & & 3.180 & 0.041 \\
3971.08 & C\,{\sc ii} & 3972.45 & 1 & 0.265 & 0.144 & 0.015 & 6604.63 & Ne\,{\sc ii} & 6607.40 & 1 & --0.303 & 0.023 & 0.003 \\
3997.69 & C\,{\sc iii} & 3999.64 & 1 & 0.258 & 0.035 & 0.008 & 6628.07 & O\,{\sc iv} & 6630.70 & 1 & --0.307 & 0.010 & 0.001 \\
4007.72 & He\,{\sc i} & 4009.26 & 1 & 0.256 & 0.191 & 0.014 & 6675.56 & He\,{\sc i} & 6678.15 & 1 & --0.313 & 4.114 & 0.055 \\
4024.64 & He\,{\sc i} & 4026.18 & 1 & 0.251 & 1.958 & 0.024 & 6712.28 & N\,{\sc ii} & 6714.99 & 1 & --0.318 & 0.032 & 0.003 \\
4074.57 & [S\,{\sc ii}] & 4076.35 & 1 & 0.237 & 0.111 & 0.031 & 6713.90 & [S\,{\sc ii}] & 6716.44 & 1 & --0.318 & 0.087 & 0.002 \\
4079.83 & O\,{\sc iii} & 4081.00 & 1 & 0.235 & 0.034 & 0.008 & 6724.89 & C\,{\sc iii} & 6727.48 & 1 & --0.319 & 0.037 & 0.003 \\
4100.13 & H6 & 4101.73 & 1 & 0.230 & 26.339 & 0.248 & 6728.30 & [S\,{\sc ii}] & 6730.81 & 1 & --0.320 & 0.133 & 0.003 \\
4119.23 & He\,{\sc i} & 4120.81 & 1 & 0.224 & 0.202 & 0.013 & 6731.54 & He\,{\sc i} & 6734.08 & 1 & --0.320 & 0.022 & 0.003 \\
4142.10 & He\,{\sc i} & 4143.76 & 1 & 0.217 & 0.156 & 0.006 & 6739.57 & C\,{\sc iii} & 6742.15 & 1 & --0.321 & 0.041 & 0.004 \\
4265.50 & C\,{\sc ii} & 4267.18 & 1 & 0.180 & 0.660 & 0.016 & 6741.76 & C\,{\sc iii} & 6744.39 & 1 & --0.322 & 0.061 & 0.003 \\
4265.51 & C\,{\sc ii} & 4267.18 & 1 & 0.180 & 0.726 & 0.013 & 6777.39 & C\,{\sc ii} & 6780.60 & 1 & --0.326 & 0.059 & 0.003 \\
4338.77 & H5 & 4340.46 & 1 & 0.157 & 46.674 & 0.307 & 6798.18 & C\,{\sc ii} & 6800.68 & 1 & --0.329 & 0.052 & 0.006 \\
4361.49 & [O\,{\sc iii}] & 4363.21 & 1 & 0.149 & 2.782 & 0.026 & 6931.26 & He\,{\sc i} & 6933.89 & 1 & --0.347 & 0.053 & 0.003 \\
4386.21 & He\,{\sc i} & 4387.93 & 1 & 0.142 & 0.517 & 0.010 & 7034.62 & C\,{\sc iii} & 7037.25 & 1 & --0.361 & 0.075 & 0.003 \\
4435.82 & He\,{\sc i} & 4437.55 & 1 & 0.126 & 0.088 & 0.013 & 7059.60 & He\,{\sc i} & 7062.28 & 1 & --0.364 & 0.026 & 0.003 \\
4469.76 & He\,{\sc i} & 4471.47 & 1 & 0.115 & 4.914 & 0.028 & 7062.49 & He\,{\sc i} & 7065.18 & 1 & --0.364 & 5.909 & 0.100 \\
4636.98 & O\,{\sc ii} & 4638.86 & 1 & 0.064 & 0.055 & 0.006 & 7092.65 & Si\,{\sc i} & 7095.49 & 1 & --0.368 & 0.015 & 0.005 \\
4640.01 & O\,{\sc ii} & 4641.81 & 1 & 0.063 & 0.033 & 0.003 & 7096.05 & N\,{\sc iv}? & 7098.60 & 1 & --0.369 & 0.020 & 0.002 \\
4701.53 & [Fe\,{\sc iii}] & 4701.53 & 1 & 0.045 & 0.022 & 0.003 & 7132.99 & [Ar\,{\sc iii}] & 7135.70 & 1 & --0.374 & 0.384 & 0.006 \\
4706.34 & N\,{\sc ii} & 4708.28 & 1 & 0.043 & 0.030 & 0.003 & 7278.53 & He\,{\sc i} & 7281.39 & 1 & --0.393 & 0.598 & 0.008 \\
4711.34 & He\,{\sc i} & 4713.14 & 1 & 0.042 & 0.672 & 0.005 & 7316.27 & [O\,{\sc ii}] & 7319.14 & 1 & --0.398 & 0.452 & 0.020 \\
4787.73 & [F\,{\sc ii}] & 4789.45 & 1 & 0.020 & 0.110 & 0.004 & 7317.32 & [O\,{\sc ii}] & 7320.19 & 1 & --0.398 & 1.340 & 0.015 \\
4859.43 & H4 & 4861.33 & 1 & 0.000 & 100.000 & 0.188 & 7326.89 & [O\,{\sc ii}] & 7329.76 & 1 & --0.400 & 0.745 & 0.015 \\
4867.28 & [F\,{\sc ii}] & 4868.99 & 1 & --0.002 & 0.030 & 0.003 & 7327.95 & [O\,{\sc ii}] & 7330.82 & 1 & --0.400 & 0.705 & 0.015
\enddata
\end{deluxetable}

\end{document}